# A Panchromatic Study of Massive Stars in the Extremely Metal-poor Local Group Dwarf Galaxy Leo A[*]

Maude Gull[1], Daniel R. Weisz[1], Peter Senchyna[2], Nathan R. Sandford[1], Yumi Choi[1,3], Anna F. McLeod[4,5], Kareem El-Badry[1], Ylva Götberg[2,18], Karoline M. Gilbert[3,6], Martha Boyer[3], Julianne J. Dalcanton[7,8], Puragra GuhaThakurta[9], Steven Goldman[10], Paola Marigo[11], Kristen B. W. McQuinn[12], Giada Pastorelli[3,13], Daniel P. Stark[14], Evan Skillman[15], Yuan-sen Ting[16], and Benjamin F. Williams[17]

[1] Department of Astronomy, University of California Berkeley, CA 94720, USA
[2] Observatories of the Carnegie Institution of Washington, 813 Santa Barbara Street, Pasadena, CA 91101, USA
[3] Space Telescope Science Institute, 3700 San Martin Drive, Baltimore, MD 21218, USA
[4] Centre for Extragalactic Astronomy, Department of Physics, Durham University, South Road, Durham DH1 3LE, UK
[5] Institute for Computational Cosmology, Department of Physics, University of Durham, South Road, Durham DH1 3LE, UK
[6] The William H. Miller III Department of Physics & Astronomy, Bloomberg Center for Physics and Astronomy, Johns Hopkins University, 3400 N. Charles Street, Baltimore, MD 21218, USA
[7] Center for Computational Astrophysics, Flatiron Institute, 162 Fifth Avenue, New York, NY 10010, USA
[8] Department of Astronomy, University of Washington, Box 351580, Seattle, WA 98195-1580, USA
[9] UCO/Lick Observatory and Department of Astronomy and Astrophysics, University of California, Santa Cruz, CA 95064, USA
[10] SOFIA Science Center, Universities Space Research Association, NASA Ames Research Center, M.S. N232-12, Moffett Field, CA 94035, USA
[11] Department of Physics and Astronomy, University of Padova, I-35122-Padova, Italy
[12] Rutgers University, Department of Physics and Astronomy, 136 Frelinghuysen Road, Piscataway, NJ 08854, USA
[13] INAF—Osservatorio Astronomico di Padova, Vicolo dell'Osservatorio 5, I-35122 Padova, Italy
[14] Steward Observatory, University of Arizona, 933 N Cherry Ave, Tucson, AZ 85721, USA
[15] Minnesota Institute for Astrophysics, University of Minnesota, 116 Church Street SE, Minneapolis, MN 55455, USA
[16] Research School of Astronomy & Astrophysics, Australian National University, Cotter Road, Weston, ACT 2611, Australia
[17] Department of Astronomy, University of Washington, Box 351580, U.W., Seattle, WA 98195-1580, USA

## Abstract

We characterize massive stars ($M > 8\ M_\odot$) in the nearby ($D \sim 0.8$ Mpc) extremely metal-poor ($Z \sim 5\%\ Z_\odot$) galaxy Leo A using Hubble Space Telescope ultraviolet (UV), optical, and near-infrared (NIR) imaging along with Keck/ Low-Resolution Imaging Spectrograph and MMT/Binospec optical spectroscopy for 18 main-sequence OB stars. We find that: (a) 12 of our 18 stars show emission lines, despite not being associated with an H II region, suggestive of stellar activity (e.g., mass loss, accretion, binary star interaction), which is consistent with previous predictions of enhanced activity at low metallicity; (b) six are Be stars, which are the first to be spectroscopically studied at such low metallicity—these Be stars have unusual panchromatic SEDs; (c) for stars well fit by the TLUSTY nonlocal thermodynamic equilibrium models, the photometric and spectroscopic values of $\log(T_{\rm eff})$ and $\log(g)$ agree to within ~0.01 dex and ~0.18 dex, respectively, indicating that near-UV/optical/NIR imaging can be used to reliably characterize massive ($M \sim 8$–30 $M_\odot$) main-sequence star properties relative to optical spectroscopy; (d) the properties of the massive stars in H II regions are consistent with constraints from previous nebular emission line studies; and (e) 13 stars with $M > 8 M_\odot$ are >40 pc from a known star cluster or H II region. Our sample comprises ~50% of all known massive stars at $Z \lesssim 10\%\ Z_\odot$ with derived stellar parameters, high-quality optical spectra, and panchromatic photometry.

*Unified Astronomy Thesaurus concepts:* Massive stars (732); Metallicity (1031); Early-type stars (430)

## 1. Introduction

Metal-poor ($Z \lesssim 10\%\ Z_\odot$) massive stars ($M \gtrsim 8\ M_\odot$) are central to a wide variety of astrophysics. In low-mass galaxies, their feedback (i.e., stellar winds, supernovae) regulates star formation, enriches the interstellar medium, and may drive the formation of dark matter cores (e.g., Mac Low & Ferrara 1999; Stinson et al. 2007; Governato et al. 2010; El-Badry et al. 2016; Emerick et al. 2018; Eldridge & Stanway 2022). At low metallicity, those processes are expected to change significantly in comparison to higher-metallicity environments (e.g., decreased line-driven winds, properties of supernovae; Heger et al. 2003; Eldridge & Tout 2004; Young et al. 2008; Smartt 2009; Vink & Sander 2021) due to changes in stellar properties (e.g., rotation, opacity, internal mixing; Maeder et al. 2005; Meynet et al. 2008, 2017; Sanyal et al. 2017; Groh et al. 2019). For example, in the early universe, ionizing photons from early generations of massive metal-poor stars may power cosmic reionization, since they experience very high effective temperatures over a sizable fraction of their lives, resulting in an extreme ionizing ultraviolet (UV) radiation field (e.g., Wang 2013; Madau & Dickinson 2014; Stark et al. 2014; Robertson et al. 2015; Robertson 2021). At all redshifts, single and binary massive metal-poor stars are thought to be the progenitors to many explosive transients, including gravitational wave sources (e.g., Smartt 2009; Smith 2014; Spera et al. 2015).

Due to their close proximity, which enables resolved star studies, nearby star-forming dwarf galaxies are among the best targets for studies of individual massive stars and their environments at extremely low metallicities (e.g., Dalcanton et al. 2009; Hunter et al. 2012; Ott et al. 2012). However, despite substantial observational investments in resolving gas and stars in nearby dwarf galaxies, our detailed knowledge of their resolved massive star content has been limited. Many massive stars are hot ($\gtrsim$15,000 K) meaning their spectral energy distributions peak at UV wavelengths. Massive stars are not always the brightest, most easily identifiable stars in the optical (e.g., Massey 2003), which limits the utility of the existing optical imaging. Older Hubble Space Telescope (HST) UV imaging (e.g., with WFPC2) exists for some local dwarf galaxies, but the data quality is modest relative to the optical, and the resulting studies are generally limited to the hottest and/or brightest objects (e.g., Bianchi et al. 2012). Other UV imaging surveys focus on the ensemble population of star-forming regions and young clusters as hosts of massive stars, but are less focused on the properties of the individual stars (e.g., LEGUS; Calzetti et al. 2015).

There is a similar paucity of spectra for massive metal-poor stars. While high-resolution optical and UV spectra of massive stars has improved our understanding of massive stars in the metal-rich environments (e.g., Bianchi & Garcia 2002; Bouret et al. 2003; Heap et al. 2006; Sota et al. 2011, 2014; Barbá et al. 2017; Russeil et al. 2017; Li 2021), data of similar quality is less abundant at lower metallicities. Much of our understanding of metal-poor massive stars is based primarily on studies of the Small Magellanic Cloud (SMC, $Z \sim 1/5Z_\odot$; Martins et al. 2004; Castro et al. 2018; Ramachandran et al. 2019). However, the metallicity of the SMC does not probe the extremely metal-poor regime. One example of the need for extremely low-metallicity spectra is the need to calibrate the relative strengths of He I absorption lines that underlie He I emission lines in nebular spectra. This is an important correction for the determination of the primordial helium abundance and currently relies entirely on simulated spectra (e.g., Aver et al. 2021). Existing stellar models for lower metallicities predict significant differences in evolutionary pathways with implications for time-integrated feedback, final fates, and compact remnants (e.g., Meynet & Maeder 2002; Eldridge et al. 2008; Chen et al. 2015; Costa et al. 2021). Lastly, spectral atmospheric models for massive metal-poor stars are currently calibrated from higher-metallicity environments. However, at lower metallicities, models predict, for example, more emission features due to increased stellar activity and deeper absorption features due to the increased effective temperature for a given spectral type, or broader absorption features at a given spectral type (e.g., Kubátová et al. 2019; Martins & Palacios 2021).

Historically, the study of metal-poor massive stars has been challenging, since most star-forming metal-poor dwarf galaxies are located at distances of $\gtrsim 1$ Mpc, meaning that even the most luminous massive stars are often faint and crowded. The limiting factors above have typically confined the study of low-metallicity massive stars to a small number of stars in the closest dwarf galaxies. While providing important insights into chemical evolution and massive stellar physics, the challenge of acquiring these data limited the sample sizes and homogeneity required for systematic progress in understanding metal-poor massive star evolution and how these stars impact their surroundings. Several recent efforts have sought to improve this situation. For example, only recently has there been a large survey published of $\sim$150 low-resolution spectra ($R \sim 1000$, signal-to-noise ratio, S/N, of $\sim$8–256 with a majority S/N $\sim$ 30–100; Lorenzo et al. 2022). The study currently only provides spectra and spectral type identifications, but no stellar parameters. This is currently the only study of its kind at such low metallicities. Other than this low resolution, optical spectroscopy is only available for a handful of massive stars in several sub-SMC dwarf galaxies within $\sim$1.5 Mpc (e.g., Venn et al. 2004; Bresolin et al. 2007; Kudritzki et al. 2008; Hosek et al. 2014; Tramper et al. 2014; Berger et al. 2018; Evans et al. 2019; Garcia et al. 2019b; Telford et al. 2021b). In these same systems, there are also concerted efforts to obtain UV spectroscopy of O-type stars with HST (e.g., García et al. 2015; Telford et al. 2021a; Wofford et al. 2021), including a dozen stars as part of the Hubble UV Legacy Library of Young Stars as Essential Standards program (e.g., Roman-Duval et al. 2020, 2021).

Motivated by the need for a panchromatic photometric and high-quality spectroscopic inventory, we use Leo A as a test-bed galaxy. We pursue the following aims: (a) to characterize the population of massive stars in Leo A and (b) test the reliability of photometric-only techniques in this metallicity regime. The latter is especially important as there are few extremely metal-poor systems in which individual massive stars can be resolved and high-quality spectra can be acquired. Leo A is an ideal test bed as it is the closest of the star-forming extremely metal-poor dwarf galaxies known in the local universe ($D \sim 760$ kpc, $Z \sim 5\%$ $Z_\odot$; e.g., van Zee et al. 2006; Nagarajan et al. 2022). Despite its favorable distance, low metallicity, and sizable massive star population (e.g., Cole et al. 2007), there has not yet been a systematic investigation of massive metal-poor star properties in Leo A. Existing studies that focus on Leo A's young stars do so in the context of H II region physics or as dynamical tracers of dark matter (e.g., Skillman et al. 1989; van Zee et al. 2006; Brown et al. 2007; Cole et al. 2007; Ruiz-Escobedo et al. 2018; Leščinskaitė et al. 2022; Ricotti et al. 2022; Stonkutė & Vansevičius 2022).

Recently, Leo A was observed as part of the HST Local Ultra Violet Infrared Treasury program (LUVIT; GO-15275, PI Gilbert and GO-16162, PI Boyer). This survey has obtained new HST UV, optical, and near-IR (NIR) imaging with HST for 19 nearby dwarf galaxies in GO-16162 and 21 nearby dwarf galaxies in GO-15275 (the overlap between the two programs consists of 19 galaxies). Combined with archival HST optical imaging, it provides broad information about the spectral energy distribution (SED) for a variety metal-poor stars across the H–R diagram.

Using LUVIT imaging for target selection, we acquired optical spectroscopy with the Low-Resolution Imaging Spectrograph (LRIS) on Keck and Binospec on MMT for two dozen of the most UV luminous stars in Leo A. This is one of the largest uniform collections of high-quality optical spectroscopy for extremely low-metallicity massive stars. Between the panchromatic HST imaging and the high-fidelity optical spectra, we undertake here a systematic study of hot, massive, metal-poor stars in Leo A and their relationship to the broader galactic environment.

This paper is organized as follows. In Section 2, we describe the multiwavelength HST observations and data reduction, along with target selection, observations, and data reduction for the Keck/LRIS and MMT/Binospec optical spectroscopy. In Section 3, we detail the process of measuring stellar properties from the HST-based SEDs. In Section 4, we expand on how we measure the stellar properties from modeling the optical spectra, and from the conventional approach of measuring equivalent widths (EWs). We present the results of each method in Section 5 and present a detailed discussion of the spectrum for each star along with a comparison of the photometric and spectroscopic parameters we measure. In Section 6, we discuss our results in the context of nebular studies of H II regions, field OB-stars, Be stars, and stars in the extremely metal-poor environments, as well as a comparison of SED fitting to full spectral fitting.

## 2. Sample Selection, Observations, and Data Reduction

### 2.1. HST Observations and Photometry

We make use of HST UV, optical, and NIR broadband imaging for selecting the spectroscopic sample and for measuring stellar properties based on HST photometry. Deep optical imaging was acquired in the Advanced Camera for Surveys (ACS)/ Wide Field Channel (WFC) F475W and F814W filters as part of GO-10590 (Cole et al. 2007). WFC3/ UVIS F275W and F336W and WFC3/IR F110W and F160W imaging was collected as part of LUVIT. We note that the ACS/WFC imaging has the largest spatial coverage of Leo A, followed by WFC3/UVIS imaging, while WFC3/IR imaging covers the smallest area. We only focus on stars for which we have at least 4-band UV+optical imaging, and include NIR when possible.

The LUVIT data are reduced by utilizing the photometric reduction pipelines that have been developed and improved for HST simultaneous N-band photometry (Williams et al. 2014, 2021). We refer the readers to the LUVIT survey paper (K. M. Gilbert 2022, in preparation) for the details of data collection and reduction for the entire LUVIT sample. Here we provide some of the Leo A specific details. We first aligned all `flc` (ACS/WFC, WFC3/UVIS) and `flt` (WFC3/IR) images to the Gaia DR2 astrometric solution following Bajaj (2017)[19] and combined images in each band weighted by exposure time using `AstroDrizzle` (Hack et al. 2020) after flagging and masking bad pixels and cosmic rays on input exposures. These aligned images, as well as the individual exposures with updated data quality extensions, are processed via the DOL-PHOT package (Dolphin 2000, 2016) for photometry. We use the combined full-depth F475W image as the astrometric reference image for our final photometry.

DOLPHOT reports measurements of point-spread function photometry for every object detected on any exposures within the reference image footprint. These include the flux, flux uncertainty, magnitude in the Vega system, S/N, and various photometry quality metrics. A raw output catalog of all objects identified by DOLPHOT is then subjected to quality cuts in each filter to include objects: (1) with S/N ⩾ 4; (2) with crowding < 1.3 and < 2.25 for UV and optical/IR filters, respectively; and (3) with square of sharpness (SHARP$^2$) < 0.15 and < 0.2 for UV/IR and optical filters, respectively, producing the "GST" catalog. These quality cuts were applied in each band independently. To remove objects that are associated with diffraction spikes after performing the GST cuts, we first identified individual tiny regions with severe diffraction spike features using the median filter technique and then culled objects belonging to those regions (K. M. Gilbert 2022, in preparation). To construct an initial spectroscopic target sample, we further required the S/N in F475W and F814W to be >15 and F475W$^2_{SHARP}$ + F814W$^2_{SHARP}$ to be ⩽ 0.3. Additional criteria for the final spectroscopic target selection are presented in Section 2.2. Even though the photometric depth reaches a couple of magnitudes fainter than the old main-sequence turnoff, we limited our stellar SED modeling to the bright targets of our interest. We note that we did not use the NIR medium-band photometry (F127M, F139M, and F153M) in this study, since it does not provide any further significant constraints in this stellar regime.

Photometric accuracy, precision, and completeness were computed from artificial star tests (ASTs). We generated ~135,000 dust-extinguished SED models in our UV, optical, and NIR bands that cover the entire H–R diagram, but more heavily populate the observed color and magnitude ranges (Choi et al. 2020). We assigned more input ASTs in higher stellar density regions of Leo A to better capture nonlinear trends in photometric characteristics as a function of source crowding. Photometric measurements for input artificial stars were made in the exact same way as we did for real objects, and then identical quality cuts were applied to determine if the injected artificial stars were recovered. Comparisons of the input and output magnitudes are used to quantify our photometry uncertainty. The 50% completeness limits for F275W, F336W, F475W, F814W, F110W, and F160W are 24.97, 25.89, 29.10, 27.89, 26.58, and 25.27, respectively.

Figure 1 shows select UV, optical, and NIR color–magnitude diagrams (CMDs) of Leo A, with the spectroscopic sample highlighted, as described in Section 2.2. The NIR CMD contains a smaller sample because the NIR imaging field of view covers a smaller spatial area of Leo A. The CMDs are close to 50% or above the completeness limit. We overplot the rotating (v/vcrit = 0.4) MESA Isochrones and Stellar Tracks (MIST; Choi et al. 2016; Dotter 2016) for reference. We shift the tracks to the Gaia eDR3-anchored distance modulus of $\mu$ = 24.40 mag reported by Nagarajan et al. (2022) and $A_v$ = 0.03 mag, consistent with the foreground maps of Schlafly & Finkbeiner (2011). We do not apply any further dust corrections, since Leo A shows little internal reddening (Skillman et al. 1989; van Zee et al. 2006).

### 2.2. Spectroscopic Target Selection

We observed Leo A for one night using Keck/LRIS in the 2020A semester. To fill the slit mask, we prioritized the stars brightest in the HST F336W − F475W CMD (Figure 1). We were able to place nine UV bright stars on the LRIS slit mask. This number was limited by crowding, LRIS slit mask alignment constraints, and science goals. The remaining slits were filled with filler stars (i.e., bright stars in ground-based photometric observations of Leo A). To achieve a sufficient S/N (see below), we integrated for a full night on this single slit mask.

We also observed Leo A on MMT/Binospec in the 2020A semester. Targets were selected using the Binospec mask software with following criteria as input: F475W − F814W ⩽ − 0.1, F275W ⩽ 20, and −0.8 ⩽ F275W − F336W ⩽ − 0.2, thereby selecting the bluest and brightest sources.

The two spectroscopic data sets have stars in common, which allows us to check for systematic changes and gauge consistency. Figure 1 shows the LRIS and Binospec spectroscopic samples overplotted on the HST CMDs. The majority of the brightest UV stars have spectroscopic coverage from at least one of the telescopes. The handful that do not were excluded due to slit mask and/or crowding constraints. We plan to target them in a future observing run.

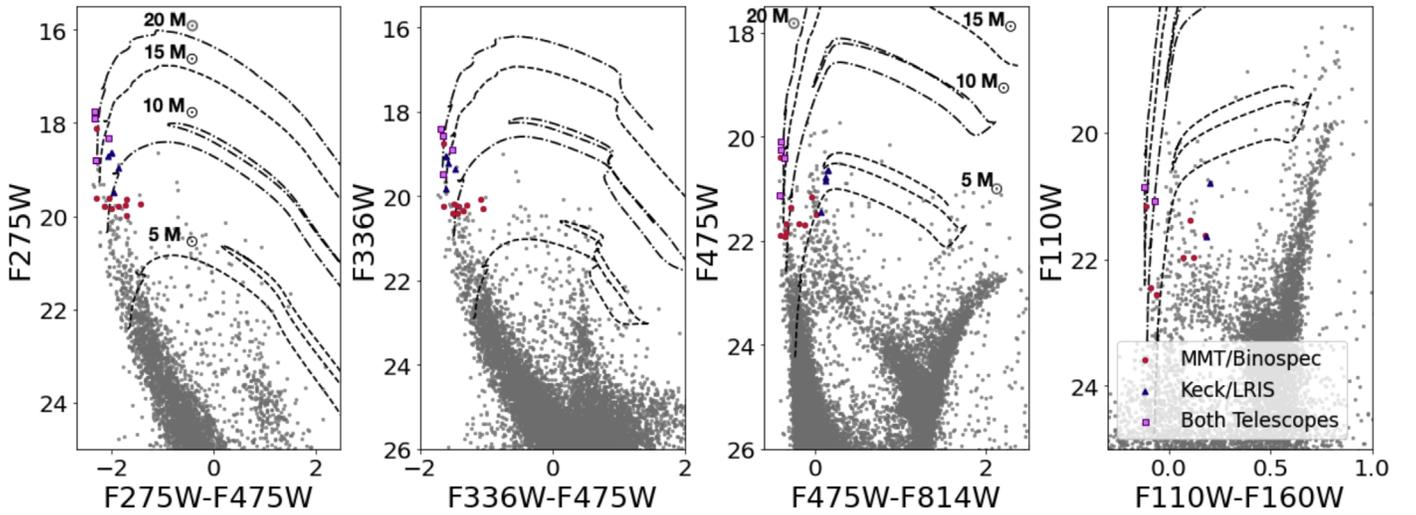

**Figure 1.** Select HST-based UV-optical (left), near-UV (NUV)-optical (left middle), optical-only (right middle), and NIR (right) CMDs of Leo A. Colored points indicate stars observed with LRIS (Keck; navy triangles), Binospec (MMT; red dots), and both telescopes (magenta squares). For reference, we overplot select MIST tracks ([Fe/H] = −1.0, v/vcrit = 0.4, $A_v$ = 0.03). There are fewer stars on the NIR CMD because WFC3/IR has a smaller field of view and does not entirely overlap the UV and optical imaging. All spectroscopic stars are massive stars (i.e., $M > 8M_\odot$). The NUV CMD shows that the stars in our spectroscopic sample are not very evolved.

Figure 2 shows zoomed-in CMDs that highlight the spectroscopic sample. We label them based on the origin of the spectra used for analysis (i.e., Keck stars: K, Binospec stars: B) and in decreasing F475W magnitude. In the UV-optical CMD (left panel), we see that the stars appear to have masses of at least 8 $M_\odot$, since they pool around the 10 $M_\odot$ track. In the optical CMD (right panel), the picture is not as clear as in the UV-optical CMD. Interestingly, a handful of stars (K4, K5, K6, K8, B5, and B8) visually appear to be main-sequence (MS) stars in the UV-optical CMD, but appear on the blue core helium burning (BHeB) sequence in the optical CMD. We discuss these stars in detail in Section 5.3. In total, we have spectroscopic observations for 18 stars. Figure 3 shows the spatial position of the spectroscopic sample overplotted on the HST ACS/F475W image.Observational properties of these stars are listed in Table 1. We now describe the details of each spectroscopic data set.

### 2.2.1. LRIS and Binospec Observation

We observed Leo A on the night of 2020 February 20 using LRIS (Oke et al. 1995) at the Keck telescope at W.M. Keck Observatory in Hawaii, and 2020 February 23, 24, and 29 using the Binospec instrument, mounted on the 6.5 m MMT telescope. For LRIS, we obtained spectra using a custom slit-mask. The gratings for the LRIS red channel were set to 1200/7500, covering 7224−8855 Å. The grism for the LRIS blue channel was set to 600/4000, covering 3330−5910 Å, and the dichroic was set to D680, yielding a nominal resolution of $R \sim 1800$. For Binospec, we observed two nights using the 1000 lpmm grating, providing spectra from 3900−5400 Å at a nominal resolution of $R \sim 3600$. The last night, we used the 270 lpmm grating and obtained spectra spanning 3900−9240 Å, with a resolution of $R \sim 1250$.

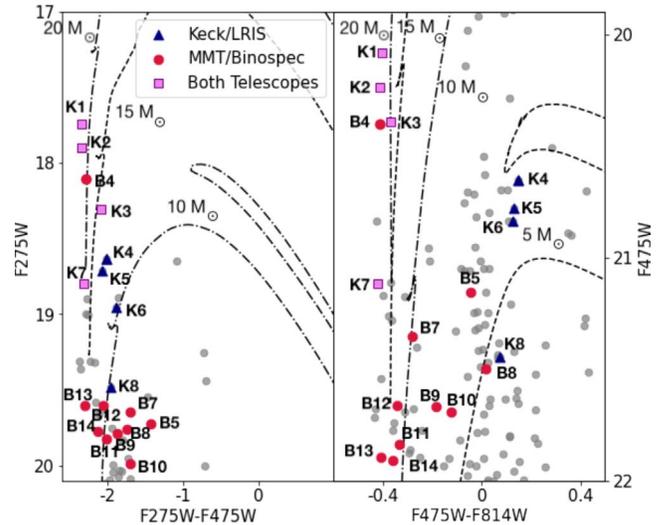

**Figure 2.** This plot is identical to Figure 1, only zoomed in on luminous blue stars selected for spectroscopy in the UV and optical CMDs. We label the stars by the names in Table 1, where K stands for stars with Keck and B for stars with Binospec spectra. For reference, we overplot select MIST tracks ([Fe/H] = −1.0, v/vcrit = 0.4, $A_v$ = 0.03). Interestingly, a handful of stars that appear on the main sequence in the UV-optical CMDs appear in the blue core helium burning region of the F475W-F814W CMD. We discuss this in Section 5.3.

For LRIS, we integrated for a total of 6.3 hr using multiple 1200–1800 s exposures. We reduced the spectra using the Pypelt pipeline (Prochaska et al. 2019) and then coadded each single epoch spectrum. The resulting S/Ns range from ∼30 to ∼83 per Å at ∼5000 Å and ∼5 to ∼17 per Å at ∼8000 Å. For Binospec, we integrated for a total of 2 hr on night 1, 1 hr on night 2, and 1 hr on night 3, in intervals of 1200 s exposures. We observed the same stars all three nights. We reduced the spectra with the Binospec Data Reduction Pipeline (Kansky et al. 2019). We coadded single-exposure spectra for each star from the first two nights. The coadded spectra from the first two nights have S/Ns ranging from ∼9 to

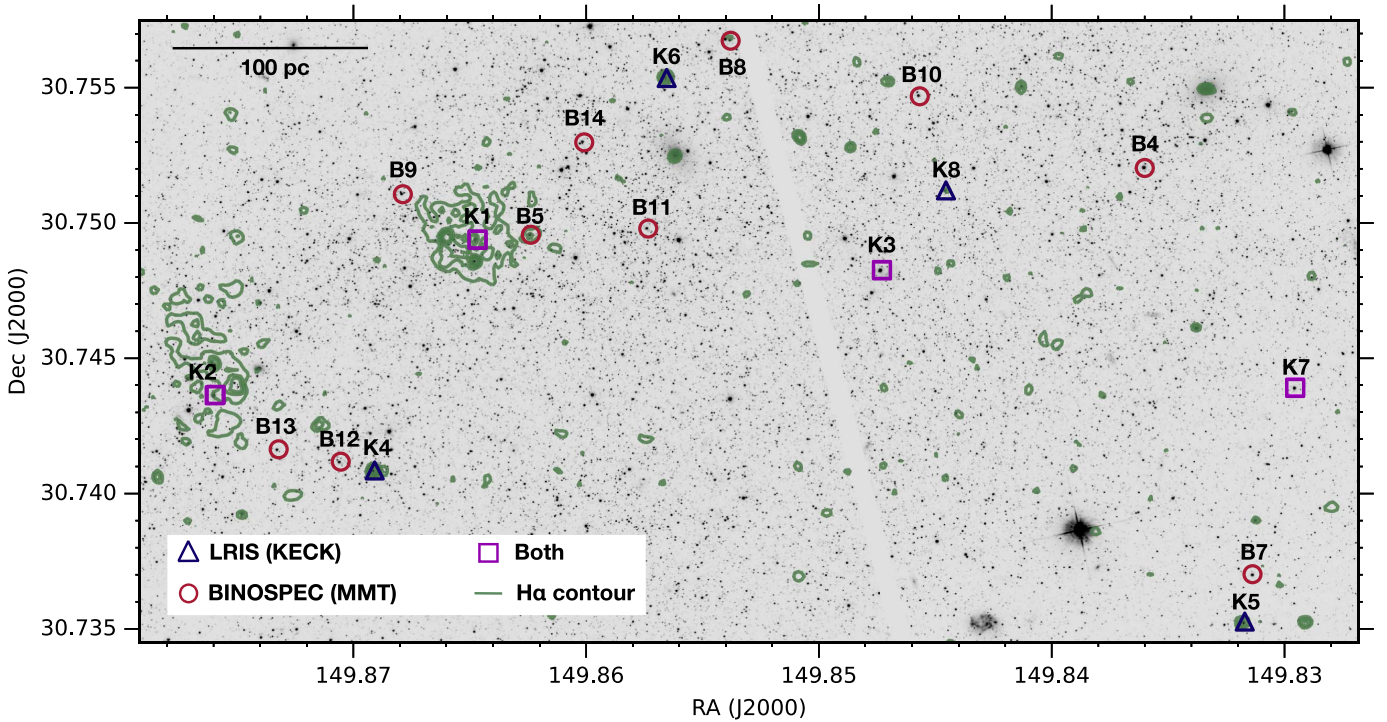

**Figure 3.** The Leo A HST ACS/F475W image zoomed in on the central region of the galaxy. Stars with spectroscopic observations are color-coded by telescope: Keck/LRIS (blue triangles), MMT/Binospec (red circles), and both (magenta squares). Contours of Hα emission from ground-based imaging (Local Volume Legacy Survey; Kennicutt et al. 2007) are overplotted in green. Note that many of the massive stars are not obviously associated with cataloged H II regions including one O-star (K7).

∼31 per Å at ∼5000 Å. Spectra from the third night have S/Ns ranging from ∼11 to ∼40 per Å.

Table 1 lists the basic properties of our LRIS and Binospec targets. As noted in Table 1, four stars were observed with both Binospec and LRIS: K1/B1, K2/B2, K3/B3, and K7/B6. For simplicity, we refer to the Keck stars as K1-K8. We note that a second star fell into the same slit K2; however, K2 is unaffected by this. This was the only slit to do so, so we do not include the second star in the sample.

### 2.3. Spectroscopic Normalization

We normalize all spectroscopic observations using PySpecKit (Ginsburg & Mirocha 2011). We exclude the parts of the spectrum showing strong absorption or emission features. We normalize all of our spectra using a fifth-order polynomial, in order to avoid over fitting. We show examples of normalized sample spectra with a detailed indication of spectral features in Figure 4.

## 3. Measuring Stellar Parameters from Panchromatic Photometry

We use the Bayesian Extinction and Stellar Tool (BEAST; Gordon et al. 2016) to infer physical properties of luminous blue stars in Leo A from the HST broadband multiwavelength imaging (F275W, F336W, F475W, F814W, F110W, and F160W). The BEAST is a probabilistic approach to simultaneously modeling the stellar and line-of-sight dust properties of individual stars in nearby galaxies by carefully considering observational uncertainties arising from photon noise and crowding. Here, we briefly summarize how the BEAST works and its application to stars in Leo A. Detailed descriptions of the BEAST technical underpinnings and its application to a nearby low-metallicity dwarf galaxy can be found in Gordon et al. (2016) and Choi et al. (2020), respectively.

Briefly, the BEAST creates a physics model grid by mapping a stellar evolution library onto a stellar atmospheric model using stellar evolutionary parameters (initial mass $M_{ini}$, age A, and metallicity Z) and stellar atmospheric parameters (effective temperature $\log(T_{eff})$, surface gravity $\log(g)$, and metallicity Z). Lower-temperature stars (i.e., A type and cooler) are paired with local thermodynamic equilibrium models from Castelli & Kurucz (2004), while hotter stars are paired with nonlocal thermodynamic equilibrium (non-LTE) atmospheres from the TLUSTY OSTAR and BSTAR grids (Lanz & Hubeny 2003, 2007). This process covers intrinsic properties for each star. Stars are then moved to the distance of the galaxy of interest (or can be left as a free parameter) and then the intrinsic light is extinguished by a dust model. The dust model used by the BEAST is a flexible mixture model that allows for smooth variation between average Milky Way–like and SMC-like attenuation curves, including the curve steepness and changes in the 2175 Å bump strength (Gordon et al. 2022). For each of the resultant dust-extinguished stellar models, an observational noise model is determined based on ASTs.

For our analysis of stars in Leo A, we adopt the following priors on various physical parameters. First, we fix the distance modulus to μ = 24.40 mag, which is the Gaia eDR3-anchored RR Lyrae distance reported by Nagarajan et al. (2022). Second, we use the MIST stellar evolutionary models with rotation (v/vcrit = 0.4) in preference over the MIST nonrotation (v/vcrit = 0.0) models and the PARSEC Bressan et al. (2012) models. This choice is primarily made due to the evidence of rotation in the spectra, which is accounted for in the MIST

**Table 1**
Observing Details of the Spectroscopic Data Taken with LRIS (KECK) and Binospec (MMT)

| Star | Alternative ID | $\alpha$ (J2000) | $\delta$ (J2000) | UT dates (yy/mm/dd) | Slit (arcsec) | $t_{exp}$ (hr) | F475W (mag) | F475W − F814W (mag) | F366W (mag) | S/N (4500 Å) | S/N (8000 Å) |
|---|---|---|---|---|---|---|---|---|---|---|---|
| (1) | (2) | (3) | (4) | (5) | (6) | (7) | (8) | (9) | (10) | (11) | (12) |
| | | | | LRIS (KECK) | | | | | | | |
| K1 | J095927.52+304457.82 | 09 59 27.52 | +30 44 57.82 | 20/02/21 | 0.7 | 6.3 | 20.1 | −0.40 | 18.4 | 82.5 | 11.5 |
| K2 | J095930.22+304437.14[A] | 09 59 30.22 | +30 44 37.14 | 20/02/21 | 0.7 | 6.3 | 20.2 | −0.41 | 18.6 | 68.9 | 16.6 |
| K3 | J095923.35+304453.76 | 09 59 23.35 | +30 44 53.76 | 20/02/21 | 0.7 | 6.3 | 20.4 | −0.36 | 18.9 | 73.5 | 10.8 |
| K4 | J095928.58+304427.08 | 9 59 28.58 | +30 44 27.08 | 20/02/21 | 0.7 | 6.3 | 20.7 | 0.15 | 19.0 | 47.8 | 4.9 |
| K5 | J095919.61+304406.98 | 9 59 19.61 | +30 44 06.98 | 20/02/21 | 0.7 | 6.3 | 20.8 | 0.12 | 19.2 | 56.1 | 13.3 |
| K6 | J095925.57+304519.29 | 9 59 25.57 | +30 45 19.29 | 20/02/21 | 0.7 | 6.3 | 20.8 | 0.13 | 19.4 | 52.2 | 5.8 |
| K7 | J095919.09+304438.11 | 9 59 19.09 | +30 44 38.11 | 20/02/21 | 0.7 | 6.3 | 21.1 | −0.42 | 18.5 | 46.3 | 5.6 |
| K8 | J095922.69+304504.34 | 9 59 22.69 | +30 45 04.34 | 20/02/21 | 0.7 | 6.3 | 21.5 | 0.0 | 19.8 | 29.7 | 5.8 |
| | | | | Binospec (MMT) | | | | | | | |
| B1/K1 | J095927.52+304457.76 | 09 59 27.52 | +30 44 57.76 | 20/01 23+24/29 | 1.0 | 2/1+1 | 20.1 | −0.40 | 18.4 | 35.1/40.6 | −/10.9 |
| B2/K2 | J095930.23+304437.08[A] | 09 59 30.23 | +30 44 37.08 | 20/01 23+24/29 | 1.0 | 2/1+1 | 20.2 | −0.41 | 18.6 | 34.0/40.0 | −/10.3 |
| B3/K3 | J095923.35+304453.70 | 09 59 23.35 | +30 44 53.70 | 20/01 23+24/29 | 1.0 | 2/1+1 | 20.4 | −0.37 | 18.9 | 30.4/36.6 | −/11.8 |
| B4 | J095920.64+304507.32 | 09 59 20.64 | +30 45 07.32 | 20/01 23+24/29 | 1.0 | 2/1+1 | 20.4 | −0.41 | 18.8 | 29.3/35.8 | −/9.3 |
| B5 | J095926.97+304458.44 | 09 59 26.97 | +30 44 58.44 | 20/01 23+24/29 | 1.0 | 2/1+1 | 21.2 | −0.04 | 20.1 | 19.8/24.0 | −/8.1 |
| B6/K7 | J095919.09+304438.05 | 09 59 19.09 | +30 44 38.05 | 20/01 23+24/29 | 1.0 | 2/1+1 | 21.1 | −0.42 | 18.5 | 19.5/22.8 | −/5.5 |
| B7 | J095919.53+304413.28 | 09 59 19.53 | +30 44 13.28 | 20/01 23+24/29 | 1.0 | 2/1+1 | 21.4 | −0.29 | 20.3 | 17.0/19.2 | −/5.3 |
| B8 | J095924.91+304524.25 | 09 59 24.91 | +30 45 24.25 | 20/01 23+24/29 | 1.0 | 2/1+1 | 21.5 | 0.02 | 20.2 | 24.2/28.1 | −/8.0 |
| B9 | J095928.29+304503.83 | 09 59 28.29 | +30 45 03.83 | 20/01 23+24/29 | 1.0 | 2/1+1 | 21.7 | −0.18 | 20.2 | 16.2/17.0 | −/6.6 |
| B10 | J095922.96+304516.86 | 09 59 22.96 | +30 45 16.86 | 20/01 23+24/29 | 1.0 | 2/1+1 | 21.7 | −0.12 | 20.4 | 12.2/14.4 | −/3.5 |
| B11 | J095925.76+304459.27 | 09 59 25.76 | +30 44 59.27 | 20/01 23+24/29 | 1.0 | 2/1+1 | 21.7 | −0.43 | 20.4 | 9.4/11.0 | −/2.2 |
| B12 | J095928.93+304428.25 | 09 59 28.93 | +30 44 28.25 | 20/01 23+24/29 | 1.0 | 2/1+1 | 21.7 | −0.35 | 20.2 | 11.9/14.0 | −/3.5 |
| B13 | J095929.57+304429.90 | 09 59 29.57 | +30 44 29.90 | 20/01 23+24/29 | 1.0 | 2/1+1 | 21.9 | −0.41 | 20.2 | 11.5/13.5 | −/3.1 |
| B14 | J095926.42+304510.73 | 09 59 26.42 | +30 45 10.73 | 20/01 23+24/29 | 1.0 | 2/1+1 | 21.9 | −0.36 | 20.4 | 14.7/17.3 | −/4.9 |

**Note.** The columns are: (1) The name of each star used throughout the paper; (2) alternative ID; (3) R.A.; (4) decl.; (5) observing date; (6) slit width; (7) integration time; (8)–(10) select HST magnitudes and colors; (11) and (12) S/N at specified wavelength. A—Same star, with different name in the observing logs. K—Keck, B—Binospec.

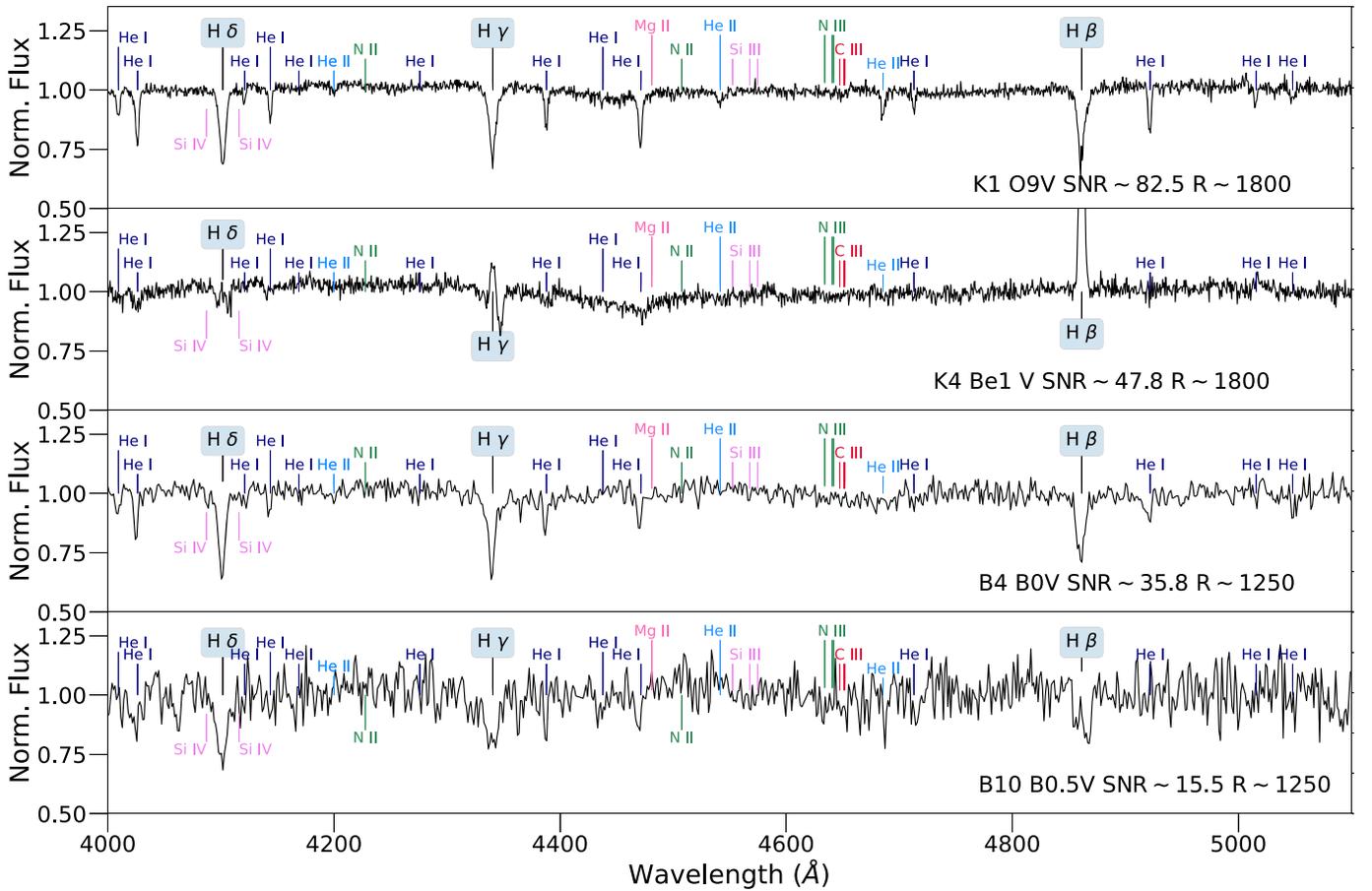

**Figure 4.** Example normalized spectra of massive stars in Leo A shown from 4000–5100 Å with key features labeled. Top two panels: Keck/LRIS ($R \sim 1800$) spectra of an O-type (top) and a Be-type star (bottom) in Leo A. The O-star exhibits He I (dark blue) and He II (light blue) absorption lines and a clear Balmer series in absorption. These are typical of a late-type O star. The Be star shows He I (dark blue) absorption but no He II (light blue). The Balmer series is primarily in emission. Bottom two panels: Binospec spectra ($R \sim 1250$) of an early B-type (top) and a B(e)-type (bottom) in Leo A. The B star shows primarily He I (dark blue) absorption features, while He II (light blue) absorption is minimal. The Balmer lines show absorption features, although there appears to be weak emission within the feature. The bottom panel shows a B0.5-type star. Though the low S/N makes identification of clear features more challenging, we still find clear He I (dark blue) absorption features. The Balmer lines are a mix of absorption and emission features. We note the star is potentially a Be star as well.

rotation model. In Appendix A, we discuss the difference between the two MIST models as well as the PARSEC models. Third, we adopt an SMC-like average extinction curve with $R_V = 2.74$ (Gordon et al. 2003), and allow $A_V$ to vary from a minimum of $A_V = 0.00$ mag to a maximum of $A_V = 0.5$ mag, which is the amount of internal dust needed to match the width of the upper MS in nearby dwarf galaxies based on fitting optical CMDs (e.g., Dolphin et al. 2003). We note that some stars can have higher $A_V$ values than the typical value set by the width of the MS. Finally, we adopt a metallicity of $Z = 0.1\ Z_\odot$ throughout the paper. This is slightly higher than the H II region metallicity of Leo A, but it is: (a) the minimum value for both of the publicly available TLUSTY OSTAR and BSTAR high-resolution spectral grids, and (b) may be more appropriate for the massive stars, which have been found to be more metal-rich (from iron absorption lines) than gas-phase metallicities (from oxygen emission lines) in other nearby dwarf galaxies (Garcia et al. 2014; Bouret et al. 2015; Telford et al. 2021b). Adopting this single metallicity ensures uniform analysis of the spectra and photometry. We report the median value (_ p50) and the uncertainty computed from the 68% confidence interval (i.e., $0.5 * ($_ p84$-$_ p16)$) of a marginalized 1D posterior probability distribution function (PDF) for each parameter. We show

example `BEAST` fits in Figure 5 and discuss them in Section 5.1.

## 4. Measuring Stellar Parameters from Optical Spectroscopy

We use two methods to characterize the stars from spectroscopy. The first, which also derives stellar parameters (e.g., $\log(T_{\rm eff})$, $\log(g)$, $v \sin i$, $A_V$, and mass), is through an adaptation of the full spectral fitting code, "The Payne" (Ting et al. 2019) for massive stars. The Payne has primarily been used for cool stars in the Galaxy, but its flexibility makes it readily adaptable to different types of stellar spectra, including massive stars (e.g., Xiang et al. 2022). The second is through the measurement of EWs. EWs are conventionally used in the field of metal-poor massive stars (e.g., Bresolin et al. 2006; Evans et al. 2007; Tramper et al. 2011, 2014; Camacho et al. 2016; Evans et al. 2019; Garcia et al. 2019b), and provide a sanity check on our quantitative fitting. The extreme metal-poor nature of Leo A means there is a lack of empirical low-metallicity templates for comparison, while the low-resolution nature of the spectroscopy means that the EWs cannot always be as cleanly determined compared to higher-

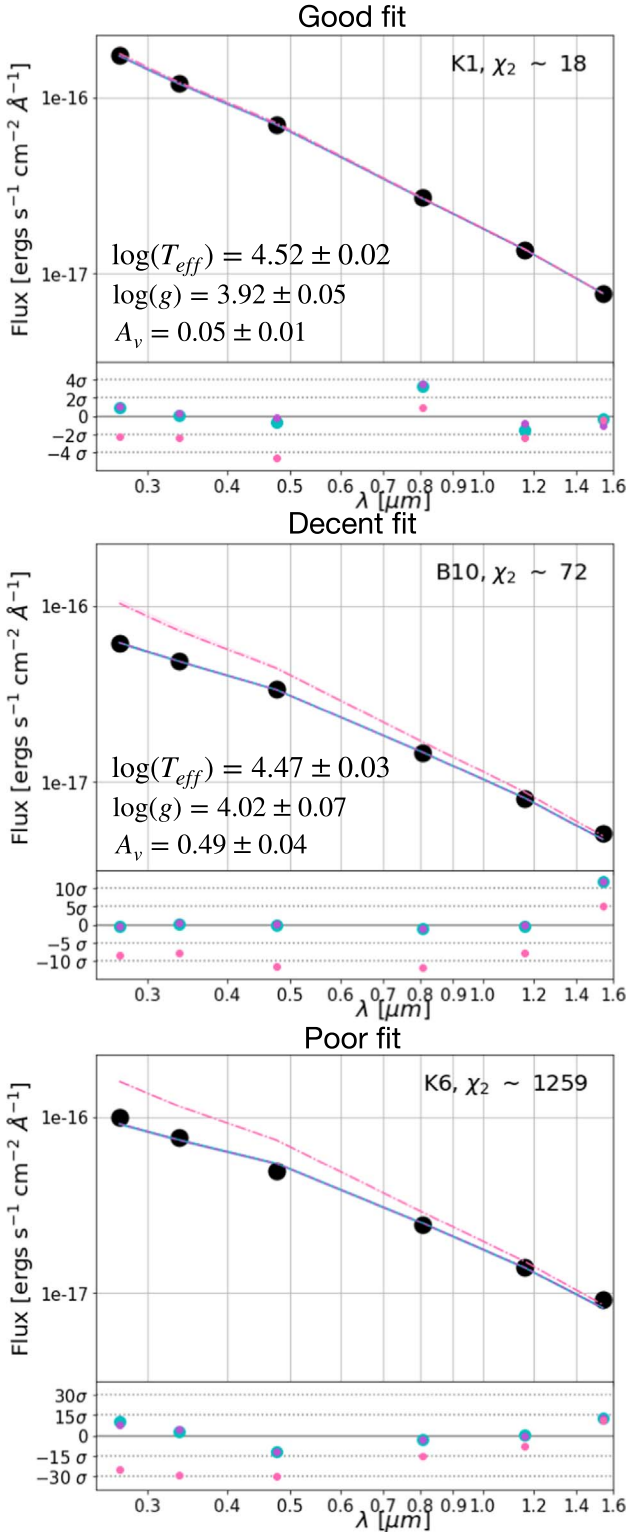

**Figure 5.** Example SED fits in each category of $\chi^2$: good (top), decent (middle), and poor (lower). The data are shown as black points. The median stellar (pink), stellar+dust (purple), and stellar+dust+bias (cyan) are overplotted. The subpanels show the residuals in units of $\sigma$. The top panel shows all residuals of the best model (cyan) within $4\sigma$. The middle panel shows residuals of the best model (cyan) within $12\sigma$ (note that for five bands, the value is closer to $4\sigma$). The bottom panel shows the largest deviations in the residual with the best model (cyan) being within $15\sigma$. For the good and decent fit, we report $\log(T_{\rm eff})$, $\log(g)$, and $A_V$.

**Table 2**
Basic Information on the TLUSTY Model Grids Used for `The Payne`

| Grid | No. of Models | $\log(T_{\rm eff})$ (dex) | $T_{\rm eff}$ (kK) | $\log(g)$ (dex) | $v_{\rm micro}$ (km s$^{-1}$) |
|---|---|---|---|---|---|
| (1) | (2) | (3) | (4) | (5) | (6) |
| OSTAR | 69 | 4.44–4.74 | 27.5–55 | 3.00–4.75 | 10 |
| BSTAR | 105 | 4.18–4.48 | 15–30 | 3.25–4.75 | 2 |

**Note.** The columns are: (1) grid name; (2) number of models in grid used to train the neural net; (3) temperature range in log-space; (4) temperature range in kilokelvin; (5) surface gravity range; and (6) fixed microturbulence value.

resolution spectroscopy. We use the EWs to provide guidance rather than rigorous parameter determination.

### 4.1. The Payne

For quantitative spectral fitting, we have adapted `The Payne` (Ting et al. 2019) for use with the TLUSTY non-LTE massive star model atmospheres (Lanz & Hubeny 2003). Specifically, we use the public OSTAR and BSTAR grids available on the TLUSTY website.[20] At its core, `The Payne` is a highly efficient, multidimensional interpolator that can generate accurate models based on sparse precomputed grid points by using a neural net (see Ting et al. 2019 for details).

We train two different neural nets that separately cover O-stars (Lanz & Hubeny 2003) and B-stars (Lanz & Hubeny 2007). Spectral features in these stars are sufficiently different, and the public TLUSTY grids are sufficiently sparse, that training nets based on separate models lead to more accurate synthetic spectra when compared to trying to interpolate across both model grids at once.

The two precomputed TLUSTY model grids are not uniform in their coverage of parameter space. For consistency, we adopt a fixed metallicity of $Z/Z_\odot = 0.1$ in both grids, similarly to the `BEAST` models chosen. Lastly, we note that a solar helium abundance is assumed and by number is fixed to He/H = 0.1. We list all other parameters in Table 2.

Beyond $\log(T_{\rm eff})$ and $\log(g)$, we also fit for radial velocity, $V_R$, and rotational velocity, $v \sin i$, of each star. For $v \sin i$, we use the Fourier method as outlined by Simón-Díaz & Herrero (2007); however, it is currently not a label we train the neural net on. Furthermore, we note that the accuracy of $v \sin i$ is limited by the resolution of the spectra. The resolution limit on $v \sin i$ is $\sim$170 km s$^{-1}$ for Keck and $\sim$80 km s$^{-1}$ for Binospec. For stars where we have both Binospec and Keck spectra, we report the Binospec derived $v \sin i$ value. We fix the macroturblent velocity to $V_{\rm macro} = 20$ km s$^{-1}$, which is motivated by various literature studies of hot, massive stars in the SMC (e.g., Mokiem et al. 2006; Penny & Gies 2009; Grassitelli et al. 2016). We further fix the instrumental resolution to $R \sim 1800$ for the Keck spectra and $R \sim 3600$ for the Binospec data. We note that having fixed resolution, $V_{\rm macro} = 20$ km s$^{-1}$ and $V_{\rm micro} = 20$ km s$^{-1}$, all might slightly bias $\log(g)$ and $v \sin i$, since they all influence the line profile.

We use `The Payne` to find best-fit stellar parameters for our spectra via $\chi^2$ minimization. We only fit LRIS and Binospec spectra blueward of 7500 Å, the red edge of the TLUSTY grid, and mask emission lines, which are not included in TLUSTY, as described in Section 5.3.

---

We use the optimization results to initialize a Markov Chain Monte Carlo (MCMC) sampler that explores the posterior probability distribution for each star. We use a Gaussian log-likelihood function and adopt uniform priors for $\log(T_{\rm eff})$, $\log(g)$, $v \sin i$, and $V_{\rm R}$. The ranges of allowed values for each are the grid edges for $\log(T_{\rm eff})$ and $\log(g)$ for the respective model grid and [0, 1000] km s$^{-1}$ for $v \sin i$ and [−50, 150] km s$^{-1}$ for $V_{\rm R}$.

We use the MCMC sampler `emcee` (Foreman-Mackey et al. 2013) to sample the posterior. We first use 512 walkers to first burn in our sample for 200 steps. We then initialize 512 walkers, applying a Gaussian scatter of 0.001 and set a maximum step number of $10^5$ steps. However, the sampling stops if one of our convergence criteria is reached: (1) the autocorrelation time, $\tau$, has changed by <1% over the last 100 steps, or (2) the sampler has run for >20$\tau$ steps. To remove any residual effects of burn-in or corrected samples, we discard the first $5\tau$ samples from each walker and thin each chain samples by a factor of $\sim\tau/2$.

For each star, we report the median value of the marginalized posterior PDFs along with the 16th and 84th percentiles as a measurement of uncertainty. If the uncertainty is less than 0.01, we round it to 0.01.

Once we have the parameters from the full spectral fitting, we use those to determine the luminosity ($\log(L)$), the evolutionary mass ($M_{\rm evo}$), the initial mass ($M_0$), and the spectroscopic mass ($M_{\rm spec}$). We note that both $M_{\rm evo}$ and $M_{\rm spec}$ are present-day masses. The luminosity is determined by using both the spectroscopic parameters as well as the photometric data. Based on the spectroscopic parameters, we choose the appropriate model to scale in order to fit to the observed and dust-corrected photometric SED from the HST photometry. We use the ratio between the observed dust-corrected SED, $f_\nu$, to the intrinsic SED (i.e., the model: $F_\nu$) and the distance to Leo A to obtain the spectroscopic radius $R$

$$R = \sqrt{\frac{f_\nu}{F_\nu} \times d^2}. \quad (1)$$

Using the spectroscopic radius, $\log(T_{\rm eff})$, and the Stefan-Boltzmann constant ($\sigma_{\rm SB}$), we calculate the luminosity

$$L = 4\pi R_{\rm spec}^2 \times \sigma_{\rm SB} \times T_{\rm eff}^4. \quad (2)$$

We use the spectroscopic radius and $\log(g)$ value to determine $M_{\rm spec}$:

$$M_{\rm spec} = 10^{\log(g)} R_{\rm spec}^2 / G. \quad (3)$$

Finally, we use the derived $\log(L)$ and spectroscopically $\log(T_{\rm eff})$ and the rotating MIST stellar models to find the mass track that the star falls on to determine $M_0$. The models are spaced at 1 $M_\odot$, and we determine the mass by visual inspection. For $M_{\rm evo}$ we use the stellar parameters to find the stellar mass of the MIST stellar models by looking at the mass at the current position on the track.

### 4.2. Equivalent Widths

To support our qualitative analysis, we measure the EWs of each spectral line from the 1D spectrum for the stars in our sample. After normalizing the spectrum as described in Section 2.3, we combine the `specutils` (an extensible spectroscopic analysis toolkit for astronomy; Earl et al. 2019) package with

`scipy` (Virtanen et al. 2020) to determine the EWs. We use the CLOUDY line list (Ferland et al. 2017) to identify the correct wavelengths. We specify the region of the spectral lines of interest and provide initial guesses for the line center, amplitude, and width. We choose to fit a Gaussian profile to all of our lines. In the scenario of a mixed absorption-emission feature, we employ a Gaussian fit to the emission component and subtract it out of the spectra before fitting the absorption feature for the EW. The results are used in Section 5.3 to support the commentary and in Table 3.

### 4.3. Analysis of Emission Line Stars

For a subset of stars (K4, K5, K6, K8, B5, B8), both the spectroscopic method and photometric method described above yield high reduced $\chi^2$ ($\geqslant$150) values. These stars also have unusual SEDs such that they are bright in the UV, red-optical, and NIR (when available), i.e., their SEDs are not well approximated by a blackbody as expected for normal stars. Their spectra usually feature strong emission, despite not being in close proximity to an H II region (except B5). They further show no or little extended emission beyond the stellar trace in the 2D spectra. We posit that these stars may have circumstellar disks (e.g., Iqbal & Keller 2013; Lazzarini et al. 2021) causing both the emission features seen in the spectra as well as the IR excess in the photometry leading to unusual SEDs. We use an alternative fitting technique to constrain the properties of the star and disk.

We fit emission lines stars in two different ways. First, we use full spectral fitting, which enables the recovery of $v \sin i$ and $\log(T_{\rm eff})$. Second, we fit the photometric SEDs, which provides $\log(T_{\rm eff})$ and $A_{\rm V}$. For spectroscopic fitting, we use `The Payne` but inflating the He I/He II features by inflating uncertainties on other features, such that they effectively only provide continuum constraints. For Be stars, we mask the Balmer lines owing to strong emission. We recover $\log(T_{\rm eff})$ and $v \sin i$ but not $\log(g)$, which is derived from the Balmer lines.

Photometrically, we opt to fit a dual model of a star and potential disk to the UV-optical-NIR SEDs. We use `pystellibs`[21] to interpolate between model SEDs, and `pyphot`[22] to calculate synthetic photometry. Only stellar models are currently publicly available, hence we opt to use stellar models for the disk as well, since the disk is a nuisance parameter. This will still allow us to capture the essential underlying parameters, temperature and extinction, although we might miss the broader physics (e.g., density, velocity, and chemical composition). For the stellar component, we continue to use the TLUSTY models, similarly to the BEAST, while for the potential disk component we use the BaSeL (Lejeune 2002) library, since we need cooler temperatures than TLUSTY can provide. We fit the following parameters: $\log(T_{\rm eff})_{\rm star}$, $R_{\rm star}$, $\log(T_{\rm eff})_{\rm disk}$, $R_{\rm disk}$, and $E(B − V)$. We adopt the same extinction curve as used in the BEAST model. We keep distance as well as mass constant ($M_{\rm star} = 10 M_\odot$ for K4, K5, K6, and K8, and $M_{\rm star} = 8 M_\odot$ for B5 and B8). We use the MCMC sampler `emcee` to determine the median value. We first use 64 walkers to first burn in our sample. We burn in the sample with 50 steps and use 128 walkers; we then set a maximum step number of 1000 steps. We determine convergence by the same criteria as

---

**Table 3**
List of Equivalent Widths Used in Analysis to Determine the Spectral Type of Each Star

| Star | He I 4121 (Å) | He I 4144 (Å) | He I 4388 (Å) | He I 4471 (Å) | He I 4713 (Å) | He I + II 4026 (Å) | He II 4200 (Å) | He II 4542 (Å) | He II 4686 (Å) | Mg II 4481 (Å) | Si II 4128 (Å) | Si III 4553 (Å) | Si IV 4089 (Å) |
|---|---|---|---|---|---|---|---|---|---|---|---|---|---|
| (1) | (2) | (3) | (4) | (5) | (6) | (7) | (8) | (9) | (10) | (11) | (12) | (13) | (14) |
| K1 | $0.16 \pm 0.10$ | $0.13 \pm 0.10$ | $0.43 \pm 0.10$ | $1.08 \pm 0.10$ | $0.61 \pm 0.10$ | $0.56 \pm 0.10$ | $0.18 \pm 0.20$ | $0.44 \pm 0.20$ | $0.71 \pm 0.10$ | ... | ... | $0.15 \pm 0.10$ | $0.15 \pm 0.10$ |
| K2 | $0.20 \pm 0.10$ | $0.37 \pm 0.05$ | $0.81 \pm 0.10$ | $0.81 \pm 0.10$ | $0.50 \pm 0.04$ | $0.65 \pm 0.10$ | ... | $\leqslant 0.30$ | $0.41 \pm 0.15$ | ... | ... | $0.21 \pm 0.15$ | $\leqslant 0.30$ |
| K3 | $0.40 \pm 0.10$ | $0.54 \pm 0.10$ | $0.76 \pm 0.10$ | $1.09 \pm 0.20$ | $0.45 \pm 0.10$ | $0.99 \pm 0.05$ | ... | ... | $\leqslant 0.17$ | $0.40 \pm 0.15$ | $0.22 \pm 0.10$ | $\leqslant 0.32$ | $\leqslant 0.21$ |
| K4 | $0.37 \pm 0.20$ | $0.43 \pm 0.20$ | $0.58 \pm 0.20$ | $0.73 \pm 0.20$ | ... | $0.64 \pm 0.20$ | ... | ... | $0.17 \pm 0.15$ | $\leqslant 0.11$ | ... | $0.19 \pm 0.10$ | $\leqslant 0.12$ |
| K5 | $\leqslant 0.13^{*}$ | $\leqslant 0.54^{*}$ | $0.62 \pm 0.25$ | $0.68^{*} \pm 0.25$ | ... | $0.46 \pm 0.20$ | ... | ... | ... | $\leqslant 0.21$ | ... | $\leqslant 0.15$ | ... |
| K6 | $0.27^{*} \pm 0.15$ | $0.42 \pm 0.20$ | $0.53 \pm 0.20$ | $0.87 \pm 0.20$ | ... | $0.84 \pm 0.20$ | ... | ... | $\leqslant 0.85^{*}$ | $\leqslant 0.27^{*}$ | $\leqslant 0.10$ | ... | $0.24 \pm 0.15$ |
| K7 | $0.42 \pm 0.10$ | $0.66 \pm 0.10$ | $0.68 \pm 0.10$ | $1.30 \pm 0.20$ | $0.60 \pm 0.20$ | $1.10 \pm 0.20$ | $0.21 \pm 0.15$ | $0.32 \pm 0.15$ | $0.34 \pm 0.15$ | $0.48 \pm 0.10$ | ... | ... | $\leqslant 0.15$ |
| K8 | ... | $0.62 \pm 0.25$ | $0.75 \pm 0.25$ | $1.21 \pm 0.15$ | ... | $0.93 \pm 0.30$ | ... | ... | ... | $\leqslant 0.28$ | ... | $\leqslant 0.30$ | $\leqslant 0.16$ |
| B4 | $0.38 \pm 0.15$ | $0.71 \pm 0.15$ | $0.73 \pm 0.15$ | $0.88 \pm 0.10$ | $0.24 \pm 0.15$ | $0.88 \pm 0.20$ | ... | ... | $0.28 \pm 0.15$ | ... | $0.17 \pm 0.10$ | $0.24 \pm 0.10$ | $0.28 \pm 0.10$ |
| B5 | $0.48 \pm 0.25$ | $0.49 \pm 0.20$ | $0.39 \pm 0.30$ | $0.92 \pm 0.20$ | $0.48 \pm 0.15$ | $1.24 \pm 0.30$ | ... | ... | ... | $0.30 \pm 0.20$ | $0.35 \pm 0.15$ | ... | $\leqslant 0.30$ |
| B7 | $0.75 \pm 0.15$ | $0.90^{*} \pm 0.20$ | $0.68^{*} \pm 0.20$ | $0.59 \pm 0.15$ | ... | $0.50 \pm 0.10$ | ... | $\leqslant 0.17$ | ... | ... | ... | $0.32 \pm 0.10$ | $\leqslant 0.16$ |
| B8 | ... | $\leqslant 0.33$ | $\leqslant 0.40$ | $0.72 \pm 0.20$ | $0.56 \pm 0.20$ | $\leqslant 0.98$ | ... | ... | ... | $\leqslant 0.28$ | ... | ... | $\leqslant 0.22$ |
| B9 | $0.83 \pm 0.20$ | $0.82 \pm 0.20$ | $0.78 \pm 0.20$ | $1.10 \pm 0.20$ | $1.02 \pm 0.20$ | $1.80 \pm 0.30$ | ... | ... | ... | $0.23 \pm 0.10$ | $\geqslant -0.48$ | $0.38 \pm 0.10$ | $\leqslant 0.26$ |
| B10 | $0.55 \pm 0.15$ | $0.80 \pm 0.25$ | $0.77 \pm 0.20$ | $1.09 \pm 0.25$ | ... | $0.80 \pm 0.25$ | ... | ... | ... | ... | ... | $\leqslant 0.55$ | $0.34 \pm 0.10$ |
| B11 | $1.26 \pm 0.30$ | $1.04 \pm 0.30$ | $1.16 \pm 0.30$ | $2.46 \pm 0.40$ | ... | $2.17 \pm 0.25$ | ... | ... | ... | ... | $0.87 \pm 0.40$ | $\leqslant 1.61$ | $0.40 \pm 0.20$ |
| B12 | $0.72 \pm 0.20$ | $0.79 \pm 0.20$ | $1.38 \pm 0.30$ | $1.20 \pm 0.30$ | ... | $2.20 \pm 0.25$ | ... | ... | ... | ... | ... | $\leqslant 0.77$ | $\leqslant 0.32$ |
| B13 | $1.04 \pm 0.30$ | $0.81 \pm 0.25$ | $0.79 \pm 0.20$ | $2.15 \pm 0.35$ | $0.83 \pm 0.15$ | $1.94 \pm 0.40$ | ... | $1.20 \pm 0.30$ | $1.46 \pm 0.40$ | ... | ... | $0.53 \pm 0.20$ | $0.63 \pm 0.20$ |
| B14 | $0.50 \pm 0.30$ | $0.37 \pm 0.10$ | $0.81^{*} \pm 0.20$ | $0.65 \pm 0.30$ | ... | $1.34 \pm 0.30$ | ... | ... | ... | $0.61 \pm 0.20$ | ... | $0.31 \pm 0.10$ | ... |

**Note.** The $^{*}$ denotes lines that have emission components that had to be masked out to determine the EW.

The Payne fitting. We use a log-Gaussian likelihood function and use a Gaussian prior for each parameter. We allow a broad range of priors so that we can simultaneously check whether the SED could also be explained by a companion star. We allow the parameters to sample over following ranges for $\log(T_{\rm eff})_{\rm star} = [4, \; 5.7]$ $\log(K)$, $R_{\rm star} = [0.8, \; 42]$ $R_\odot$, $\log(T_{\rm eff})_{\rm disk} = [2, \; 4]$ $\log(K)$, $R_{\rm disk} = [0.8, \; 127]$ $R_\odot$, and $E(B-V) = [0, \; 0.5]$ dex. We note that we use the same extinction value for both components, since we are only interested in the stellar component. Lastly, we initialize the sampler by using the spectroscopic values. The results are used in Section 5.3.3 to support the commentary.

## 5. Results

### 5.1. Stellar Parameters from Photometry

For comparison with our spectroscopic sample, we only ran the BEAST on stars with F475W < 22 and F475W − F814W < 0.5 (i.e., $M \gtrsim 5 M_\odot$ on the MS). For select science cases (e.g., Section 5.1), we used the BEAST on slightly fainter stars.

To facilitate discussion and comparison, we divide the resulting photometric SED fits into three groups based on the $\chi^2$ values. Fits with $\chi^2 < 20$ are considered good fits, $20 \leqslant \chi^2 < 80$ are decent fits, and $\chi^2 \geqslant 80$ are poor fits. These $\chi^2$ divisions are based on qualitative assessment of the fits, following other BEAST papers in the literature (e.g., Van De Putte et al. 2020). They are only meant to be used qualitatively to facilitate discussion. Furthermore we note that the uncertainties are likely underestimated, since they take the models at face value.

Figure 5 shows three example SEDs in each of our $\chi^2$ categories. These examples all have spectroscopic data enabling further evaluation. The top panel illustrates the case of a good SED fit for star K1 ($\chi^2 \sim 18$ for the stellar, dust, and crowding bias fit). All SED points for the median model (combined star, dust, and bias model) are consistent with the observed SED within $4\sigma$. The derived physical parameters ($\log(T_{\rm eff})$, $\log(g)$, $A_{\rm V}$, initial mass $M_{\rm ini}$, present-day mass $M_{\rm act}$, and $\log(L)$; see Section 5.3.1) all appear reasonable given the star's location on the observed CMDs. Overall, they indicate that K1 is a late-type O dwarf. We compare the SED and spectroscopically determined parameters in Section 5.3.

The middle panel of Figure 5 illustrates a decent SED fit using star B10 ($\chi^2 \sim 72$). The residuals show that the median model SED is within ∼4$\sigma$ for five bands and ∼12$\sigma$ of the observed SED for one NIR band. Like the good $\chi^2$ for K1, the stellar parameters for B10 are in reasonable agreement with its position on the CMD and a B0.5 dwarf. The star, however, appears to have quite high extinction ($A_{\rm V} = 0.49$), which is puzzling. We compare the SED and spectroscopically determined parameters in Section 5.3.

The bottom panel of Figure 5 displays an example of a poor SED fit ($\chi^2 \sim 1259$) for star K6. In general, the median model values range in agreement from ∼5$\sigma$ for the two UV SED points to ∼15$\sigma$ for the F475W SED point. K5 falls into the category of stars that show an unusual SED, bright in the UV and red-optical, possibly due to the presence of a disk (see Section 5.3.3). Hence, it is not surprising that a single star model cannot be well fit to the SED.

Having illustrated our SED fitting with a few examples, we now consider general trends in the SED fits of luminous stars in Leo A. Figure 6 shows observed CMDs in Leo A color-coded by the results of our photometric SED fitting for select stellar parameters ($\log(T_{\rm eff})$, $\log(g)$, $M_{\rm ini}$, $A_{\rm V}$) along with the $\chi^2$ value of the fit. The values assigned to these points reflect the median of the posterior distributions reported by the BEAST.

In general, as expected, the temperature decreases for redder CMD colors (second row of Figure 6). This is particularly evident in the UV-optical CMDs, where the hottest stars are located at the blue edge of the MS. We find that $\log(T_{\rm eff})$ only weakly depends on $\chi^2$; poor fits seem to have a qualitatively reasonable correspondence between UV-optical color and $\log(T_{\rm eff})$ (i.e., the redder stars, the cooler the stars).

However, the situation changes slightly when considering optical-only colors. Here, while the majority still follow the expected trend, we find several outliers. Namely, several decently fit stars that appear on the MS in the UV-optical CMDs and have hot SED-based values of $\log(T_{\rm eff})$ appear to be cooler stars (i.e., BHeB region stars) in the optical only CMD. In the NIR CMD we see, similarly to the optical-only CMD, that the outliers are primarily the same stars that are displaying odd temperature to color combinations.

For a subset of these color outliers, we have spectra that we use to cross-check the reasonability of the photometrically determined parameters. In many cases, these color outliers appear to have emission features that suggest they are not nebular in origin and instead may indicate these are Be stars. We compare spectroscopic and photometric parameters for this subset in Section 5.3 and discuss the candidate Be stars in Section 6.3. Similar to $\log(T_{\rm eff})$, trends in $\log(g)$ (third row of Figure 6) are generally consistent with expectations. The bluest stars should have the highest values of $\log(g)$, which is what we find. For redder stars, we expect and observe a decrease in $\log(g)$. We find that several of the stars causing discrepancies in $\log(T_{\rm eff})$ in the optical-only CMD show questionable values for $\log(g)$ as well. In the UV CMD, we observe the same outliers as in the optical CMD. In the NIR CMD, we similarly observe that the outliers in $\log(T_{\rm eff})$ are outliers in $\log(g)$.

We observe a broad trend between $M_{\rm ini}$ and $\log(g)$, expected from stellar models, with a few exceptions. The most-massive stars are generally the brightest and hottest, while fainter stars are less massive. Stars with poor fits in $\log(g)$ are exceptions. Their values of $M_{\rm ini}$ are higher than one expects from their location on the panchromatic CMDs. Stars in the BHeB region of the optical CMD have $M_{\rm ini} \lesssim 13 M_\odot$, but are all poor fits, suggesting we should use caution in taking their masses at face value.

Several studies of Leo A find it has very little dust (e.g., van Zee et al. 2006; Ruiz-Escobedo et al. 2018). This is reflected in the BEAST fitting as well. The MS stars all show little to no dust ($A_{\rm V} \lesssim 0.08$), though the BHeB sequence shows higher $A_{\rm V}$ values. This difference can potentially be explained by elevated dust production in supergiants (e.g., Waters 2010; de Wit et al. 2014), but we note that several of the high $A_{\rm V}$ stars also have poor fits, as is especially visible in the NIR CMD. Several of the previously mentioned outliers (i.e., the stars with peculiar colors) may have higher extinction, since the BEAST is trying to fit the excess IR emission created by the disk as we discuss in Section 5.3. Another highly probable explanation for the poor fits of stars appearing on the BHeB sequence in the optical CMD with unphysically high dust is the lack of calibration of stellar models for BHeB stars at the extreme low metallicity (e.g., McQuinn et al. 2012).

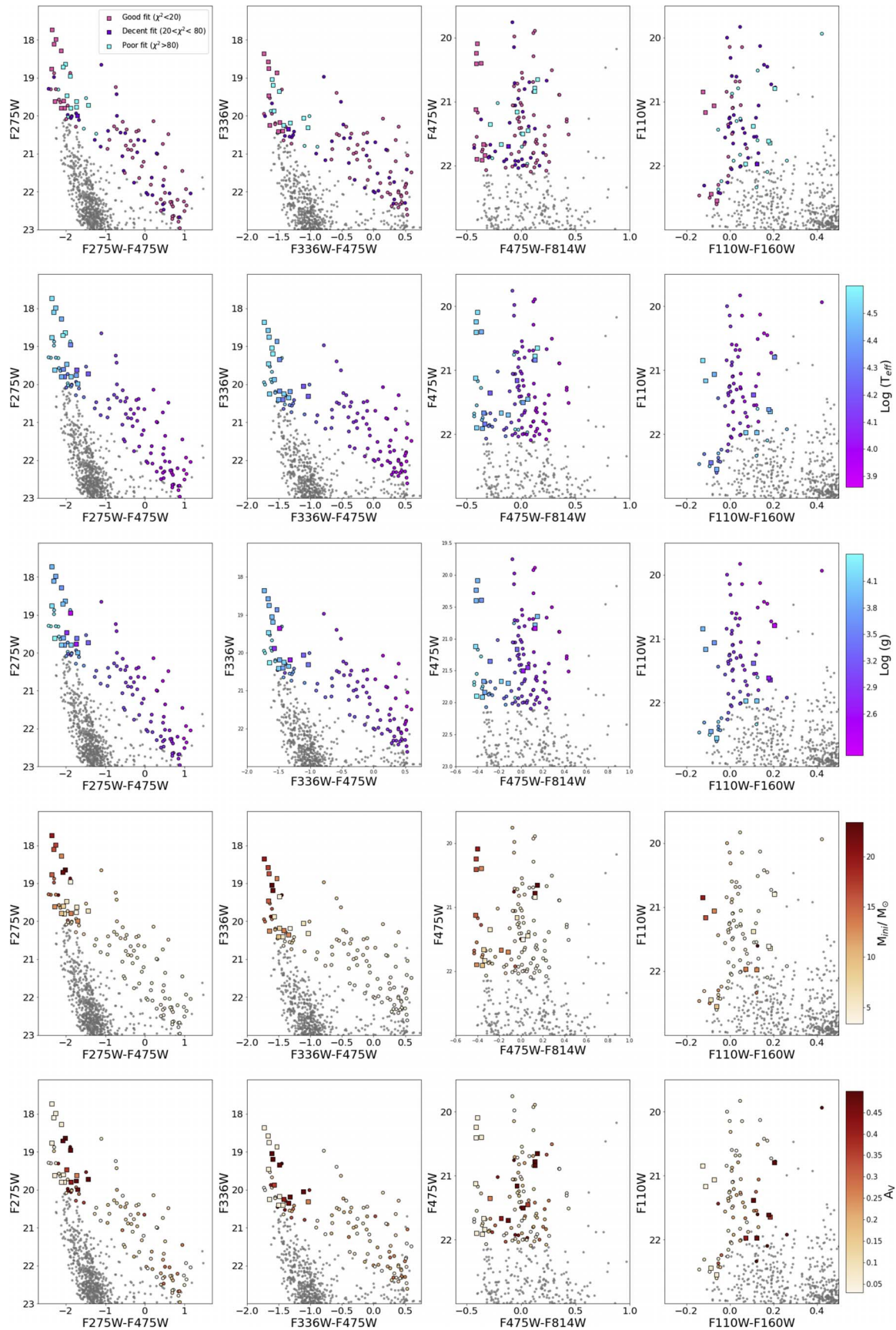

**Figure 6.** A panchromatic gallery of CMDs for Leo A ranging from UV-optical on the left to NIR on the right. Points in each row are color-coded by results from SED fitting with the BEAST. From top to bottom, the rows reflect: $\chi^2$, $\log(T_{\rm eff})$, $\log(g)$, $M_{\rm ini}$, and $A_V$. For $\chi^2$, we divide the fits into three categories: good (pink), decent (purple), and poor (cyan). Stars that have spectroscopic data are plotted as squares.

Table 4
Spectroscopic Parameters Determined Using `The Payne`

| Star | $M_{\rm spec}$ $M_\odot$ | $M_0$ $M_\odot$ | $M_{\rm evo}$ $M_\odot$ | $\log(T_{\rm eff})$ | $\log(g)$ | $\log(L/L_\odot)$ | $v_{\sin i}$ (km s$^{-1}$) | $V_R$ (km s$^{-1}$) | $V_{\rm RGauss}$ (km s$^{-1}$) | $\chi^2$ | Spectral Type |
|------|------|------|------|------|------|------|------|------|------|------|------|
| (1) | (2) | (3) | (4) | (5) | (6) | (7) | (8) | (9) | (10) | (11) | (12) |
| K1 | 17 ± 3 | 19 ± 1 | 19 ± 1 | 4.49 ± 0.01 | 3.69 ± 0.01 | 4.90 ± 0.09 | <80 | 21 ± 2 | 18 ± 8 | 15 | O9V |
| K2 | 27 ± 6 | 17 ± 1 | 17 ± 1 | 4.50 ± 0.01 | 4.10 ± 0.02 | 4.72 ± 0.09 | 95 ± 20 | −13 ± 2 | −25 ± 13 | 21 | O9.7 V |
| K3 | 20 ± 5 | 13 ± 1 | 13 ± 1 | 4.42 ± 0.01 | 3.75 ± 0.05 | 4.53 ± 0.12 | <80 | 37 ± 2 | 27 ± 21 | 17 | B1.5 V |
| K7 | 13 ± 3 | 13 ± 1 | 13 ± 1 | 4.48 ± 0.01 | 4.04 ± 0.02 | 4.38 ± 0.08 | 103 ± 20 | 24 ± 3 | 19 ± 21 | 18 | O9.5 V |
| B4 | 15 ± 2 | 15 ± 1 | 15 ± 1 | 4.46 ± 0.01 | 3.75 ± 0.04 | 4.65 ± 0.04 | <80 | 29 ± 3 | 33 ± 8 | 67 | B0V |
| B9 | 15 ± 3 | 12 ± 1 | 12 ± 1 | 4.45 ± 0.02 | 4.18 ± 0.06 | 4.25 ± 0.04 | 115 ± 20 | 20 ± 8 | 5 ± 32 | 47 | B1.5 V |
| B10 | 11 ± 6 | 11 ± 1 | 11 ± 1 | 4.43 ± 0.02 | 3.92 ± 0.11 | 4.24 ± 0.04 | 141 ± 25 | 31 ± 11 | 31 ± 24 | 100 | B0.5 V / Be0.5 V |
| B12 | 8 ± 4 | 11 ± 1 | 11 ± 1 | 4.44 ± 0.05 | 4.00 ± 0.10 | 4.07 ± 0.05 | <80 | 30 ± 8 | 27 ± 28 | 48 | B1.5V |
| B13 | 19 ± 4 | 12 ± 1 | 12 ± 1 | 4.50 ± 0.02 | 4.64 ± 0.06 | 4.07 ± 0.08 | <80 | 24 ± 8 | 18 ± 29 | 141 | O9.7 V |
| B14 | 24 ± 5 | 8 ± 1 | 8 ± 1 | 4.39 ± 0.02 | 4.41 ± 0.08 | 3.87 ± 0.08 | 80 ± 15 | 20 ± 5 | 32 ± 19 | 78 | B3 V |

**Note.** The columns are: (1) star name; (2) spectroscopic mass; (3) initial mass; (4) evolutionary mass; (5) effective temperature; (6) surface gravity; (7) luminosity; (8) rotational velocity; (9) heliocentric corrected radial velocity from full spectral fitting; and (10) heliocentric corrected radial velocity measurements using Gaussian profile fitting. Columns (4), (5), (7), and (8) are the stellar parameters that directly resulted from the full spectral fitting with `The Payne`. Columns (2) and (6) are derived using the previous stellar parameters in addition to the SED. Column (3) lists the evolutionary mass determined by eye.

### 5.2. Radial Velocities from Spectroscopy

We determined the radial velocity ($V_R$) of the coadded spectra for each star using two methods. First, we model the full spectrum of each star using `The Payne`, for which $V_R$ is a free parameter as described Section 4. Second, we measure the Doppler displacement of specific spectral lines using a Gaussian-fitting method to obtain the radial velocity. The latter method is more commonly used in the literature (e.g., Sana et al. 2013; Camacho et al. 2016; Evans et al. 2019), and we use it as a check on the full spectral fitting approach. We follow the cited literature studies and fit Gaussian line profiles to a collection of Balmer absorption lines (Hα, Hβ, Hγ, Hδ) and He absorption lines (He I 4387, He I 4713, and He II 4686) that have been identified as only mildly impacted by winds in the temperature regime. We exclude any lines that show strong emission. If the emission is small relative to the absorption feature, we mask out pixels with an emission component. Hα is only measured for stars in the Binospec sample, since it is not in the wavelength range of the Keck spectra. We report the radial velocity of each star as the error-weighted mean of the velocities measured for each line. We apply heliocentric corrections of $v_{\rm helio} = −5.08$ km s$^{-1}$ to our Keck measurements and $v_{\rm helio} = −6.49$ km s$^{-1}$ to our Binospec measurements.

We find that all stars agree within 1σ between the two methods (see Table 4). This confirms that full spectral fitting can provide accurate $V_R$ measurements for S/Ns as low as ~10 (lowest S/N in sample). $V_R$ measurements may be accurately measured at lower S/Ns as well, but will be investigated in future work. We further find that the measurements yielded by the full spectral fitting are more precise ($\sigma_{\rm Gauss} \lesssim 32$ km s$^{-1}$, $\sigma_{\rm Payne} \lesssim 11$ km s$^{-1}$). For the remainder of the analysis, we use the $V_R$ measurements derived from the full spectral fitting and reported in Table 4.

Figure 7 shows the distribution of measured radial velocities with select literature measurements overplotted for comparison. For our eight Keck stars, we find a mean radial velocity of $V_R = 23.7 ± 16.3$ km s$^{-1}$. For our 10 Binospec stars, we find a mean radial velocity of $V_R = 29.4 ± 16.2$ km s$^{-1}$. They are consistent with each other and agree well with radial velocities of supergiants in Leo A ($V_R = 22.9 ± 10.2$ km s$^{-1}$; Brown et al. 2007) and of the old red giant branch stars ($V_R = 26.2 ± 9.0$ km s$^{-1}$; Kirby et al. 2017). Our combined Keck and Binospec samples for nonoverlapping stars yield a mean $V_R = 26.9 ± 12.3$ km s$^{-1}$. The stellar measurements are in line with previous gas velocity studies (Young & Lo 1996; Hunter et al. 2012).

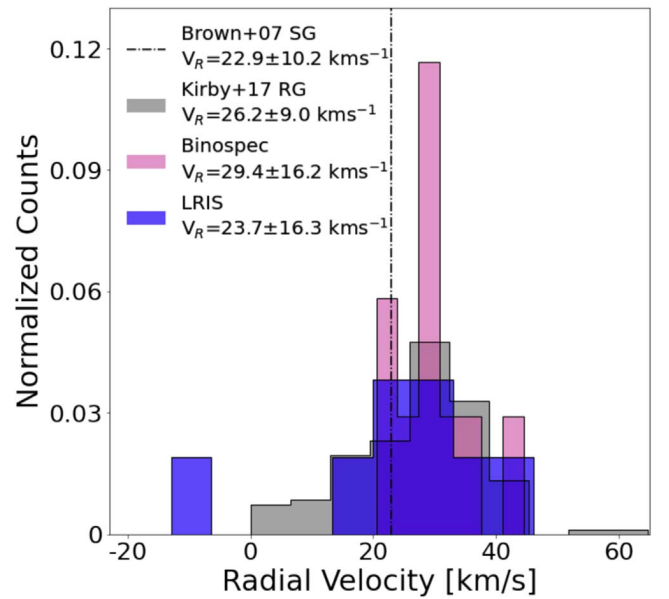

**Figure 7.** The distribution of radial velocities for stars in Leo A. Our LRIS (blue) and Binospec (pink) velocities are consistent with the Keck/DEIMOS-based velocities of red giants (RG) measured by Kirby et al. (2017), and the average velocity measured from the Supergiants (SG) by Brown et al. (2007). Each sample is normalized by counts.

Three of the four stars observed with both telescopes yield consistent velocities. The Keck spectrum of the fourth star (K2) yields $V_R = −13 ± 2$ km s$^{-1}$, whereas the Binospec spectrum yields $V_R = 16 ± 4$ km s$^{-1}$. There is no obvious issue with the data that would cause this level of discrepancy. One possibility is that this is a binary star, and that the different velocities are the result of observing the system at different points in the orbital phase. We are in the process of obtaining time resolved spectroscopy to better quantify the degree of binarity in the massive stars of Leo A.

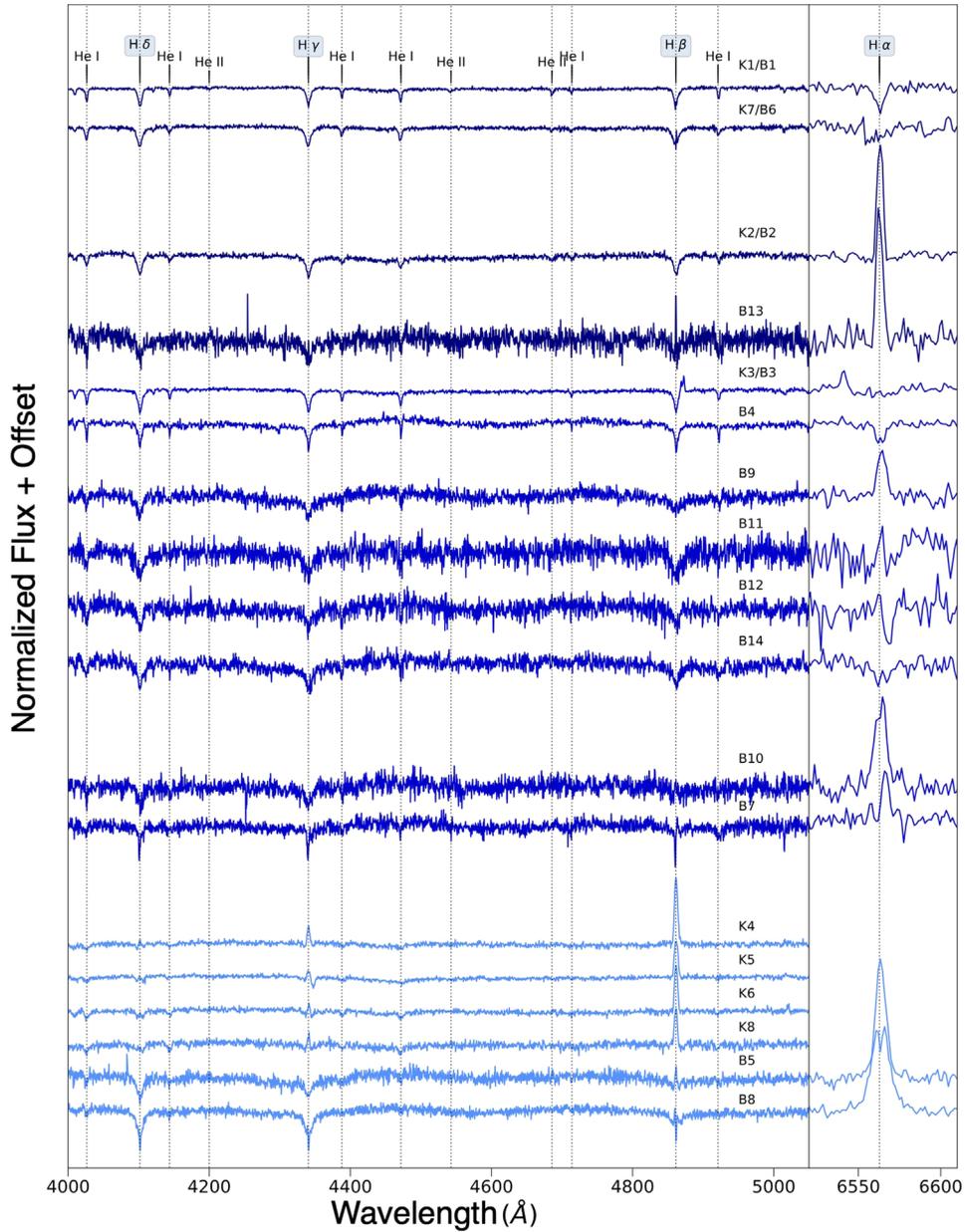

**Figure 8.** Keck/LRIS and MMT/Binospec normalized spectra (with an arbitrary offset added for clarity) for all 18 stars in the spectroscopic sample. For stars with both Keck and MMT data, we show the Keck spectra in the 4000–5100 Å region. If there is no LRIS spectrum, we show the Binospec (1000 lpmm) spectrum. The spectra in the 6520–6610 Å region are from the Binospec 270 lpmm grating and are shown for all stars that have Binospec spectra. The spectra are sorted by spectral type from O-type stars at the top to B-type and Be-type stars at the bottom. This is one of the largest set of optical spectra for sub-SMC metallicity massive dwarf stars.

### 5.3. Stellar Parameters from Spectroscopic Analysis

We present a discussion of our spectroscopic sample. This section includes an example of each stellar type found in the study; a detailed individual discussion of all stars can be found in the Appendix B. We determine these stellar parameters for the stars using full spectral fitting via The Payne and via photometric SED fitting with the BEAST. We also use EWs to provide spectral types using conventional stellar classification schemes in the literature. We discuss agreement between the methods. The spectrum of each star is shown in Figure 8, and Table 3 provides a full list of lines we detect and measure in each spectrum. For guiding the discussion, we group the stars into broad categories (e.g., O-stars and B(e)-stars) and, for brevity, focus only on the most salient features in the spectra and photometry.

#### 5.3.1. O-type Stars

We classify four stars as O-type stars in our sample (K1/B1, K2/B2, K7/B6, and B13) based on their EW measurements. Overall, the parameters derived from spectroscopy and photometry yield consistent values with the EW classification. We present a detailed example of the analysis of K1 and highlight interesting features of the other stars.

K1/B1: K1/B1 is the brightest star in our Leo A sample (F475W$_{VEGA}$ = 20.09 and F336W$_{VEGA}$ = 18.36). Its locations in the optical and UV CMDs are typical of an O-star. Its

**Table 5**
Photometric Parameters Determined Using the BEAST and MIST Models with $v/v_{crit} = 0.4$

| Star | $M_{ini,Exp}$ ($M_\odot$) | $M_{act,Exp}$ ($M_\odot$) | $\log(T_{eff})$ (kK) | $\log(g)$ | $\log(L/L_\odot)$ | $A_v$ | $\log(Age)$ | $\chi^2$ | No. of Filters | Spectral Type |
|------|------|------|------|------|------|------|------|------|------|------|
| (1) | (2) | (3) | (4) | (5) | (6) | (7) | (8) | (9) | (10) | (11) |
| K1 | 20.1 ± 1.29 | 19.99 ± 1.19 | 4.53 ± 0.02 | 3.92 ± 0.05 | 4.88 ± 0.05 | 0.05 ± 0.01 | 6.9 ± 0.08 | 11 | 6 | O9V |
| K2 | 16.81 ± 0.21 | 16.74 ± 0.30 | 4.50 ± 0.01 | 3.85 ± 0.01 | 4.74 ± 0.01 | 0.06 ± 0.02 | 7.0 ± 0.07 | 4 | 4 | O9.7 V |
| K3 | 12.89 ± 0.17 | 12.87 ± 1.24 | 4.41 ± 0.01 | 3.66 ± 0.02 | 4.50 ± 0.01 | 0.07 ± 0.02 | 7.2 ± 0.08 | 7 | 6 | B1.5 V |
| K7 | 14.69 ± 1.06 | 14.68 ± 1.0 | 4.52 ± 0.02 | 4.23 ± 0.05 | 4.45 ± 0.04 | 0.06 ± 0.03 | 6.8 ± 0.15 | 17 | 4 | O9.5 V |
| B4 | 16.38 ± 0.18 | 16.32 ± 0.27 | 4.50 ± 0.01 | 3.91 ± 0.01 | 4.68 ± 0.01 | 0.04 ± 0.01 | 7.0 ± 0.07 | 10 | 6 | B0V |
| B9 | 10.87 ± 6.13 | 9.96 ± 7.5 | 4.49 ± 0.05 | 4.12 ± 0.68 | 4.39 ± 0.14 | 0.42 ± 0.08 | 7.0 ± 0.85 | 83 | 6 | B1 V |
| B10 | 12.2 ± 1.21 | 12.19 ± 1.24 | 4.47 ± 0.03 | 4.02 ± 0.07 | 4.35 ± 0.1 | 0.49 ± 0.04 | 7.1 ± 0.11 | 73 | 6 | B0.5 V / Be0.5 V |
| B11 | 6.26 ± 0.69 | 6.26 ± 0.64 | 4.27 ± 0.04 | 3.81 ± 0.06 | 3.45 ± 0.13 | 0.07 ± 0.01 | 7.8 ± 0.13 | 2 | 6 | B2 V |
| B12 | 6.61 ± 1.21 | 6.61 ± 1.14 | 4.34 ± 0.05 | 3.86 ± 0.08 | 3.73 ± 0.19 | 0.05 ± 0.03 | 7.6 ± 0.16 | 3 | 4 | B1.5V |
| B13 | 12.45 ± 0.5 | 12.45 ± 0.42 | 4.50 ± 0.01 | 4.37 ± 0.03 | 4.11 ± 0.02 | 0.06 ± 0.02 | 6.5 ± 0.22 | 3 | 4 | O9.7 V |
| B14 | 9.22 ± 0.99 | 9.22 ± 0.91 | 4.42 ± 0.02 | 4.12 ± 0.06 | 3.92 ± 0.08 | 0.06 ± 0.02 | 7.3 ± 0.16 | 15 | 6 | B3 V |

**Note.** The columns are: (1) star name; (2) initial mass; (3) actual mass; (4) temperature; (5) surface gravity; (6) luminosity; (7) extinction parameter; (8) stellar age in log-space; (9) $\chi^2$ yielded by the BEAST; (10) number of HST bands used for the SED fitting (when the value is 4, we observed F275W, F336W, F475W, and F814W, and when the value is 6, we additionally observed F110W and F160W); and (11) spectral type derived from EW analysis. Columns (2), (7), and (8) list primary parameters, while columns (4), (5), and (6) list derived parameters.

spectrum is shown in Figure 8. We list its EW measurements in Table 3 and comment only on a subset of those features.

The spectrum shows the characteristics of a typical late-type O-star: strong absorption of He I and He II. Most of the Balmer series is in absorption, though Hβ appears to exhibit weak emission, likely due to K1's location in an H II region (see Figure 3).

We first estimate the stellar type of K1 using the relevant EWs, which are listed in Table 3. We use both Galactic (Conti & Alschuler 1971; Sota et al. 2011; Martins 2018) and subsolar metallicity classification schemes (Castro et al. 2008), prioritizing the latter whenever possible. From the EW(He I λ4471)/EW(He II λ4542) ratio, we find that K1 is of spectral type O9, or perhaps later, in the traditional spectral classification scheme of Conti & Alschuler (1971). We also consider EW ratios of the (He I λ4471/He II λ4542) and (He I λ4388/He II λ4542) in the classification scheme of Martins (2018). They indicate a star of type O8.5-9. The He II λ4542 to He I λ4388 comparison yields a spectral type O9.5 in the Sota et al. (2011) classification scheme. Lastly, using the Castro et al. (2008) classification scheme, we find the star to be an O9. K1 could be either a dwarf or a subgiant star based on the comparison with the Martins (2018) classification scheme using (log(EW(He II λ4686))+log(EW(He I λ4388))), since its value falls within both ranges. However, using Castro et al. (2008), we determine the star is likely a dwarf. Our best estimate from spectroscopy is that K1 is an O9V star. We employ the same EW classification scheme for our other O stars.

We fit the full Keck spectrum of K1 using The Payne. The median parameters for the spectral fitting are listed in Table 4. For K1, we find $\log(T_{eff}) = 4.49 \pm 0.01$ dex and $\log(g) = 3.69 \pm 0.01$ dex, which, like the EW ratios, indicates that K1 is a late-type O star. Using Equations (1)–(3), we derive $M_{spec} = 17 \pm 3 M_\odot$ and $M_0 = 19 \pm 1 M_\odot$, and $\log(L/L_\odot) = 4.90 \pm 0.09$ dex. All of these values agree with the classifications derived from the EW measurements and location in the CMD. The photometric analysis yields $\log(T_{eff}) = 4.52 \pm 0.02$ dex and $\log(g) = 3.92 \pm 0.05$ dex. While $\log(T_{eff})$ is consistent, $\log(g)$ shows a 0.23 dex difference. This difference may be explainable by three different reasons: (1) keeping the instrument resolution, and micro- and macroturbulence constant might impact $\log(g)$ since they all affect the line widths; (2) subtraction of the nebular emission during data reduction; if the Balmer lines are slightly over/under-subtracted, then the line width and $\log(g)$ can be affected; and (3) assumptions within stellar atmosphere models give rise to systematic differences (~0.2 dex; Markova et al. 2018; Dufton et al. 2019). Computing our own systematic uncertainties requires delving into stellar atmosphere physics for very metal-poor stars, and is beyond the scope of this paper. We find that $M_0$ and $M_{ini}$ are consistent with each other, as seen in Tables 4 and 5.

We show the spectrum of K1 with the models generated from the two different methods in Figure 9; we overplot synthetic spectra of K1 from our spectroscopy (pink) and photometric (blue) parameters. Both synthetic spectra provide a similar match to the He I/He II feature, as expected since our $\log(T_{eff})$ is in agreement. We find the difference in $\log(g)$ is best displayed in the Hβ feature, primarily in the wings. The Balmer lines are well fit by the models, which we highlight by showing a zoom in on Hβ. However, while the model appears to match the He I/He II ratio, it struggles with the recovering the depth of the He I lines, in that the model is shallow. This is clear from He I λ4471. A lower temperature may improve the He I fit, but this would make the fit to He II worse. This difference could be due to the fixed resolution of the spectra or, alternatively, could be due to the H to He ratio that is expected to change at lower metallicity (e.g., Martins & Palacios 2021), but is currently fixed at the solar value. Despite some modest disagreements, the model spectrum appears to provide a good overall characterization of K1. Our formal fitting consistently suggests K1 is a late-type O dwarf.

K2/B2: K2/B2 is an O9.7 V type star. We find a small mismatch in $\log(g)$ between spectroscopy and photometry. The emission line contamination of the Balmer lines may affect our spectroscopic determination of $\log(g)$. The discrepancy in $\log(g)$ is likely why $M_{spec}$ is in disagreement with $M_{act}$. We further note that $M_0 < M_{spec}$, which is physically impossible.

We find a >3σ discrepancy between the RV measured for K2/B2 with Keck (−13 ± 2 km s$^{-1}$) and Binospec (16 ± 4 km s$^{-1}$). We do not observe this effect in any other star in our sample. We suggest that K2 may be a binary star system. This is potentially further supported by the odd present-day-to-initial-mass relation (see Section 17). Currently it is not possible to verify the binary status with our limited time information. We are in the process of acquiring time series spectra with Keck to further investigate this possibility.

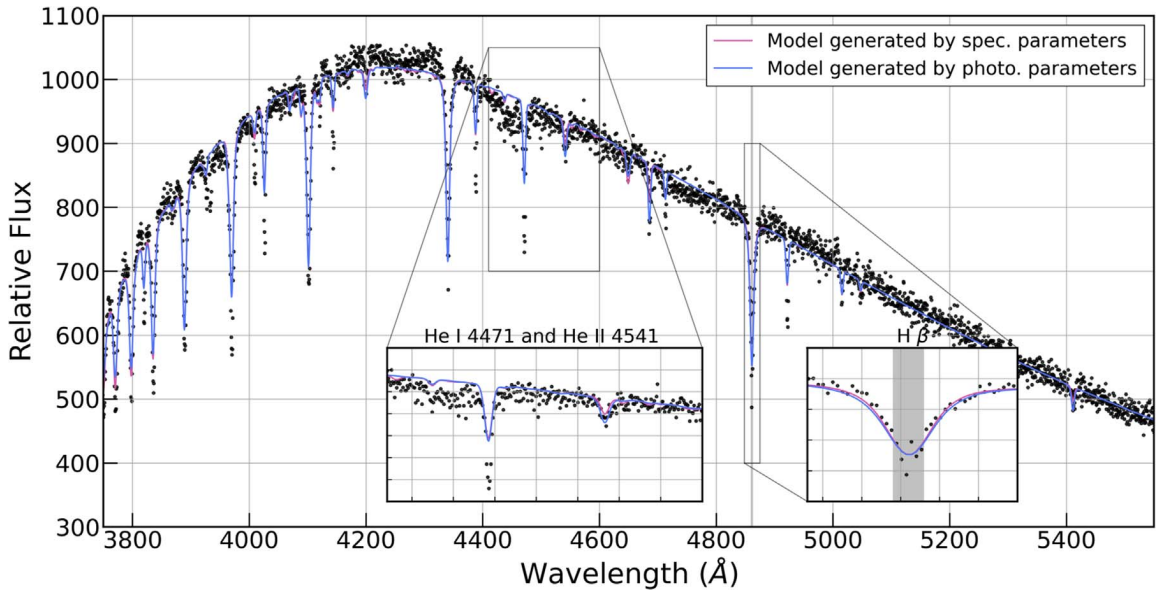

**Figure 9.** Comparison between the parameters derived for K1 using full spectral fitting and SED fitting. We show a plot of the observed spectra (black dots), the TLUSTY-based model generated by the median parameters derived from the full spectral fitting (pink), and the TLUSTY-based model generated by the median parameters derived from the SED fitting (blue). While both models reasonably replicate the observed He I 4471 and He II 4541 fits, both struggle to fully fill the absorption feature of the helium lines. This shortcoming is potentially due to the fixed resolution, metallicity, or macroturbulence. We see the effect of the different $\log(g)$ values between the methods most prominently in the Balmer lines (see the H$\beta$ feature). The gray bar shows the part of the spectrum that we masked.

K7/B6: K7/B6 is an O9.5 V type star and yields consistent parameters from photometry and spectroscopy. Although K7 is classified as an O-type star, it is not near any of the known H II regions in Leo A. We discuss the possibility of it being a runaway star in Section 6.2.

B13: B13 yields an S/N of ∼12 making the classification challenging. The star is likely an O9.7 V type star, but could be consistent with a B0 V type star as well. Its Hα feature is fully in emission, and the remaining Balmer lines are in absorption with some light emission within the feature. Given that the star shows narrow emission features, poor spectroscopic fitting, and extended emission in the 2D spectra, we postulate that the star must have some nebular contamination.

### 5.3.2. B-type Stars

We classify eight stars as B-type star in our sample (K3/B3, B4, B7, B9, B10, B11, B12, and B14) based on their EW measurements. Overall the parameters derived from spectroscopy and photometry yield consistent values with the EW classification. We present a detailed example of the analysis of K3/B3 and highlight interesting features of the other stars.

K3/B3: K3 is the third brightest star in our sample (F336W$_{VEGA}$ = 18.87 and F475W$_{VEGA}$ = 20.4). Its locations in the optical and UV CMDs are typical of an MS star. Its spectrum is shown in Figure 8. We list its EW measurements in Table 3 and comment only on a subset of those features.

The spectrum broadly suggests the star is a B-type star: He I, Mg II, Si II, and Si III are in absorption. The Balmer series shows no Hα, while Hβ is primarily in absorption. For Hβ, there is emission in the red wing, which could either be consistent with the wavelength of two O II features at λ4865Å and λ4872Å and the forbidden [Fe IV] emission at λ4868Å, or the kinematic components of some Hβ emission. The remaining Balmer lines appear to have little to no emission in their red wings. The Hβ in the 2D spectra shows that the emission is spatially extended with a slope toward redder wavelengths, as opposed to perpendicular to the trace. The features appear to spatially extend to ∼5.5 pc. We find Fe II at λ6540Å in emission to the blue side of where Hα should be. We further find Fe II in emission at λ8617Å and [S IV] at λ8575Å. Based on the EW, we use the Castro et al. (2008) and Evans et al. (2015) classification schemes and find that K3 is most likely a B1.5 V star.

From full spectral fitting, we find K3 has median values of $\log(T_{eff}) = 4.42 \pm 0.01$ dex and $\log(g) = 3.75 \pm 0.05$ dex, which, like the EW ratios, indicate that K3 is an early-type B star. Using Equations (1)–(3), we derive $M_{spec} = 20 \pm 5 M_\odot$, $M_0 = 13 \pm 1 M_\odot$, and $\log(L/L_\odot) = 4.53 \pm 0.12$ dex. These are all in line with the classification derived from the EW measurements and location in the CMD. The spectroscopic parameters are consistent with the results from SED fitting, which yield $\log(T_{eff}) = 4.41 \pm 0.01$ dex, $\log(g) = 3.66 \pm 0.01$ dex, and $M_{ini} = 12.9 \pm 0.17 M_\odot$.

We postulate that the majority of the emission is due to the nebula southwest of the star, which appears to fall into our slit, based on the HST imaging. The nebula can explain the emission features near Hβ as well as Fe II and other emission at redder wavelengths. The 2D spectra further support this scenario, since the emission primarily extends to only one side of the stellar trace. Furthermore, the nebula shape suggests that we may be observing a bow shock nebula (e.g., Gull & Sofia 1979; van Buren & McCray 1988; Meyer et al. 2014). This is supported by the following: K3 is a runaway candidate (see

Section 6.2), the nebula is the most visible in the F110W and F160W imaging, and the absence of Hα. To capture the intrinsic characteristics of the nebula, its formation, and the star-nebula interaction, follow-up observations are necessary, ideally with an integral field spectrograph.

B4: B4 is a B0 V type star and yields consistent parameters from photometry and spectroscopy. The presence of emission lines coupled with the star not being near an H II region suggest the emission is not gaseous in origin. Instead, this may indicate the presence of a circumstellar nebula or of stellar winds. Given that Hα shows only weak emission and He II 4686 in absorption, the stellar wind scenario seems unlikely (e.g., Lamers & Cassinelli 1999). Alternatively, the star could have a companion that provides the emission components to the spectra.

B7: B7 is a P-Cyg B1 V or a Be1 V type star. The star resembles the traditional P Cyg star (Rivet et al. 2020, and references therein). We do not observe metal lines with P-Cygni profiles. This is not surprising given Leo A's low metallicity. However, the lines are narrower than what would be expected from stellar wind emission lines. Alternatively, the star could be a Be star, whose disk is oriented so that we coincidentally capture the emission resembling the P-Cygni profile. This could also explain the lower $v \sin i$ value, since Be stars typically have higher $v \sin i$ values (e.g., Iqbal & Keller 2013; Rivinius et al. 2013; Arcos et al. 2018). Potentially this could also explain why, in comparison to the other Be stars in this sample (see Section 5.3.3), the star does not show signs of potential NIR excess in the optical CMD. Extended spectroscopic coverage into the UV could help shed light on wind parameters, or photometric coverage into the NIR to shed light on a potential disk. We discuss a more clear example in Section 5.3.3.

B9: B9 is a B1.5 V type star. Hα appears in emission, and all other Balmer lines are in absorption. B9 yields consistent parameters from photometry and spectroscopy.

B10: B10 is a B0.5/Be0.5 V type star. The photometric parameters and spectroscopic parameters are consistent. Both methods yield decent to high $\chi^2$ values (see Tables 4 and 5) suggesting that potentially a circumstellar disk or nebulae could affect the star. The stellar emission and the high $v \sin i$ value would support a Be star scenario. We discuss more clear examples in Section 5.3.3.

B11: B11 is a B2 V type star. Most Balmer lines are in absorption with variable amounts of emission. Since the B11 spectrum shows an S/N of <10, we cannot reliably recover the spectroscopic parameters. Given the S/N is ∼10, we refrain from making any further speculations as to the origin of the emission.

B12: B12 is a B1.5 V type star. The star shows signs of stellar emission, but the low S/N (∼12) prohibits detailed analysis. We do note that we only have four bands in the photometry. B12 is one of the few stars that we do not find consistency between the two methods, although both $\chi^2$ values would suggest they are good or decent fits.

B14: B14 is a B3 V type star. Hα is in absorption, while the other Balmer lines show potential emission in their wings. Due to the low S/N (∼15) of the spectrum, the weak emission in the Balmer lines could be artifacts of the data. log($T_{\text{eff}}$) is consistent between the two methods, while log($g$) shows a discrepancy likely due to the affected Balmer lines.

### 5.3.3. Classical Be Stars

Our sample contains several objects that appear to be classical Be stars: rapidly rotating, main-sequence B stars with emission lines that originate in a circumstellar accretion disk (Rivinius et al. 2013). The key spectroscopic feature is emission in the Balmer lines (usually strongest in Hα), while the key photometric feature is an IR excess due to the extended circumstellar disk. We provide the analysis of the clearest example (K6) in this section. In total we find six stars (K4, K5, K6, K8, B5, and B8) for which we can identify both key features. They all show strong Balmer emission lines, and their particular SEDs cannot be reproduced by a "normal" single star model.

K6: K6 (F475W$_{\text{VEGA}}$ = 20.84 and F336W$_{\text{VEGA}}$ = 19.35) is located on the BHeB sequence in the optical CMD, but appears on the MS in the UV-optical CMD. Its spectrum is shown in Figure 8. We list its EW measurements in Table 3 and comment only on a subset of those features. Stars K4, K5, K8, and B5 share similar CMD characteristics. The Keck spectrum of K6 is shown in Figure 8.

The spectral features suggest K6 is a fast rotating early B-type star. K6 shows He I in absorption and emission, and Mg II, S II, and Si IV in absorption. He II appears to potentially be a mix of emission with absorption. We find strong emission in Hβ. We find an emission feature within an absorption feature for Hγ and Hδ. However, Hε appears to be mostly in absorption. All of the features are very shallow, which is typical of fast rotators. The signs of strong rotation and emission mean that detection of EWs might be harder to recover for certain features (e.g., Si, Mg, and He II), since they become so shallow that they are indistinguishable from the noise. Based on Lennon (1997), Castro et al. (2008), and Evans et al. (2015), we classify it as a Be2 V star.

The strong emission appears to be associated with the star, as opposed to being gaseous in origin. K6 is not near the known H II regions (see Figure 3). Moreover, the lack of typical nebular emission lines further suggests it is not from an H II region. Leščinskaité et al. (2022) identified K6 as a strong Hα emitter based on the Subaru/Suprime-Cam narrowband imaging, postulating that it might be an emission star.

Due to the strong emission in the Balmer lines, the likely presence of a disk and the high $\chi^2$ in the photometry, we adopt here the alternative fitting methods introduced in Section 4.3.

Using this adapted spectroscopic method, we determine that the star has log($T_{\text{eff}}$) = 4.43 ± 0.02 dex and a $v \sin i$ = 220 ± 10 km s$^{-1}$. This is in line with an early-type Be star (Rivinius et al. 2013). The high $v \sin i$ is a key indicator that the star has to be a rotating MS star as opposed to a BHeB star. An evolved star would break apart at such high $v \sin i$. As previously mentioned, the star is among the bluest in the UV CMD but among the reddest in the optical CMD.

Using our composite photometry fitting (see Section 4.3), we determine that log($T_{\text{eff}}$)$_{\text{star}}$ = 4.48$^{+0.04}_{-0.03}$ dex, which is in line with our spectroscopic measurement. The disk log($T_{\text{eff}}$)$_{\text{disk}}$ = 3.73 ± 0.03 dex. This serves primarily as an estimate, since we use the model of a cool star to mimic the disk spectra.

Figures 10 and 11 show the median photometric model along with samples from the posterior. In Figure 10 we can clearly see that the UV photometry is fitted well by a rotating MS B-type star. In the NIR, the fit is not as good. However, since we are not actually fitting a disk model, this could explain

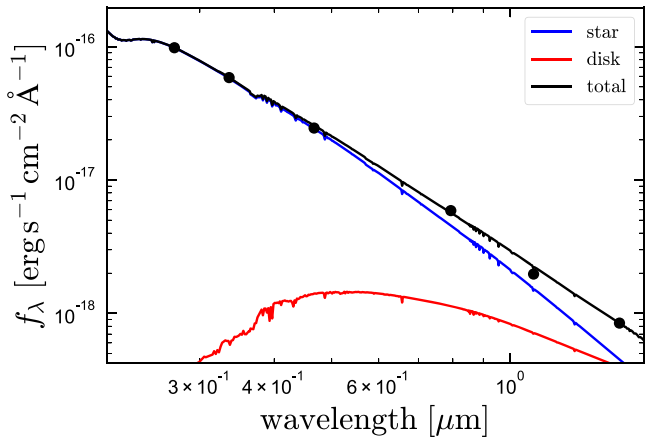

**Figure 10.** Models for K6 from adapted composite SED fitting (see Section 4.3). The data (black dots) are matched by scaling and combining the stellar models. The star (blue line) and the disk (red line) are combined for the composite SED (black line). The stellar and the disk models are both varied to find the median SED. We note that the error bars of the data are smaller than the dots representing the data point.

the discrepancy. Overall, the fit yields a good $\chi^2$ value (see Table 6).

We considered other explanations for the unusual photometric SED. The first possibility could be dust giving rise to NIR excess. However, given how blue the star is in the UV CMD, we believe dust is unlikely. A second possibility is an unresolved stellar companion. We tried fitting two stellar models simultaneously. However, one of the stars always yields nonphysical parameters (i.e., a temperature and radius combination that is not expected for any star). Given that there is no evidence for a red star in the spectra, and we are not able to fit a physically reasonable star to the photometric data, this scenario is unlikely.

We note here that this does not mean that the star is necessarily a single star, simply that there is no evidence for a secondary polluting our current data.

Based on the present evidence, we suggest this is a Be star. This provides the first spectroscopic confirmation for Be stars in Leo A.

K4: K4 is a Be1 V type star. Due to the strong emission in the Balmer lines and the high $\chi^2$ (>300) in the BEAST fitting, we postulate it is a star surrounded by a disk. Another possibility is an unresolved red stellar companion. For this particular star, we cannot completely exclude this possibility, since we do not have the NIR photometry to determine the nature of the excess. However, given that for other Be stars in our sample we find the same characteristics as for K4, we postulate that extinction occurring in the photometry is due to a disk.

K5: K5 is a Be2 V type star. Similar to K4, we assume that the star is a Be star whose photometry is affected by extinction/reddening due to the disk, as opposed to extensive dust or a red companion.

B5: B5 is a Be3 V type star. The 2D spectra show extended emission, which suggest it is due to nebular emission from the nearby H II region. However, the H$\alpha$/ H$\beta$ emission ratio suggests that the H$\alpha$ emission cannot be purely gaseous. While our data cannot exclude a binary influencing the star (e.g., we need time series spectra), the spectroscopic features and SED fit are consistent with a circumstellar disk.

K8: K8 is a Be2 V type star. Similarly to K6, the star is a Be star whose photometry is affected by extinction/reddening due to the disk as opposed to extensive dust or a red companion.

B8: B8 is a shell Be3 V type star. H$\alpha$ and H$\beta$ are in emission and both show a double-peaked line profile. The distinct Balmer profiles suggest that the star is most likely a shell star (i.e., a Be star viewed edge-on). Despite the low $v \sin i$, the strong emission and spectral shape support the classification of the star as a shell star.

### 5.4. The Utility of Panchromatic Photometry for Massive Star Studies

We have shown the first extensive and quantitative comparison between stellar parameters for massive stars derived from optical spectroscopy and UV-optical-NIR broadband photometry at low metallicity.

Figure 12 shows the comparison between photometric and spectroscopic $\log(T_{\rm eff})$, $\log(g)$, and $\log(L/L_{\odot})$ for stars that have $\chi^2 < 100$ in both methods. We find that a median absolute deviation (MAD) between the two $\log(T_{\rm eff})$ of <0.01 dex and a median difference of $-0.03$ dex (1600 K). The 16th and 84th percentiles of the difference span 0.05 dex (i.e., $\pm 0.025$ dex around the median), indicating that the observed differences are slightly smaller than scatter in the data. For $\log(g)$, the MAD is 0.18 dex and the median difference is $0.06 \pm 0.20$ dex. Lastly, for $\log(L/L_{\odot})$, we find an MAD of 0.05 dex and a median difference of $-0.03 \pm 0.06$ dex. The clear outlier in each panel is B12. B12 is among the lowest S/N stars in our spectroscopic sample (S/N~ 12) and has modest emission, though not as strong as other stars, which could explain the larger differences between the photometric and spectroscopic parameters. In general, these findings show that for "normal" O/B star MS stars in our sample (i.e., those with no or little emission), photometric and spectroscopic parameters are consistent with each other.

This comparison clearly shows that panchromatic photometric studies can yield reliable stellar parameters at least for "normal" single stars, which is encouraging especially when spectra are not available. For example, most of the LUVIT sample consist of galaxies too far away for acquiring optical spectra with the current generation of telescopes. Poor photometric fits with the BEAST are often due to stellar activity (e.g., binarity, disks) that are not captured by single star models. We find that with full UV through NIR broadband SEDs it is possible to identify such candidates by their poor $\chi^2$ values, providing an invaluable way to identify unusual objects for future follow-up (e.g., with JWST, ELT, GMT, and TMT). We note that the BEAST is in development of including more models.

### 6. Discussion

#### 6.1. Stars in H II Regions

Leo A hosts several known H II regions (see Figure 1 of van Zee et al. 2006), two of which are included in the HST and the spectroscopic data set. Previous analysis of these H II region ionized gas abundance patterns have placed indirect constraints on the properties of the ionization fields, and in turn, the underlying types of ionizing sources (e.g., the spectral type of the OB stars; Skillman et al. 1989; van Zee et al. 2006; Ruiz-Escobedo et al. 2018). There have been a few direct studies of the massive stars that power H II regions at such low metallicities (e.g., Garcia et al. 2014; Camacho et al. 2016; Evans et al. 2019; Telford et al. 2021b). In most cases, these studies have found that a singular star is responsible for powering most of the H II region.

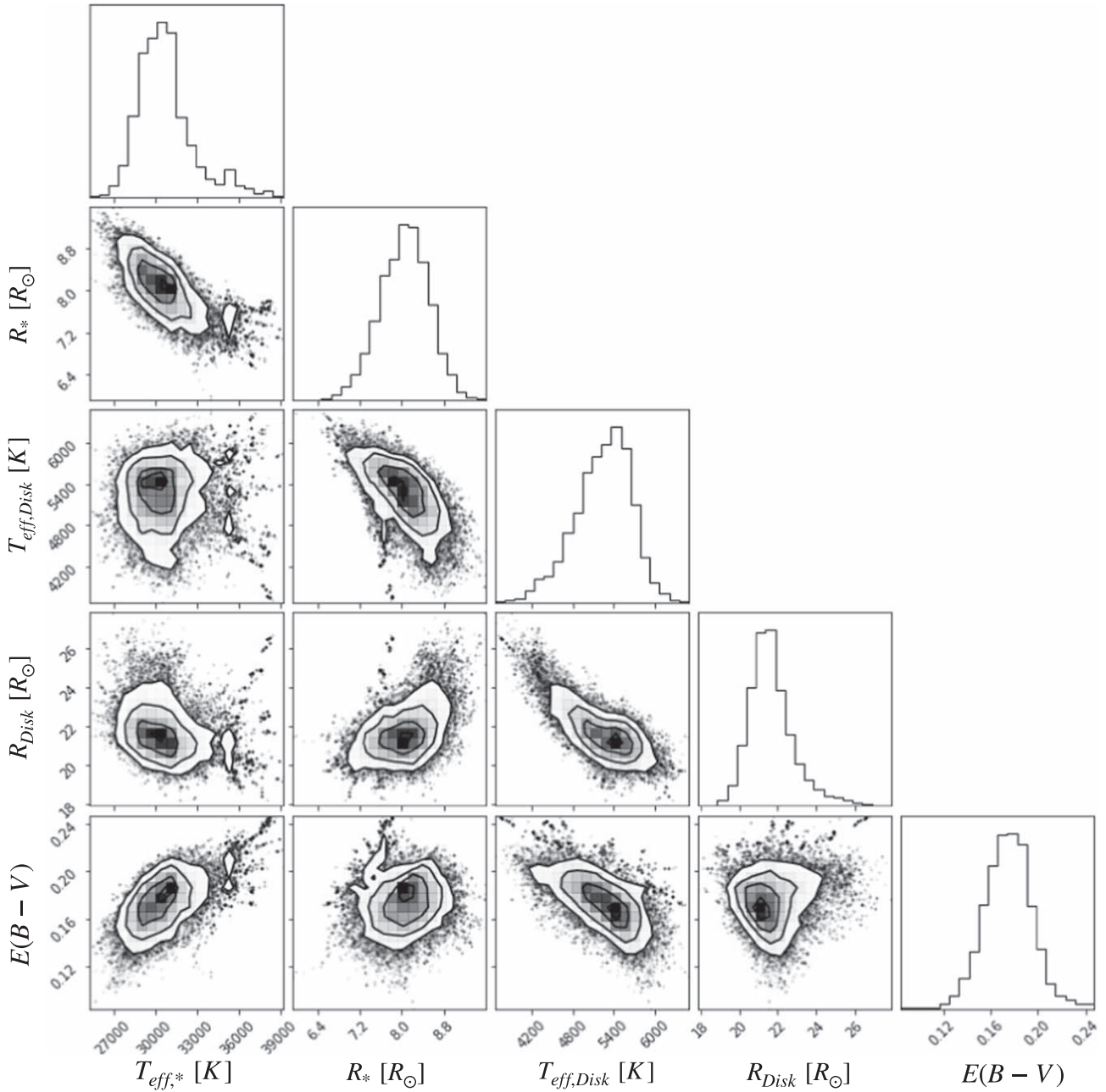

**Figure 11.** Corner plot of stellar parameters from the adapted double SED fitting of K6. The first two columns show the contours for the derived $\log(T_{\mathrm{eff}})$ and radius of the star. The next two columns show the contours for the derived $\log(T_{\mathrm{eff}})$ and radius of the disk. The last column shows $E(B-V)$.

In this section, we discuss the results of our stellar spectroscopic and SED fitting in the context of literature nebular line analysis of the H II regions in Leo A.

Figure 13 shows the two H II regions that are covered by our UV and optical HST data along with their corresponding optical CMDs. Stars located approximately within each H II region are color-coded by their age from our SED fitting, while the CMD of the entire Leo A field population is shown for reference in gray. For clarity of discussion, we label the two H II regions: Region A and Region B. We have six band coverage for Region B, and four band (two UV and two optical) coverage for Region A.

### 6.1.1. Region A

Region A is the left H II region in Figure 13. It is listed as Leo A+069 018 in van Zee et al. (2006). Region A is a diffuse H II region with weak [O III]$\lambda$5007, which suggests a low ionization scenario (van Zee et al. 2006). After correcting for extinction ($E(B-V) = 0.021$; Schlegel et al. 1998), they determine a B0—O9.5 zero-age MS star to be the likely source of ionizing photons.

**Table 6**
Stellar Parameters Determined Using the Adapted Methods

| Star | $\log(T_{\mathrm{eff,spec}})$ | $\log(T_{\mathrm{eff,SED*}})$ | $R_{\mathrm{SED*}}$ | $v_{\sin i}$ (km s$^{-1}$) | $E(B-V)$ | $\log(T_{\mathrm{eff,DISK}})$ | $R_{\mathrm{DISK}}$ | $M_{\mathrm{model}}$ ($M_\odot$) | $V_{\mathrm{R\,helio,corr}}$ (km s$^{-1}$) | $V_{\mathrm{R\,Gauss}}$ (km s$^{-1}$) | $\chi^2$ | Spectral Type |
|---|---|---|---|---|---|---|---|---|---|---|---|---|
| (1) | (2) | (3) | (4) | (5) | (6) | (7) | (8) | (9) | (10) | (11) | (12) | (13) |
| K4 | $4.45 \pm 0.01$ | ... | ... | $339 \pm 10$ | ... | ... | ... | 10 | $29 \pm 10$ | $16 \pm 16$ | ... | Be1 V |
| K5 | $4.46 \pm 0.02$ | ... | ... | $370 \pm 90$ | ... | ... | ... | 10 | $46 \pm 11$ | $31 \pm 25$ | ... | Be2 V |
| B7 | $4.44 \pm 0.02$ | ... | ... | $125 \pm 11$ | ... | ... | ... | 10 | $44 \pm 6$ | $36 \pm 30$ | ... | P-Cyg B1 V / Be1 V |
| K6 | $4.43 \pm 0.02$ | $4.48^{+0.04}_{-0.03}$ | $8.0^{+0.6}_{-0.6}$ | $220 \pm 5$ | $0.17 \pm 0.02$ | $3.73^{+0.03}_{-0.03}$ | $21.5^{+1.2}_{-1.0}$ | 10 | $18 \pm 6$ | $18 \pm 18$ | 3 | Be2 V |
| K8 | $4.44 \pm 0.03$ | $4.51^{+0.04}_{-0.06}$ | $5.9^{+0.5}_{-0.2}$ | $290 \pm 10$ | $0.18 \pm 0.04$ | $3.64^{+0.10}_{-0.04}$ | $12.6^{+1.9}_{-1.2}$ | 10 | $28 \pm 8$ | $38 \pm 17$ | 19 | Be2 V |
| B5 | $4.40 \pm 0.01$ | $4.38^{+0.18}_{-0.04}$ | $9.0^{+0.8}_{-1.1}$ | $165 \pm 15$ | $0.13 \pm 0.03$ | $3.74 \pm 0.17$ | $12.0^{+8.7}_{-1.7}$ | 8 | $36 \pm 10$ | $22 \pm 15$ | 12 | Be3 V |
| B8 | $4.42 \pm 0.03$ | $4.37^{+0.15}_{-0.03}$ | $6.9 \pm 1.0$ | $135 \pm 80$ | $0.12 \pm 0.06$ | $3.74^{+0.06}_{-0.10}$ | $13.7 \pm 1.7$ | 8 | $28 \pm 4$ | $23 \pm 24$ | 20 | shell Be3 V |

**Note.** The table lists the results from the adapted spectroscopic and photometric analysis. The columns are: (1) name; (2) spectroscopic temperature of the star; (3) photometric temperature of the star; (4) photometric radius of the star; (5) rotational velocity; (6) reddening value from the SED fitting; (7) photometric temperature of the disk; (8) photometric radius of the disk; (9) present-day mass picked for the combined SED fitting; (10) heliocentric corrected radial velocity from full spectral fitting; (11) heliocentric corrected radial velocity measurements using Gaussian fitting; and (12) reduced $\chi^2$ value from the composite photometric fitting. Columns (2), (5), (10), and (11) are the stellar parameters derived from the adapted spectroscopic analysis. Columns (3), (4), (6), (7), (8), and (12) are the stellar parameters derived from the composite photometric analysis.

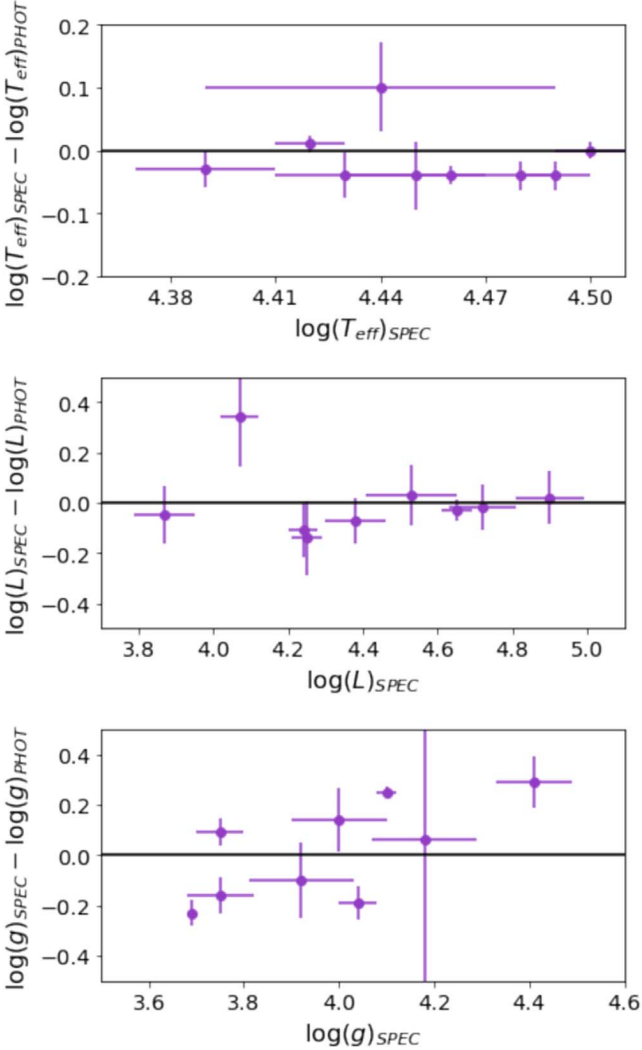

**Figure 12.** A comparison between the parameters derived from the `BEAST` using the models with v/vcrit = 0.4 (PHOT) and the parameters derived from the `PAYNE` (SPEC). The top panel compares $\log(T_{\rm eff})$. The middle panel compares $\log(L/L_\odot)$. The bottom panel compares $\log(g)$. The comparison between the two methods shows that panchromatic photometry can be used to recover reliable stellar parameters at least for "normal" single stars.

The brightest star in Region A is K2. From our spectra and photometry, we find that K2 is an O9.7V type star with a $\log(T_{\rm eff}) = 4.50 \pm 0.01$ dex, which is very similar to what van Zee et al. (2006) suggested. Furthermore, K2 has an age of $\log(t) = 7.0 \pm 0.07$ (yr), $M_{\rm ini} = 16.81 \pm 0.21 M_\odot$, and $A_{\rm V} = 0.06 \pm 0.02$. For K2, we are able to calculate the intrinsic ionizing photon production rate using the `BEAST` (Choi et al. 2020). We derive a median rate of $10^{47.6\pm0.1}$ photons s$^{-1}$. Using the H$\alpha$ value reported in van Zee et al. (2006), we compute the photoionization rate required to produce the H$\alpha$ luminosity (Hummer & Storey 1987; Storey & Hummer 1995; Kennicutt 1998; Choi et al. 2020). We derive a rate of $10^{47\pm0.05}$ photons s$^{-1}$. This is consistent with the value derived for K2 from the `BEAST`. This suggests that K2 could be responsible for a majority of the ionization in Region A. Furthermore, this finding supports the, thus far, low extinction from dust along the line of sight, since we would otherwise expect our derived stellar value to be larger than the value reported in literature.

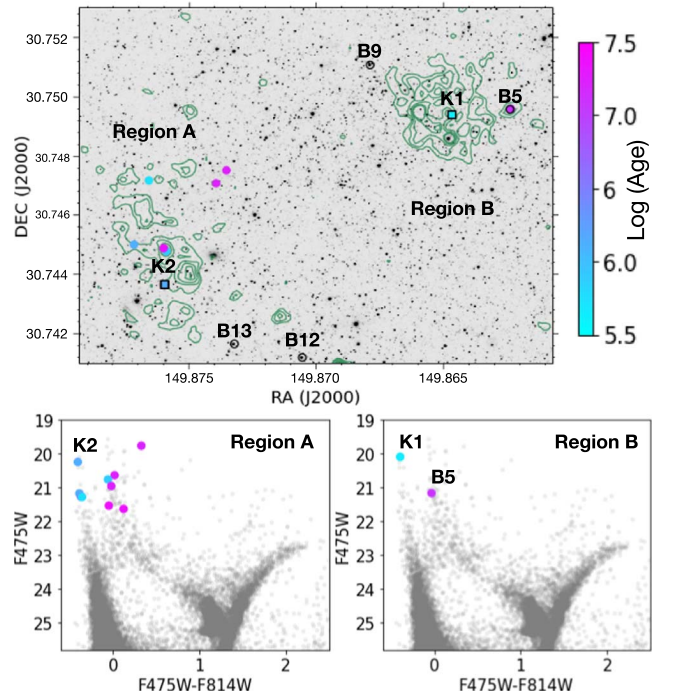

**Figure 13.** Top panel: the Leo A HST ACS/F475W image zoomed in on two H II regions. Stars with spectroscopic observations are labeled by their reference names. Contours of H$\alpha$ emission from ground-based imaging are overplotted in green. The H II regions are overplotted with the stars color-coded by log(Age) <7.5 dex from the `BEAST` fits and include only good fits from the UV-optical catalog. Bottom panel: the optical CMDs are overplotted with the stars shown in the top panel and use the same color-coding for both regions.

### 6.1.2. Region B

Region B is the right region in Figure 13. It is described as the east H II region in Ruiz-Escobedo et al. (2018), who also noted that it is faint and appears to have low extinction ($c$ (H$_\beta$) = 0.08 $\pm$ 0.05).

Ruiz-Escobedo et al. (2018) detected only the hydrogen Balmer series in emission as well as [O II] $\lambda3727$Å, [N II] $\lambda6583$Å, and [S II] $\lambda6717, 6731$Å. Since they were only able to obtain an upper limit for [O III]$\lambda5007$Å, they determined that a star with $\log(T_{\rm eff}) \geqslant 4.45$ is necessary and an ionizing photon production rate of $10^{48.0}$ photons s$^{-1}$ is required to produce the observed H II region size of 1.5 pc, assuming an electron density of 100 cm$^{-3}$.

Star K1 is located in Region B. From our spectra, we find K1 is an O9V type star. We find $\log(T_{\rm eff}) = 4.49 \pm 0.01$ from our spectroscopic analysis and $\log(T_{\rm eff}) = 4.52 \pm 0.02$ from SED fitting. Both are in reasonable agreement with the estimate of $\log(T_{\rm eff}) \geqslant 4.45$ from Ruiz-Escobedo et al. (2018). Our SED fitting shows that K1 has $\log(t) = 6.9 \pm 0.08$ (yr), $M_{\rm ini} = 20.1 \pm 1.29 M_\odot$, and $A_{\rm V} = 0.05 \pm 0.01$. The age and mass of this star are comparable to what is expected for an H II region, while the extinction is similar to what is derived from the nebular emission analysis. We find that B5 is in the H II region as well. However, B5 is a Be star, which makes the age and mass determination more difficult. For K1, we are able to calculate the intrinsic ionizing photon rate using the `BEAST`. We derive a median rate of $10^{48.1\pm0.1}$ photons s$^{-1}$, which is consistent with the derivation from Ruiz-Escobedo et al. (2018). This supports that the majority of the ionization is likely coming from K1. Again, our results suggest that

Table 7
Parameters of Isolated Stars in Leo A

| Star | $M_{ini}$ | $M_0$ | Spectral Type | Age | $V_R$ | Dist. to Nearest H II Region/Cluster | Velocity (Dist./Age) |
|------|-----------|-------|---------------|-----|-------|--------------------------------------|----------------------|
| | $(M_\odot)$ | $(M_\odot)$ | | (Myr) | (km s$^{-1}$) | (pc) | (pc/Myr) |
| (1) | (2) | (3) | (4) | (5) | (6) | (7) | (8) |
| K3 | $12.9 \pm 0.17$ | ... | B1.5V | $15.8 \pm^{2.7}_{2.4}$ | $27 \pm 2$ | >200 | >11–15 |
| K7 | $15.3 \pm 1.06$ | ... | O9.5V | $6.3 \pm^{2.6}_{1.8}$ | $24 \pm 23$ | >200 | >22–45 |
| B4 | $16.2 \pm 0.18$ | ... | B0V | $10 \pm^{1.7}_{1.5}$ | $30 \pm 3$ | >200 | >17–23 |
| B11 | $5.9 \pm 0.69$ | ... | B2V | $63 \pm^{32}_{16.3}$ | ... | >40 | >1 |
| B12 | $7.3 \pm 1.21$ | ... | B1.5V | $39.8 \pm^{17.7}_{12.3}$ | $30 \pm 8$ | ~40 | ~1–2 |
| B14 | $9.3 \pm 0.99$ | ... | B3V | $20 \pm^{8.9}_{6.1}$ | $20 \pm 5$ | ~50 | ~2–3 |
| B10 | $12.8 \pm 1.21$ | ... | B0.5V/Be0.5V | $12.6 \pm^{3.6}_{2.8}$ | $31 \pm 11$ | >200 | >12–20 |
| B7 | ... | $11 \pm 2$ | B1V/Be1V | 5–50 | $44 \pm 6$ | ~70 | ~1–14 |
| K4 | ... | $13 \pm 2$ | Be1V | 5–50 | $30 \pm 10$ | ~60 | ~1–12 |
| K5 | ... | $14 \pm 2$ | Be2V | 5–50 | $35 \pm 11$ | ~80 | ~2–16 |
| K6 | ... | $12 \pm 2$ | Be2V | 5–50 | $18 \pm 6$ | ~100 | ~2–20 |
| K8 | ... | $17 \pm 2$ | Be2V | 5–50 | $28 \pm 8$ | ~200 | ~4–40 |
| B8 | ... | $10 \pm 2$ | Be3V | 5–50 | $28 \pm 4$ | ~230 | ~5–46 |

**Note.** The columns are: (1) star name; (2) initial mass (photometry); (3) initial mass (spectroscopy); (4) spectral type; (5) stellar age in megayears; (6) radial velocity; (7) approximate distances to the closest H II region or cluster for each star; and (8) approximate velocity derived from taking the value in column (7) and dividing it by the values in column (5). When available we list the initial mass; otherwise, we list the evolutionary mass. 1—The star has a really high $\chi^2$-value making the age unreliable and therefore the velocity unreliable as well.

dust along the line of sight in Leo A has to be low to negligible.

Overall, we find consistency between previous studies of the H II regions in Leo A and our study of massive stars in those regions. van Zee et al. (2006) considered whether the lack of higher O-type stars can be explained by either a truncated initial mass function (IMF) or that the H II regions are evolved. To explore this, we consider the ages derived by the BEAST of the stars (K1, K2, and B5) in those regions (Figure 13). Given that those stars are high-mass stars, they are already at the end of their lifetime on the MS. In Figure 13, we can clearly see that Region B is a more evolved H II region than Region A, since the average age of the massive stars in the region is older. However, we also find very few massive stars in both H II regions. Lastly, the results suggest that in both H II regions a majority of the ionization is coming from a singular O-star.

### 6.2. Field OB-stars

It is well established that most stars form in clustered environments (e.g., Lada & Lada 2003; Krumholz et al. 2019). Accordingly, OB-stars are expected to be found in or near H II regions or star clusters, as their short lifetimes imply they do not have time to wander to significant distances before ending their stellar lives.

However, a subset (20%–30%) of massive stars are found in the "field" with no clear association to clusters or H II regions (e.g., Dorigo Jones et al. 2020). The existence of those stars poses a dilemma for stellar kinematics, star formation theory, IMF sampling, etc. (e.g., Blaauw 1961; Clarke & Pringle 1992; Hoogerwerf et al. 2000; Oey et al. 2004; de Wit et al. 2005; Renzo et al. 2019). This has led to major differing theories to the formation of those stars: in situ star formation (Oey et al. 2013) and ejection (Oey et al. 2018; Dorigo Jones et al. 2020). Recent studies (e.g., Oey et al. 2018; Dorigo Jones et al. 2020; Vargas-Salazar et al. 2020) favor the latter scenario in most cases, which is further broken down into two different type of ejections: a dynamical ejection scenario (DES) and a binary supernova scenario (BSS). While both require massive stellar binaries, in the DES, a close encounter with a binary system ejects the star (or binary) at high velocity out of the cluster or H II region. In the BSS scenario, a core-collapse supernova of the more evolved star ejects its OB companion or sometimes the entire system (Dorigo Jones et al. 2020).

The majority (~70%, 13 stars) of the stars in our spectroscopic sample are not within $d \sim 40$ pc of any H II region. To obtain a rough understanding of which mechanism is responsible for the position of the stars, we derive estimated required velocities from the age estimate yielded by the BEAST and distances from the H II region map (Figure 3). We summarize the derived values in Table 7. In this discussion, we take advantage of the coincidence that 1 pc Myr$^{-1}$ is 0.98 km s$^{-1}$.

K3, K7, B4, and B10 can be considered runaway candidates, since their lower limit for velocity suggest that the stars could have been ejected. All four stars are at least 200 pc from the nearest H II region or cluster. They are among the youngest field stars (with the following ages in megayears: K3: 15.8 $\pm^{2.7}_{2.4}$, K7: 6.3 $\pm^{2.6}_{1.8}$, B4: 10 $\pm^{1.7}_{1.5}$, and B10: 12.6 $\pm^{3.6}_{2.8}$). Although recent studies (e.g., Renzo et al. 2019; Schoettler et al. 2020; Vargas-Salazar et al. 2020) suggest a cutoff for runaways around ~30 km s$^{-1}$, we still can consider those four stars runaway candidates, since we only report lower limits. Due to the lack of precise velocity measurements, it is currently not possible to distinguish whether the stars were ejected through DES or BSS.

B11, B12, and B14 are more likely walkaway candidates (de Mink et al. 2014; Renzo et al. 2019; Schoettler et al. 2020; Vargas-Salazar et al. 2020). Overall, the stars are on average ~40–70 pc to the closest H II region or nearest H II region. In comparison to the other group, they are older in age (with the following ages in megayears: B11: 63 $\pm^{22}_{16.3}$, B12: 39.8 $\pm^{17.7}_{12.3}$, and B14: 20 $\pm^{8.9}_{6.1}$). Their estimated velocities are of the order a couple of parsecs per megayear. This is in line with intermediate-mass runaway velocities (Renzo et al. 2019). However, these types of walkaways remain rare (Renzo et al. 2019), and therefore are impossible to find without precise velocity measurements, much less to associate with different mechanisms.

Five of our Be stars (K4, K5, K6, K8, and B8) and B7 are also in the field. Since we do not have age measurements, we can only present rough velocity ranges in Table 7. All five stars show velocity limits consistent with runaway stars. In the literature, Be stars in the field have been associated with BSS (Dorigo Jones et al. 2020). However, we cannot exclude that DES could be responsible for the ejection and velocity of these stars.

Further studies of OB-stars in Leo A would likely provide a better understanding on the overall distribution of the mechanism responsible for runaway stars at extremely low metallicity.

### 6.3. Purity of the Blue Core Helium Burning Sequence

BHeB stars provide important constraints on stellar evolution and serve as excellent tracers of recent star formation (Langer & Maeder 1995; Dohm-Palmer et al. 2002; Dolphin et al. 2003; Larsen et al. 2011; Leščinskaitė et al. 2022). In particular, the extent of the blue loop and the ratio of blue to red core helium burning stars are sensitive to several poorly constrained aspects of massive stars physics (e.g., rotation, convective overshoot; Alongi et al. 1991; Ritossa 1996; McQuinn et al. 2011; Jie et al. 2015; Georgy et al. 2021; Eldridge & Stanway 2022; Farrell et al. 2022). The luminosities and spatial positions of BHeBs in nearby galaxies have been used to construct spatially resolved maps of star formation over the past few hundred megayears (Dohm-Palmer et al. 2002; McQuinn et al. 2012; Leščinskaitė et al. 2022).

However, some of the data and findings presented in this paper suggest that not all stars that reside in the BHeB region of an optical CMD are bona fide BHeBs. This has been previously suggested by Stonkutė & Vansevičius (2022) and Leščinskaitė et al. (2022). In our combined photometric and spectroscopic sample, we find that indeed, six stars in our sample out of seven that appear to be in the BHeB region on the optical CMD (K4, K5, K6, K8, B5, and B8) are, in fact, spectroscopically Be-type stars. This provides the first spectroscopic confirmation for Be stars in Leo A.

We compare the UV and optical CMDs to see if any other stars that appear to be in the MS region in the UV CMD are in the BHeB region in the optical. We compare stars that are brighter than 23.5 mag in the F475W filter. We further apply the following cuts: F475W-F814W >0.2, F475W-F814W < −0.15, and F275W-F475W <0.0. Lastly, we apply cuts by eye on stars that are on the MS region. These cuts are visualized in Figure 14. We find that ∼5%–10% of the stars in the BHeB region above 23.5 mag in F475W exhibit this behavior. This finding suggests that when taking the BHeB/RHeB-ratio from photometric surveys, we have to take into account that the BHeB region could be contaminated by Be stars. However, since we can disentangle the Be star candidates in the UV, corrections can be applied when multiwavelength coverage is available. Lastly, the findings propose a new avenue to potentially search for and detect Be star candidates through photometry.

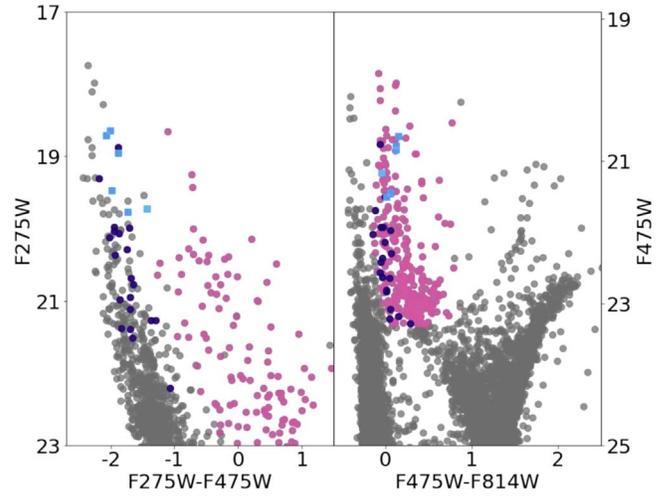

**Figure 14.** UV-optical (left) and optical (right) HST-based CMDs of Leo A. The pink colored circles are stars that appear in the BHeB region both in the UV and the optical CMD. The blue colored circles are stars that are on the MS in the UV CMD, but appear on or near the BHeB region in the optical CMD. The cyan squares are the stars identified as Be stars in our spectroscopic sample. This shows candidate Be stars could possibly pollute the BHeB in the optical CMD, and that we can filter for candidate Be stars in panchromatic photometry studies.

### 6.4. Extremely Metal-poor Environments

#### 6.4.1. Spectral Type vs. Effective Temperature

Spectroscopic studies of sub-SMC metallicity massive stars continues to be challenging, because of the large distances to the nearest galaxies that host them. Though tremendous efforts have gone into characterizing metal-poor massive stars in nearby dwarf galaxies (e.g., Kaufer et al. 2004; Venn et al. 2004; Bresolin et al. 2006, 2007; Evans et al. 2007; Garcia & Herrero 2013; Hosek et al. 2014; Tramper et al. 2014; Camacho et al. 2016, 2017; Garcia 2018; Evans et al. 2019; Garcia et al. 2019c, 2019a; Telford et al. 2021b; Lorenzo et al. 2022), most our knowledge remains based on larger studies in the SMC (e.g., Heap et al. 2006; Mokiem et al. 2006; Hunter et al. 2009; Penny & Gies 2009; Dunstall et al. 2011; Bouret et al. 2013; Dufton et al. 2019; Ramachandran et al. 2019; Bouret et al. 2021).

In Figure 15, we compare our spectroscopic sample stellar types to other samples of massive stars with spectra at SMC and sub-SMC metallicities with published temperature and spectral type. We compare the four O-stars to the best fit of the Massey et al. (2009) sample and the Dufton et al. (2019) $\log(T_{\rm eff})$ to spectral type relation. We further compare the stars to two stars in Sextans A: S3 Telford et al. (2021b) and OB521 Camacho et al. (2016). We note that two of our stars (K2 and B13: $\log(T_{\rm eff}) = 4.50$ dex and SpT is O9.7V) fall on top of each other in the $\log(T_{\rm eff})$ versus spectral type plot, which is why we only see three O-stars in the plot.

We find that our O-stars are consistent with the sub-SMC stars S3 and OB521. When comparing to the SMC trend lines, we see that our parameters are consistent with Massey et al. (2009), but find lower $\log(T_{\rm eff})$ for the same spectral type than Dufton et al. (2019). However, O-stars in Sextans A and Leo A appear to have similar $\log(T_{\rm eff})$ to the SMC sample from Massey et al. (2009). They are also all lower in $\log(T_{\rm eff})$ than the relation derived by Dufton et al. (2019), contradictory to expectation (see, e.g., Mokiem et al. 2004). One possibility is the difference in the underlying models. While Telford et al. (2021b) and our work use the `TLUSTY` models, other groups use other models (e.g., `CMFGEN`, Hillier & Lanz 2001; `FASTWIND`, Santolaya-Rey et al. 1997), introducing at least some uncertainty.

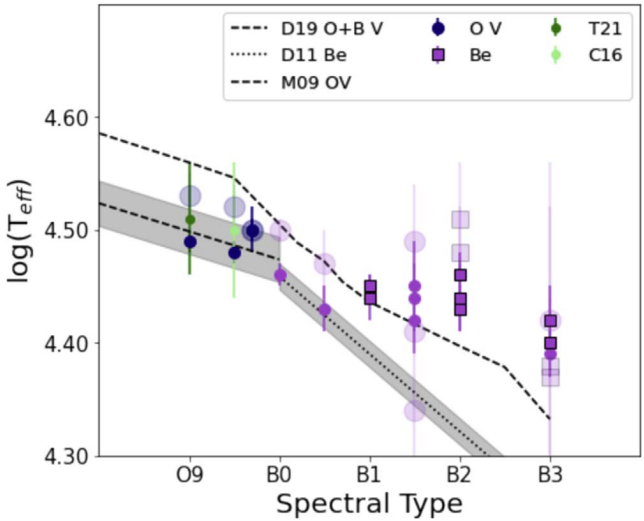

**Figure 15.** Properties of the OB-stars in Leo A. $\log(T_{eff})$ is plotted versus spectral type for early-type (O-B3) stars in our sample. The spectroscopically derived $\log(T_{eff})$ are indicated by the solid markers, while the photometrically derived $\log(T_{eff})$ are the fainter bigger markers. The O-stars in Leo A are blue circles, the B-stars are purple circles, and the Be stars are purple squares. We compare our stars ($\chi^2 \leqslant 100$) with SMC results and Sextans A late-type O stars. For the SMC O-stars, we fit a best line to the V type stars in Massey et al. (2009; leftmost black dashed line, M09). For the B-type stars, we plot the O V and B V spectral types to the $\log(T_{eff})$ scale derived by Dufton et al. (2019; dashed line, D19). We also plot the fit to the Be stars in the SMC from Dunstall et al. (2011; dotted line, D11). Furthermore, we plot the two late-type O stars observed by Telford et al. (2021b; dark green, T21) and Camacho et al. (2016; light green, C16) in Sextans A, for which there is a $\log(T_{eff})$ measurement for the spectral type.

For the photometrically determined values of $\log(T_{eff})$, we find that for two of the stars (K2 and B13), we derived hotter $\log(T_{eff})$, while for the other two (K1 and K7), the photometry yields the same $\log(T_{eff})$.

We compare our B-stars to the SMC relation for B-stars from Dufton et al. (2019) for the dwarfs and to Dunstall et al. (2011) for the Be stars. For the B0-B1 stars, we find that we are cooler than the Dufton et al. (2019) derived relation. For stars after B1, the stars seem to be closer to the derived relation and hotter. We find that all of our Be stars are hotter than the derived relationship from the SMC for Be stars, both spectroscopically and photometrically. We did not find any sub-SMC B V stars to compare our sample to. While, for the O-stars, the slope between our points appears similar to the literature, we find that we have a flatter fall off for our B-stars. Since the B stars in our sample represent the first MS stars we have both spectral type and photometry for, they provide the first insight into the scale at sub-SMC. Overall, we find that at later spectral types $\log(T_{eff})$ decreases in our sample, as expected. However, there is a large spread, which is likely due to decreasing data quality. Increasing our sample of sub-SMC stars is crucial to obtain a full understanding of the relation between spectral type and $\log(T_{eff})$.

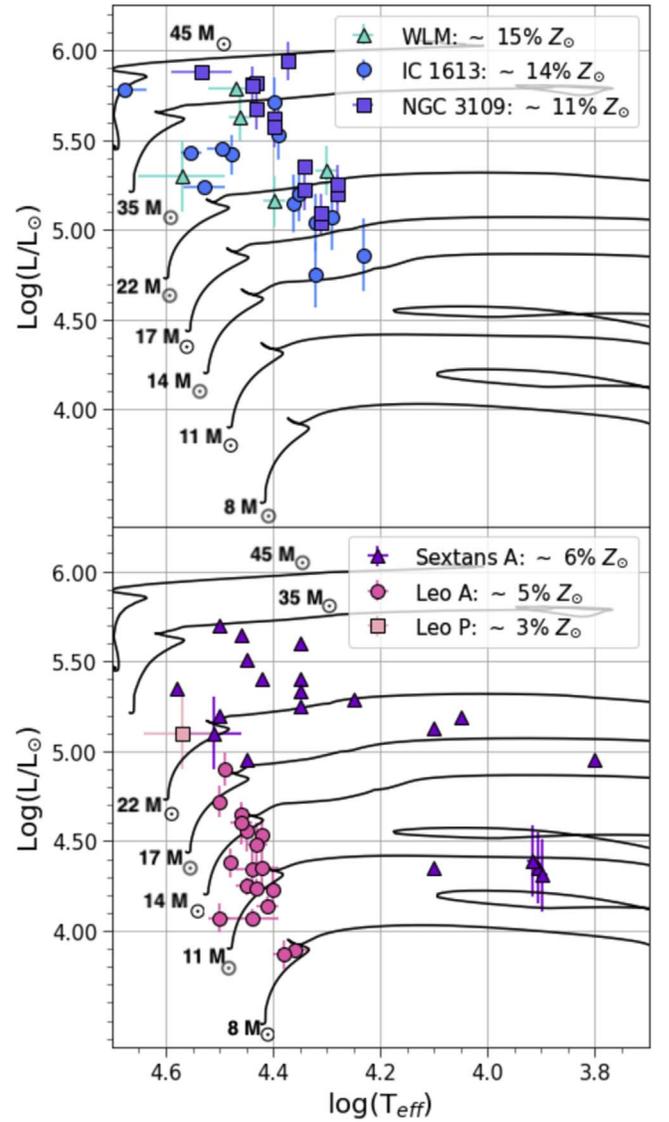

**Figure 16.** The HR diagrams of massive sub-SMC metallicity stars in the Local Group with derived stellar parameters. We overplot MIST tracks. Top: all of the stars with temperatures and luminosities in the metal-poor galaxies: WLM (∼15% $Z_\odot$; aqua triangle), IC 1613 (∼14% $Z_\odot$; blue circles), and NGC 3109 (∼11% $Z_\odot$; purple-blue squares). The WLM stars are from Bresolin et al. (2006), Tramper et al. (2011, 2014), and Telford et al. (2021b); the IC 1613 stars are from Bresolin et al. (2007), and the NGC 3109 stars are from Evans et al. (2007), Tramper et al. (2014), and Hosek et al. (2014). Bottom: all stars with temperatures and luminosities in the extremely metal-poor galaxies Sextans A (∼6% $Z_\odot$; purple triangle) and Leo P (∼3% $Z_\odot$; rose squares). As well as the stars in our spectroscopic sample (∼5% $Z_\odot$; pink circles). The Sextans A stars are from Kaufer et al. (2004), Camacho et al. (2016) and Telford et al. (2021b); the Leo P star is from Telford et al. (2021b). Our sample represents the first extended sample of MS stars at sub-SMC metallicities with derived stellar parameters. Furthermore it is the largest sample at sub-10% $Z_\odot$ with derived stellar parameters.

### 6.4.2. Current Observational Knowledge of Metal-poor Massive Stars

In Figure 16, we plot HR diagrams for the entire sub-SMC metallicity star sample in the literature with temperatures and luminosities that have been derived from spectroscopy. We note that recently Lorenzo et al. (2022) collected ∼150 spectra of metal-poor massive stars; however, currently there are no stellar parameters for this sample, and we cannot include them on this plot. We overplot the MIST stellar tracks at select masses for reference. We indicate the gas-phase metallicities for each galaxy after placing them on a common scale using Grevesse et al. (2010) for the solar abundance ($12 + \log(O/H) = 8.69$).

The top panel shows metal-poor galaxies WLM ($Z \sim 15\%$ $Z_\odot$; Bresolin et al. 2006; Tramper et al. 2011, 2014 and Telford et al. 2021b), IC 1613 ($Z \sim 14\%$ $Z_\odot$; Bresolin et al. 2007) and NGC 3109 ($Z \sim 13\%$ $Z_\odot$; Evans et al. 2007; Tramper et al. 2014 and Hosek et al. (2014)). The bottom panel shows extremely metal-poor galaxies Sextans A ($Z \sim 10\%$ $Z_\odot$; Kaufer et al. 2004; Camacho et al. 2016; Telford et al. 2021b) and Leo P ($Z \sim 3\%$ $Z_\odot$; Telford et al. 2021b), along with stars in our Leo A spectroscopic sample.

We clearly see that the literature thus far has populated mostly the very upper area of the HR diagram. This is not surprising, since the most luminous stars are more accessible to observe. In comparison, our sample extends to fainter luminosities, later types, and less evolved stars.

We double the number of OB-stars thus far with spectroscopically derived stellar parameters at sub-10% solar metallicity. It also contains the first sample of spectroscopically confirmed sub-SMC Be stars and the largest sample of sub-SMC B-type dwarf stars.

### 6.4.3. Initial and Present-day Masses across Analysis Methods

The mass of a star is among its most fundamental parameters. Here, we examine the masses we have measured from photometric SED fitting and from spectroscopy. SED fitting yields an initial mass of $M_{ini}$ and a present-day mass of $M_{act}$. Spectroscopic fitting provides an initial mass of $M_{evo}$ and present-day masses of $M_{evo}$ and $M_{spec}$. As a reminder, $M_{evo}$ is determined by using $\log(T_{eff})$, $\log(L/L_\odot)$ and evolutionary tracks, whereas $M_{spec}$ is determined using Equations (1)–(3) as explained in Section 2.2. The exact derivation of each of these masses is given in Sections 5.1 and 4. The different masses are listed in Tables 4 and 5 and comparison plots are presented in Figure 17. We note that the comparison includes stars that have $\chi^2 < 100$ in both methods.

For the initial masses, $M_{ini}$ and $M_0$, we find an MAD between the two masses of $0.59 M_\odot$ and a median difference of $1.1 \pm 1.1 M_\odot$. Overall, the $M_{ini}$ values are slightly larger than $M_0$, but they are all roughly within the uncertainty except for B12 (which was pointed out as an outlier in Section 5.4). B12 shows a difference ∼3 times larger than the uncertainty and shows a significantly higher initial mass ($M_0 \sim 11$ $M_\odot$) for spectroscopy. For the present-day masses of the stars in the sample, we find overall that $M_{act}$ and $M_{evo}$ are consistent with each other. We find an MAD between the two masses of $0.69 M_\odot$ and a median difference of $0.99 \pm 1.42 M_\odot$. Similar to the initial masses, the $M_{act}$ values are slightly larger than $M_{evo}$, but within the uncertainty. We find B12 remains an outlier.

In Figure 17, we also compare $M_{evo}$ and $M_{spec}$. We do not include $M_{act}$, since the values agree well with $M_{evo}$. In principle, $M_{evo}$ and $M_{spec}$ should be the same, and differences between them are used as a consistency check for evolutionary models. For example, as first noted by Herrero et al. (1992), there have been multiple studies (e.g., Herrero et al. 2002; Massey et al. 2005; Weidner & Vink 2010; Mahy et al. 2015; Markova & Puls 2015; McEvoy et al. 2015; Ramírez-Agudelo et al. 2017; Sabín-Sanjulián et al. 2017; Markova et al. 2018; Schneider et al. 2018; Castro et al. 2021) that discuss the tension between $M_{evo}$ and $M_{spec}$ for O-type stars. The tension rises from $M_{evo}$ being larger than $M_{spec}$, although they should be the same. Most of these studies suggest that the discrepancy is the most significant at $M > 40 M_\odot$. We do not have any stars this massive, but we can examine differences in the ∼10 − 20 $M_\odot$ range. For all of the stars in our sample, we find an MAD between the two masses of $3.0 M_\odot$ and a median difference of $< 0.01 \pm 5.3 M_\odot$. To get a sense of typical differences for stars that agree, we consider stars that agree within the uncertainty measurement, and we find an MAD between the two masses of $1.5 M_\odot$ and a median difference of $0.0 \pm 2.2 M_\odot$. We find three outliers in our comparison, K2 (∼1.5 times larger), K3 (∼1.5 times larger), and B14 (∼3 times larger). In Section 5.3, we noted that these stars show a slightly higher-than-expected $\log(g)$ value, which is likely driving up the mass, in comparison to the stars' photometry. We note that these stars also show $M_{spec} > M_0$, which is not physical, and is likely due to a high $\log(g)$ value or $\log(L/L_\odot)$ value. It is interesting that K2 and K3 are outliers, since the K2 observations suggest it could be a binary candidate, while K3 has a peculiar spectra, which can be explained by a binary scenario. Overall, we find that for "normal" single stars, the evolutionary model derived masses and spectra derived masses are in agreement at the level of 1–2 $M_\odot$.

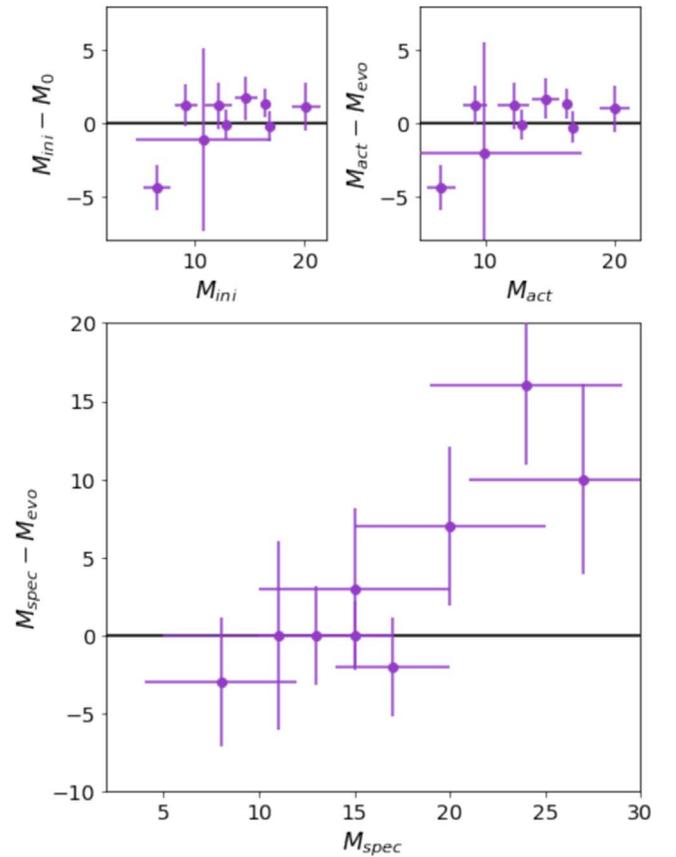

**Figure 17.** Three panels comparing the different masses derived throughout the study: $M_{ini}$ and $M_{act}$ from the photometry, $M_0$, $M_{evo}$, and $M_{spec}$ from the spectroscopy. The top-left panel compares the initial masses derived from each method ($M_{ini}$ vs. $M_0$). The top-right panel compares the present-day masses derived from each method using evolutionary tracks and isochrones ($M_{act}$ vs. $M_{evo}$). The bottom panel compares the present-day masses derived from spectroscopy ($M_{spec}$ vs. $M_{evo}$). We show all stars with $\chi^2 < 100$ in both photometry and spectroscopy.

### 6.5. Finding Be Stars in Low-metallicity Environments

We provide the first confirmation of Be stars in Leo A. We also spectroscopically observed the lowest-metallicity Be stars to date. We confirm six Be stars (K4, K5, K6, K8, B5, and B8) and label two B stars as potential Be stars (B7 and B10). Excluding the last two stars, we find that for our spectroscopic sample, ~45% of B-type stars are Be stars. This is high compared to the Be fraction in the Milky Way (≈17%; Zorec & Briot 1997), LMC (≈20%; Martayan et al. 2006), and SMC (≈30%; Martayan et al. 2007a, 2007b), and appears in line with the expected trend of an increased Be star fraction at low metallicity due to weaker winds and faster rotation rates. However, our sample is also biased toward early spectral types, where the Be occurrence rate is higher; further study is needed to confirm this trend.

Furthermore, we show that Be stars in Leo A can be photometrically identified by a strong IR excess relative to stars near the main sequence in UV colors, even without spectroscopic follow-up. We show that the panchromatic SED allows us to put some constraints on both the stellar and disk parameters. While we explore the simplest parameters for the Be stars, the combination of our obtained spectra and photometry presents a unique opportunity to gain first insights on the disk physics at low metallicity. Lastly, the identification through an NIR excess is especially promising with JWST. HST UV and Hα imaging surveys can now be combined with the NIR imaging to identify and characterize Be stars.

## 7. Conclusions

We have presented the first combined stellar spectroscopy and HST photometry analysis of the extremely metal-poor OB-stars in Leo A, a low-luminosity, dwarf galaxy at a distance modulus of $\mu = 24.40$ mag. Our findings from the LUVIT HST, LRIS (KECK), and Binospec (MMT) observations are as follows:

1. We analyze the brightest OB-stars (F475W < 22 and F475W − F814W < 0.5) in Leo A using the LUVIT HST photometry. We spectroscopically observe and analyze a total of 18 stars within the photometric sample. We find four O-type dwarfs, eight B-type dwarfs, and six Be stars.

2. We derive stellar parameters of the photometric sample using the `BEAST`. We find that for good fits ($\chi^2 < 80$) log($T_{eff}$) and log($g$) increase for bluer F475W−F814W colors as expected from stellar models. For stars on the MS region of the CMD, $M_{ini}$ trends as expected; the brightest stars (e.g., F275W <18 ) have the highest $M_{ini}$ with masses up to $M_{ini} \sim 20 M_\odot$. Similarly, $A_V$ matches findings from previous studies, which suggest little to no dust in Leo A ($A_V \lesssim 0.08$). However, for stars on the BHeB region of the CMD, the trends do not appear quite as clear, which we expand upon below.

3. We use `The Payne` and two different publicly available TLUSTY spectroscopic grids for O-stars and B-stars to measure spectroscopic parameters for 18 stars. We measure log($T_{eff}$), log($g$), $V_R$, and $v \sin i$ for each star. We use MCMC sampling to obtain the uncertainties on all parameters. We find our spectroscopic sample has a log($T_{eff}$) range of 4.39 to 4.50, a log($g$) range of 3.69 to 4.64, and a $v \sin i$ range of <80 to 370. We find $V_R$ variation for K2 with Keck ($-13 \pm 2$ km s$^{-1}$) and Binospec ($16 \pm 4$ km s$^{-1}$), making it a viable binary system candidate.

4. The photometrically and spectroscopically derived parameters (log($T_{eff}$), log($g$), and log($L/L_\odot$)) are consistent to within $1\sigma$ for stars with $\chi^2 < 100$ in both methods. We find an MAD of <0.01 dex for log($T_{eff}$), 0.18 dex for log($g$), and 0.05 dex for log($L/L_\odot$). When we find disagreement, it is usually due to stars with signs of emission, peculiar SED colors (i.e., IR excess), or high $\chi^2$ values. This is a promising avenue to study more distant low-metallicity galaxies for which resolved stellar spectra cannot be obtained.

5. We compute and compare both initial masses and present-day masses with photometry and spectroscopy. The initial stellar masses ($M_{ini}$ (photometry) and $M_0$ (spectroscopy)) determined by the two techniques are consistent within $1\sigma$. $M_{ini}$ ranges from $6.26 M_\odot$ to $20.1 M_\odot$, and $M_0$ ranges from $8.0 M_\odot$ to 19 $M_\odot$. Similarly, the present-day masses derived from the stellar parameters in combination with evolutionary tracks ($M_{act}$ (photometry) and $M_{evo}$ (spectroscopy)) are consistent within $1\sigma$. $M_{act}$ ranges from $6.26 M_\odot$ to $19.99 M_\odot$, and $M_{evo}$ ranges from $8.0 M_\odot$ to 19 $M_\odot$. We further compare $M_{evo}$ to the spectroscopic present-day mass ($M_{spec}$), and find that the stars are consistent to $1\sigma$ when excluding stars with $M_{spec} > 20 M_\odot$ (K2, K3 and B14). These stars show disagreements larger than $5 M_\odot$. $M_{spec}$ ranges from $8.0 M_\odot$ to $27 M_\odot$.

6. We find that the gas, RGB, and stellar velocities are consistent within $1\sigma$. We further are able to show that full spectral fitting is reliable at recovering RV measurements to higher precision.

7. Overall, we find that our study of massive stars in two H II regions is consistent with previous gas-phase studies of these H II regions in Leo A. The O-type stars (K1 and K2) in the H II regions provide the expected ionization power for each ($10^{48.1\pm0.1}$ photons s$^{-1}$ and $10^{47.6\pm0.1}$ photons s$^{-1}$, respectively), and suggest little dust along the line of sight.

8. We find a total of 13 isolated stars in our spectroscopic sample, i.e., they are not within $d \sim 40$ pc of any H II region. For seven stars, we are able to use the derived ages and approximate distances to the nearest H II region to estimate velocities. Four stars yield velocities consistent with runaway stars (i.e., $V \gtrsim 30$ km s$^{-1}$), while three are consistent with walkaway stars (i.e., $V < 30$ km s$^{-1}$).

9. We confirm that six of our spectroscopically observed stars are Be stars; these are the first at sub-10% solar metallicity. We further find that these stars are photometrically contaminating the BHeB region in the optical CMD. We unexpectedly find that ~5%–10% of stars in the BHeB region above 23.5 mag (F475W) in the optical CMD appear to be on the MS region in the UV CMD.

10. We find the observed sub-SMC metallicity late-type O dwarfs and our O dwarfs to be consistent in log($T_{eff}$) within $1\sigma$ for a given spectral type. We increase the sample of O dwarfs in the sub-10% solar metallicity regime with derived stellar parameters from five to nine. We provide the largest sub-SMC metallicity B dwarfs sample. We also spectroscopically observe the first sub-10% solar metallicity Be stars. We present the largest spectroscopic sample of MS stars at sub-SMC metallicity with derived stellar parameters.

M.G. thanks Miriam Garcia, Chris Evans, Sally Oey, Nate Bastian, Morgan Fouseneau, Dietrich Baade, and the referee for helpful discussions, feedback, and/or comments. M.G. acknowledges support of the "Schweizerische Studienstiftung" and UC Berkeley Cranor Fellowship. Support for this work was provided by NASA through grants GO-15275, GO-15921, GO-16162, GO-16717, AR-15056, AR-16120, and HST-HF2-51457.001-A from the Space Telescope Science Institute, which is operated by AURA, Inc., under NASA contract NAS5-26555.

The authors wish to recognize and acknowledge the very significant cultural role and reverence that the summit of Maunakea has always had within the indigenous Hawaiian community. We are most fortunate to have the opportunity to conduct observations from this mountain.

Part of the data presented herein were obtained at the W. M. Keck Observatory, which is operated as a scientific partnership among the California Institute of Technology, the University of California, and the National Aeronautics and Space Administration. The Observatory was made possible by the generous financial support of the W. M. Keck Foundation. Further, this data was made accessible by the Keck Observatory Archive (KOA), which is operated by the W. M. Keck Observatory and the NASA Exoplanet Science Institute (NExScI), under contract with the National Aeronautics and Space Administration.

Part of the data presented here were obtained at the MMT Observatory, a joint facility of the University of Arizona and the Smithsonian Institution.

This work is based on photometric observations made with the NASA/ESA Hubble Space Telescope, obtained from the data archive at the Space Telescope Science Institute (STScI). STScI is operated by the Association of Universities for Research in Astronomy, Inc. under NASA contract NAS 5-26555.

This work made extensive use of NASA's Astrophysics Data System Bibliographic Services.

## Appendix A
## The Effects of Stellar Model Choice on Photometric Fitting Results

We compare the stellar parameters derived from the `BEAST` using the different stellar models models it has available. They are: (i) MIST models with no rotation (MIST$_{NR}$; Choi et al. 2016); (ii) MIST models with rotation (MIST$_R$; Choi et al. 2017); and (iii) the PARSEC models that do not have rotation (PARSEC; Bressan et al. 2012). We limited this comparison to massive stars that fall within our spectroscopy color and magnitude limits (F475W < 22 and F475W − F814W < 0.5). This selects MS stars with $M \gtrsim 5 M_\odot$.

We compare parameters derived for the same stars from two different models. We divide the photometric SED fits into three groups based on the reduced $\chi^2$ values. Fits with $\chi^2 < 20$ for models being compared are considered good fits. Values of $\chi^2 \geqslant 80$ in either of the evolutionary models are considered poor fits. All other fits are fits of decent quality. These categories are only meant to be used qualitatively to facilitate discussion. When comparing the agreement between different parameters, we only use stars deemed to have good or decent fits.

Figure 18 shows a comparison between MIST$_R$ and MIST$_{NR}$ in the left column and the between MIST$_R$ and PARSEC in the right column. The y-axis is the difference between parameter values for the two models, and the x-axis is the MIST$_R$ parameter values.

For log($T_{eff}$) (top row), the MIST$_R$ versus MIST$_{NR}$, we find an MAD of 0.01 dex and median difference of −0.01 dex. The 16th and 84th percentiles of the difference span 0.04 dex (i.e., ±0.02 dex around the median). This indicates that the differences are smaller than scatter in the data. In general, these findings show that for decent and good-fit stars, the rotational and nonrotational models are generally consistent with each other. We note that the scatter is larger for stars with log($T_{eff}$) >4.0, where we find an MAD of 0.03 dex and scatter of 0.05 dex. As discussed in Choi et al. (2017), at this log($T_{eff}$), the MIST models change mass-loss prescriptions, which offers the possibility for increased scatter. For MIST$_R$ versus PARSEC, we find an MAD of 0.01 dex and a median difference of 0.00 ± 0.02 dex. The two models provide consistent values of log($T_{eff}$).

The second row shows log($g$). For the MIST$_R$ versus MIST$_{NR}$, we find an MAD of 0.05 dex and a median of −0.11 ± 0.09 dex. This implies that MIST$_{NR}$ has systematically larger log($g$) values than MIST$_R$. It is possible that rotation may affect log($g$) (e.g., due to line broadening), which could contribute to the disagreement. For the MIST$_R$ versus PARSEC, we find an MAD of 0.02 and a median difference of 0.00 dex. The 16th and 84th percentiles of the difference span 0.08 dex (±0.04 dex), indicating that MIST$_R$ versus PARSEC are consistent with each other.

The third row shows $M_{ini}$. For the MIST$_R$ versus MIST$_{NR}$, we find an MAD of 0.46$M_\odot$ and a median of −0.78 ± 0.97$M_\odot$. The two models are consistent, but MIST$_{NR}$ returns slightly larger values of $M_{ini}$ values for more massive stars. For the MIST$_R$ versus PARSEC, we find an MAD of 0.06$M_\odot$ and median difference of 0.05 ± 0.29$M_\odot$. MIST$_R$ and PARSEC are consistent with each other. There is one clear outlier, which appears to be a good fit (i.e., $\chi^2 < 20$) for both model fits. Interestingly, for all other parameters, the fits are consistent with each other. We currently do not have spectra to further investigate this star.

The fourth row shows log($L/L_\odot$). For the MIST$_R$ versus MIST$_{NR}$, we find an MAD of 0.04 dex and median of 0.15 ± 0.14 dex. In this case MIST$_R$ returns slightly higher luminosities than MIST$_{NR}$. For the MIST$_R$ versus PARSEC, we find an MAD of 0.02 dex and median difference of −0.01 ± 0.09 dex. Showing that MIST$_R$ versus PARSEC are consistent.

The fifth row shows log($A$). For the MIST$_R$ versus MIST$_{NR}$, we find an MAD of 0.10 dex and median of 0.20 ± 0.42 dex. We find the two models are consistent. For the MIST$_R$ versus PARSEC, we find an MAD of 0.00 dex and median difference of 0.00 ± 0.06 dex. The MIST$_R$ versus PARSEC models also yield values that are consistent.

The bottom row shows $A_V$. For the MIST$_R$ versus MIST$_{NR}$, we find an MAD of 0.04 dex and median of −0.01 ± 0.05 dex. Interestingly, most of the high $A_V$ values in either models are also associated with poor fits. The presence of larger $A_V$ values in a galaxy with minimal dust may be an indicator that we are not observing "normal" stars. Overall, the $A_V$ values from both models are consistent. For the MIST$_R$ versus PARSEC, we find an MAD of 0.02 dex and median difference

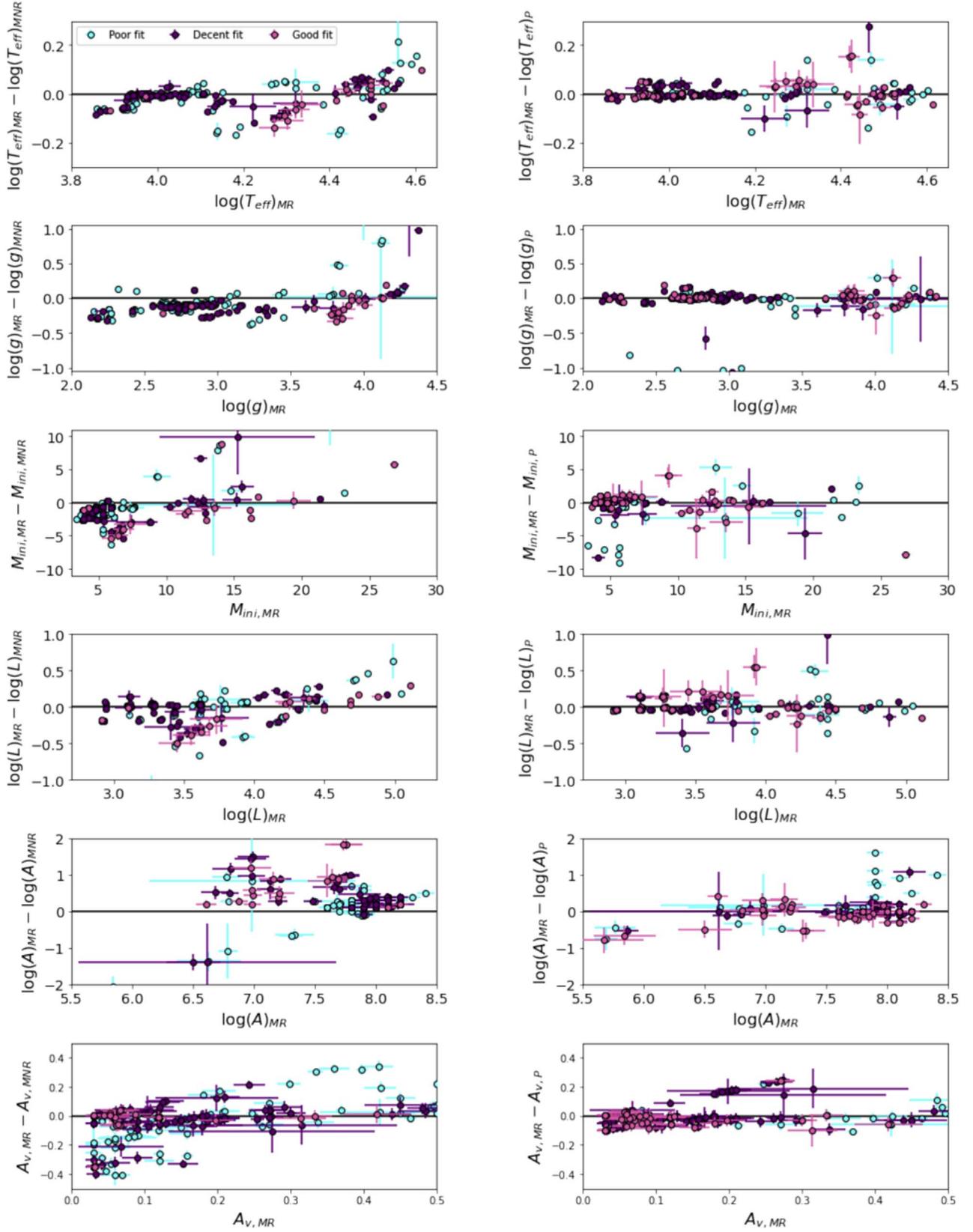

**Figure 18.** The figure shows a direct comparison between the parameters derived from the BEAST using the different available evolutionary models. The left panels compare the MIST models with v/vcrit = 0.4 (MR) and v/vcrit = 0.0 (MNR). The right panels compare the MIST models with v/vcrit = 0.4 (MR) to the PARSEC models (P). We plot the values for the stars in both our spectroscopic and photometric sample. The top two panels compare $\log(T_{eff})$. The second two panels compare the $\log(g)$. The third two panels compare $M_{ini}$. The fourth panels compare $\log(L/L_\odot)$. The fifth panels compare $\log(A)$. The bottom panels compare the $A_V$. Stars with $\chi^2 < 20$ for both fits are pink (good fits), stars with $\chi^2 > 80$ in either fit are cyan (poor fits), and all other stars are purple (decent fits).

**Table 8**
Photometric Parameters Determined Using the BEAST and the PARSEC Models

| Star | $M_{ini}$ $M_\odot$ | $\log(T_{eff})$ | $\log(g)$ | $\log(L/L_\odot)$ | $A_v$ | $\log(Age)$ | $\chi^2$ | No. of Filters | Notes |
|---|---|---|---|---|---|---|---|---|---|
| (1) | (2) | (3) | (4) | (5) | (6) | (7) | (8) | (9) | (10) |
| K1 | 24 ± 3.67 | 4.59 ± 0.05 | 4.08 ± 0.14 | 5.02 ± 0.12 | 0.06 ± 0.03 | 6.80 ± 0.18 | 17 | 6 | O9V |
| K2 | 16.8 ± 0.16 | 4.50 ± 0.01 | 3.88 ± 0.01 | 4.74 ± 0.01 | 0.03 ± 0.02 | 7.00 ± 0.07 | 13 | 4 | O9.7 V |
| K3 | 13.3 ± 1.05 | 4.42 ± 0.02 | 3.67 ± 0.06 | 4.54 ± 0.07 | 0.10 ± 0.04 | 7.15 ± 0.09 | 7 | 6 | B1.5 V |
| K7 | 15.9 ± 0.17 | 4.54 ± 0.01 | 4.24 ± 0.01 | 4.50 ± 0.02 | 0.08 ± 0.03 | 6.8 ± 0.07 | 21 | 4 | O9.5 V |
| B4 | 16.3 ± 0.15 | 4.50 ± 0.01 | 3.93 ± 0.01 | 4.69 ± 0.01 | 0.03 ± 0.02 | 7.0 ± 0.07 | 13 | 6 | B0V |
| B7 | 8.6 ± 0.08 | 4.34 ± 0.01 | 3.70 ± 0.01 | 3.99 ± 0.01 | 0.33 ± 0.07 | 7.5 ± 0.07 | 371 | 4 | P-Cyg B1 V / Be1 V |
| B9 | 15.9 ± 0.15 | 4.54 ± 0.01 | 4.24 ± 0.01 | 4.50 ± 0.01 | 0.48 ± 0.02 | 7.00 ± 0.07 | 104 | 6 | B2 V |
| B10 | 7.5 ± 0.16 | 4.33 ± 0.02 | 3.72 ± 0.01 | 3.86 ± 0.01 | 0.38 ± 0.03 | 6.8 ± 0.07 | 80 | 6 | B0.5 V |
| B11 | 4.7 ± 0.29 | 4.22 ± 0.02 | 3.70 ± 0.03 | 3.23 ± 0.06 | 0.08 ± 0.03 | 7.60 ± 0.08 | 16 | 6 | B2 V |
| B12 | 6.5 ± 2.05 | 4.30 ± 0.08 | 3.81 ± 0.14 | 3.57 ± 0.29 | 0.04 ± 0.04 | 7.70 ± 0.25 | 6 | 4 | B1.5V |
| B13 | 10.9 ± 0.11 | 4.47 ± 0.01 | 4.28 ± 0.01 | 4.05 ± 0.01 | 0.03 ± 0.03 | 7.00 ± 0.07 | 7 | 4 | O9.7 V |
| B14 | 5.3 ± 0.90 | 4.27 ± 0.04 | 3.83 ± 0.08 | 3.37 ± 0.14 | 0.03 ± 0.02 | 7.9 ± 0.17 | 16 | 6 | B3 V |

**Note.** The columns are: (1) star name; (2) initial mass; (3) temperature; (4) surface gravity; (5) luminosity; (6) extinction parameter; (7) stellar age in log-space; (8) reduced $\chi^2$ yielded by the BEAST; (9) number of HST bands used for the SED fitting; when the value is 4 we observed F275W, F336W, F475W, and F814W. When the value is 6, we additionally observed F110W and F160W. Columns (2), (6), and (7) list primary parameters, while columns (3), (4), and (5) list derived parameters.

**Table 9**
Photometric Parameters Determined Using the BEAST and the MIST Models with $v/vcrit = 0.0$

| Star | $M_{ini}$ $M_\odot$ | $\log(T_{eff})$ | $\log(g)$ | $\log(L/L_\odot)$ | $A_v$ | $\log(Age)$ | $\chi^2$ | No. of Filters | Notes |
|---|---|---|---|---|---|---|---|---|---|
| (1) | (2) | (3) | (4) | (5) | (6) | (7) | (8) | (9) | (10) |
| K1 | 19.0 ± 0.4 | 4.49 ± 0.01 | 3.87 ± 0.02 | 4.74 ± 0.01 | 0.03 ± 0.02 | 6.6 ± 0.07 | 18 | 6 | O9V |
| K2 | 16.0 ± 0.2 | 4.44 ± 0.01 | 3.79 ± 0.01 | 4.57 ± 0.01 | 0.03 ± 0.02 | 6.8 ± 0.07 | 12 | 4 | O9.7 V |
| K3 | 14.6 ± 1.1 | 4.41 ± 0.03 | 3.71 ± 0.10 | 4.48 ± 0.08 | 0.07 ± 0.05 | 6.9 ± 0.11 | 20 | 6 | B1.5 V |
| K7 | 15.0 ± 0.3 | 4.47 ± 0.01 | 4.13 ± 0.02 | 4.31 ± 0.01 | 0.03 ± 0.02 | 5.6 ± 0.08 | 36 | 4 | O9.5 V |
| B4 | 17.0 ± 0.5 | 4.48 ± 0.01 | 3.93 ± 0.05 | 4.59 ± 0.03 | 0.04 ± 0.02 | 6.6 ± 0.12 | 13 | 6 | B0V |
| B7 | 10.0 ± 0.2 | 4.33 ± 0.01 | 3.72 ± 0.01 | 3.98 ± 0.01 | 0.18 ± 0.03 | 7.2 ± 0.07 | 230 | 4 | P-Cyg B1 V / Be1 V |
| B9 | 14.4 ± 0.5 | 4.46 ± 0.01 | 4.13 ± 0.02 | 4.27 ± 0.04 | 0.40 ± 0.04 | 5.9 ± 0.28 | 107 | 6 | B1.5 V |
| B10 | 13.4 ± 0.8 | 4.43 ± 0.02 | 4.03 ± 0.05 | 4.22 ± 0.05 | 0.45 ± 0.04 | 6.5 ± 0.25 | 73 | 6 | B0.5 V |
| B11 | 11.3 ± 0.2 | 4.41 ± 0.01 | 4.15 ± 0.01 | 3.94 ± 0.01 | 0.13 ± 0.03 | 5.9 ± 0.10 | 9 | 6 | B2 V |
| B12 | 10.6 ± 0.9 | 4.38 ± 0.03 | 4.03 ± 0.06 | 3.90 ± 0.10 | 0.05 ± 0.04 | 6.8 ± 0.43 | 3 | 4 | B1.5V |
| B13 | 6.0 ± 0.1 | 4.59 ± 0.02 | 3.39 ± 0.02 | 4.34 ± 0.01 | 0.05 ± 0.03 | 7.9 ± 0.07 | 52 | 4 | O9.7 V |
| B14 | 11.0 ± 0.2 | 4.40 ± 0.01 | 4.13 ± 0.01 | 3.91 ± 0.01 | 0.13 ± 0.03 | 6.2 ± 0.07 | 200 | 6 | B3 V |

**Note.** The columns are: (1) star name; (2) initial mass; (3) temperature; (4) surface gravity; (5) luminosity; (6) extinction parameter; (7) stellar age in log-space; (8) reduced $\chi^2$ yielded by the BEAST; (9) number of HST bands used for the SED fitting; when the value is 4 we observed F275W, F336W, F475W, and F814W. When the value is 6, we additionally observed F110W and F160W. Columns (2), (6), and (7) list primary parameters, while columns (3), (4), and (5) list derived parameters.

of $-0.02 \pm 0.04$ dex, indicating that $MIST_R$ versus PARSEC derived parameters are consistent.

We list all of the parameters for the stars covered by the spectroscopic sample in Tables 8 and 9. We note that between all three evolutionary models, the $MIST_R$ models yield the best reduced $\chi^2$ values for all stars. Overall, $MIST_R$ versus PARSEC yields consistent results, meaning that they can be used interchangeably.

# Appendix B
## Detailed Discussion of Individual Stars

### B.1. O-Type Stars

**K2/B2:** K2/B2 is the second brightest star in our sample (F475W$_{VEGA}$ = 20.24 and F336W$_{VEGA}$ = 18.57). Like K1, its locations on the UV and optical CMDs indicate it is likely an O star. Its spectrum is shown in Figure 8. We list its EW measurements in Table 3 and comment only on a subset of those features.

The spectrum shows the characteristics of a typical late-type O-star. We find emission line contamination in the Balmer series, which is not surprising given K2's location in the middle of an H II region. The narrow emission lines are particularly clear for Hα and Hβ and weaker in the Hγ. The 2D spectrum shows that emission is spatially extended, consistent with a nebular origin. In most cases, we can only derive upper limits of EWs for the absorption features. Thus, we can only roughly estimate the spectral type. Following the EW classification process described for K1, we find that K2 is an O9.7 V type star in all of the schemes (Castro et al. 2008; Sota et al. 2011; Martins 2018).

When fitting the full spectrum using The Payne, we mask lines clearly affected by emission (e.g., the Hα feature and the center of the Hβ feature). The spectroscopic analysis yields log($T_{eff}$) = 4.50 ± 0.01 dex and log($g$) = 4.10 ± 0.02 dex, which is consistent with a late O-type MS star. We find a $v \sin i$ = 95 ± 20 km s$^{-1}$, which is reasonable for a late O-type star. Using Equations (1)–(3), we derive $M_{spec}$ = 27 ± 6$M_\odot$ and $M_0$ = 17 ± 1$M_\odot$, and log($L/L_\odot$) = 4.72 ± 0.09 dex. Overall, this agrees well with the EW measurements and the CMD location. We find the SED fitting yields log($T_{eff}$) = 4.50 ± 0.01 dex and log($g$) = 3.85 ± 0.02 dex, which are also consistent with a late O-type star.

**K4:** K4 is among the brighter stars (F475W$_{VEGA}$ = 20.65 and F336W$_{VEGA}$ = 19.04) in our spectroscopic sample. On the CMD, it is located on the BHeB sequence in the optical, but appears on the MS in the UV-optical CMD. Its spectrum is shown in Figure 8. We list its EW measurements in Table 3 and comment only on a subset of those features. It is not located near a known H II region (Figure 3).

Several of the spectral features suggest a Be-type star: He I, Mg II, Si III, and Si IV are in absorption. The Balmer lines are strong in emission for Hβ and Hγ, and we observe emission features within absorption for Hδ and Hε. Based on the EWs, we classify K4 as Be1 V (Castro et al. 2008; Evans et al. 2015).

K4 only has measurements in four bands (two UV and two optical), since the stars falls outside of the NIR footprint. This makes the composite photometry method not viable, since we cannot constrain the disk. Using the adapted spectroscopy method, we determine that log($T_{eff}$)$_{star}$ = 4.45 ± 0.01 dex and $v \sin i$ = 339 ± 10 km s$^{-1}$. This is in line with an early-type Be star. The high $v \sin i$ is a key indicator that the star has to be a rotating MS star as opposed to a BHeB star. An evolved star would break apart at such high $v \sin i$ (e.g., Rivinius et al. 2013). Furthermore, the scenario is supported by Leščinskaitė et al. (2022), whose photometric analysis shows that K4 is a strong Hα emitter, similar to K6 and K8.

**B4:** B4 is the third brightest star in the sample (F475W = 20.40 and F336W = 18.75). Its locations in the optical and UV CMDs are typical of an MS star. Its spectrum is shown in Figure 8. We list its EW measurements in Table 3 and comment only on a subset of those features.

The spectra shows the characteristics of a late O-type/early B-type star: He I, He II, Si II, Si III, and Si IV in absorption and emission. B4 shows weak emission in the wings of Hβ, Hγ, Hδ, and Hε absorption lines. Hα is in absorption with weak narrow emission at the central wavelength. Based on the EW and the schemes by Lennon (1997), Evans et al. (2015), and Castro et al. (2008), we classify this as a B0V type star.

We model the full spectrum of B4, but mask out pixels clearly affected by emission lines, as they are not captured by the TLUSTY models. Our spectroscopic fitting yields log($T_{eff}$) = 4.46 ± 0.01 dex and log($g$) = 3.75 ± 0.04 dex, which is consistent with an early B-type star. Using Equations (1)–(3), we derive $M_{spec}$ = 15 ± 2$M_\odot$ and $M_0$ = 15 ± 1$M_\odot$, and log($L/L_\odot$) = 4.65 ± 0.05 dex. Overall, this aligns well with the EW measurements and the CMD location.

The results from SED fitting are consistent with the spectroscopic parameters. The SED fitting yields log($T_{eff}$) = 4.50 ± 0.01 dex and log($g$) = 3.91 ± 0.01 dex. The photometric parameters are also consistent with a late-type O star. However, given the cooler log($T_{eff}$) from spectroscopy and the EW analysis from the spectra, we conclude the star is likely an early B-type star.

**K5:** K5 (F475W$_{VEGA}$ = 20.78 and F336W$_{VEGA}$ = 19.19), like the previous stars, is located on the BHeB sequence in the optical, but appears on the MS in the UV-optical CMD. Its spectrum is shown in Figure 8. We list its EW measurements in Table 3 and comment only on a subset of those features.

The spectral features suggest an early Be-type star: He I is in absorption and emission. Mg II and Si III are in absorption. It is possible that He II is in absorption, but the features are nearly indistinguishable from the continuum. We do not detect any other absorption features for any other elements. This is likely because the fast rotation of the star has washed out these features. For the Balmer lines, we find Hβ in emission, and we note that the star is not near any H II regions. Hγ shows emission within the absorption profile. We further note that the Hδ feature appears to have emission nearly canceling the absorption feature. Based on the presence of Si IV and the He I lines, using the classification scheme of Castro et al. (2008) we determine the star is likely an early Be2 V star. The Be-star scenario is supported by Leščinskaitė et al. (2022), who found that K5 is a strong Hα emitter, similar to K6 and K8.

Strong emission in the Balmer lines and the high reduced $\chi^2$ (>300) in the photometry lead us to adopt the adapted methods. We only have observations in four bands, making the composite photometric fitting not possible, since we cannot constrain the disk properties. Using the adapted spectroscopic method, we determine log($T_{eff}$)$_{star}$ = 4.46 ± 0.02 dex, and $v \sin i$ = 370 ± 90 km s$^{-1}$. This is consistent with an early-type Be star.

**B5:** B5 (F475W$_{VEGA}$ = 21.5 and F336W$_{VEGA}$ = 20.18) is located on the BHeB sequence in the optical CMD, but appears on the MS in the UV-optical CMD. Its spectrum is shown in Figure 8. We list its EW measurements in Table 3 and comment only on a subset of those features.

The spectrum shows the characteristics of a typical early-to-mid-type Be star. He I, Si II, and Mg II are in absorption Hα is in emission, while Hβ shows a strong emission component within its absorption feature. The remaining Balmer lines are all in absorption with weak emission features. We estimate the stellar type of Be3 V using its EWs (Castro et al. 2008).

Balmer line emission and the large BEAST reduced $\chi^2$ (>200) hint at the presence of a disk. Using the adapted spectroscopic method (Section 4.3), we determine $\log(T_{\rm eff}) = 4.40 \pm 0.01$ dex and $v \sin i = 165 \pm 15$ km s$^{-1}$. This is in line with an early-to-mid-type Be star. Using the composite photometric method, we determine that the star has a $\log(T_{\rm eff})_{\rm star} = 4.38^{+0.13}_{-0.04}$ dex and $\log(T_{\rm eff})_{\rm disk} = 3.74 \pm 0.17$, which is in line with the spectroscopic findings. Leščinskaitė et al. (2022) identified B5 as a strong H$\alpha$ emitter based on photometry.

**K7/B6:** K7/B6 (F336W$_{\rm VEGA}$ = 18.57, F475W$_{\rm VEGA}$ = 19.46) appears to be an O-type star based on its CMD location. Its spectrum is shown in Figure 8. We list its EW measurements in Table 3 and comment only on a subset of those features.

We observe both He I and He II in absorption. We further measure Mg II in absorption. Most Balmer lines appear in absorption; though the spectral region around H$\alpha$ does not show a clear Gaussian absorption feature. However, due to the low resolution ($R \sim 1250$), it is not clear if this is a data artifact or if there are other physical processes at work (e.g., stellar winds). Based on the EW ratios suggest that K7 is an O9.7 (Sota et al. 2011) or O9.5 (Martins 2018) type star, with a luminosity class of V (Martins 2018). Lastly, using Castro et al. (2008), we determined the star to be an O9.5 V star. This fits well with the star's position in the CMD (see Figure 1).

The median parameters from full spectral fitting for K7 are $\log(T_{\rm eff}) = 4.48 \pm 0.01$ dex and $\log(g) = 4.04 \pm 0.02$ dex (Table 4). Like the EW ratios, these values indicate the K7 is a late-type O star. The derived $\log(g)$ is consistent with K7 being an MS star. We find a $v \sin i = 97 \pm 20$ km s$^{-1}$. Using Equations (1)–(3), we derive $M_{\rm spec} = 13 \pm 3 M_\odot$ and $M_0 = 13 \pm 1 M_\odot$, and $\log(L/L_\odot) = 4.38 \pm 0.08$ dex. These values are in agreement with the classification derived from the EW measurements and its location on the CMD. The spectroscopic and photometric parameters for K7 are in broad agreement, although we observe an elevated $\log(g)$ value similarly to K1. The photometric parameters are $\log(T_{\rm eff}) = 4.52 \pm 0.02$ dex and $\log(g) = 4.23 \pm 0.05$ dex; both analyses are consistent with this being an O-type star. We find that $M_0$ and $M_{\rm ini}$, as well as $M_{\rm spec}$ and $M_{\rm act}$, are consistent with each other.

**B7:** B7 (F475W$_{\rm VEGA}$ = 21.35 and F336W$_{\rm VEGA}$ = 20.31) is among the bluest stars in the optical CMD, but in the UV, the star appears to be among the reddest stars. Its spectrum is shown in Figure 8. We list its EW measurements in Table 3 and comment only on a subset of those features.

The star shows emission resembling a P-Cygni profile in each of its Balmer lines. The clearest P-Cygni feature is the H$\beta$ line, since H$\alpha$ shows a strong emission feature. The central wavelength of H$\alpha$ falls in the blue wing of the emission feature hinting at a P-Cygni profile. H$\gamma$ and H$\delta$ on the other hand show narrow absorption features with very weak emission on the red side. We find He I, Mg II , Si II, Si III, Si IV, and C I are in absorption (see Table 3). We also find He I in emission, as well as NO III, O II, and C II. Based on the EW ratios and following Castro et al. (2008), we determined the star to be an early B1-type/Be1-type star.

The strong emission appears to be associated with the star, as opposed to being gaseous in origin. B7 is not near the known H II regions (see Figure 3). Due to the strong emission in the

Balmer lines and the high reduced $\chi^2$ ($\chi^2 > 300$) in the photometry, we adopt here the alternative fitting methods introduced in Section 4.3. B7 only has measurements in four bands (two UV and two optical), since the stars falls outside of the NIR footprint. This makes the composite photometry method not viable, since we cannot constrain the NIR to determine the nature of the star's SED profile.

Using this adapted spectroscopic method, we find $\log(T_{\rm eff}) = 4.44 \pm 0.02$ dex, and $v \sin i = 125 \pm 11$ km s$^{-1}$ (see Table 6).

**K8:** K8 (F475W$_{\rm VEGA}$ = 21.45 and F336W$_{\rm VEGA}$ = 19.88), like the previous star, is located on the BHeB sequence in the optical CMD, but appears on the MS in the UV-optical CMD. Its spectrum is shown in Figure 8. We list its EW measurements in Table 3 and comment only on a subset of those features.

The Keck spectrum of K8 is shown in Figure 8. The spectral features suggest an early-type fast rotating Be star. We find He I, Si III, and Si IV in absorption and He II in emission. The Balmer lines shows emission in H$\beta$. H$\gamma$ appears to show an emission feature within its absorption feature. Interestingly, the H$\delta$ feature is a very shallow absorption feature, and lastly, we find H$\epsilon$ in absorption. All spectral features are quite shallow, indicating that there is significant rotation. Based on the detected upper limits for both He II and Si, we type the star as an early Be2 V. Similarly to K6, Leščinskaitė et al. (2022) identified K8 as a strong H-$\alpha$ emitter. Due to the stars similarity to K6, both in spectroscopy and photometry, we use the described alternative method of fitting the data (see Section 4.3).

Using our adapted spectroscopic method, we determined that $\log(T_{\rm eff}) = 4.44 \pm 0.03$ dex and $v \sin i = 290 \pm 10$ km s$^{-1}$. This is consistent with an early-type Be star. Using the composite photometric method, we determined that $\log(T_{\rm eff})_{\rm star} = 4.51^{+0.04}_{-0.06}$ dex and $\log(T_{\rm eff})_{\rm disk} = 3.64^{+0.10}_{-0.04}$ dex, which is in line with the spectroscopic method.

**B8:** B8 is a faint star (F475W$_{\rm VEGA}$ = 21.5 and F336W$_{\rm VEGA}$ = 20.18) and is the reddest of those we observed with MMT. Its spectrum is shown in Figure 8. We list its EW measurements in Table 3 and comment only on a subset of those features.

The spectrum contains several unusual features. We observe only the upper limit for He I in B8, and we find Mg II and Si II in absorption.

H$\alpha$ and H$\beta$ are in emission and both show a double-peaked line profile. For the other Balmer lines with combined absorption and emission features, we observe a widened wing and narrow absorption within the emission lines. It is likely that emission in the wings is making line profiles shallower. Using the EWs, we classify the star as a Be3 V (Castro et al. 2008). Lastly, the distinct Balmer profiles suggest the star is most likely a shell star (i.e., a Be star viewed edge-on).

The unusual pattern of emission and absorption in the Balmer lines and the high $\chi^2$(>200) in the BEAST fitting suggest B8 may have a circumstellar disk.

Using the adapted spectroscopic method (Section 4.3), we determine that the star has a $\log(T_{\rm eff}) = 4.42 \pm 0.03$ dex and $v \sin i = 135 \pm 80$ km s$^{-1}$. This is in line with an early-to-mid-type Be star. Using the composite photometric method, we determine that $\log(T_{\rm eff})_{\rm star} = 4.37^{+0.15}_{-0.03}$ dex and $\log(T_{\rm eff})_{\rm disk} = 3.74^{+0.06}_{-0.10}$ dex, which is in line with our previous findings. Interestingly, we would expect the $v \sin i$ to be higher given that we are likely looking at the star equator-on based on the doubled peaked profile in Hα. However, if the star is really viewed edge-on, the $v \sin i$ value is surprisingly low. However, the uncertainty on our measurement is quite large, suggesting that the star could have a higher $v \sin i$ value more in line with expectation (e.g., Iqbal & Keller 2013; Rivinius et al. 2013; Arcos et al. 2018). Leščinskaitė et al. (2022) identified B8 as a strong Hα emitter. Despite the low $v \sin i$, the strong emission and spectral shape supports the classification of the star as a shell star.

B9: B9 is one of our fainter spectroscopic targets (F475W = 21.67, F336W = 20.25). Figure 1 shows that B9 is located between the MS and the BHeB sequences. Its spectrum is shown in Figure 8. We list its EW measurements in Table 3 and comment only on a subset of those features.

The spectrum shows the characteristics of an early B-type star. We find He I, Si II, Si III, and Mg II in absorption. The Hα appears in emission, while all other Balmer lines are in absorption. The Hα is marginally broader than what might be expected from purely gaseous emission, suggesting there may be a stellar contribution. The EW suggests that B9 is a B1.5 V in multiple classification schemes (Lennon 1997; Evans et al. 2015, and Castro et al. 2008).

We fit the full spectrum of B9 using The Payne and find $\log(T_{\mathrm{eff}}) = 4.45 \pm 0.02$ dex and $\log(g) = 4.18 \pm 0.08$ dex, and $v \sin i = 115 \pm 20$ km s$^{-1}$. Using Equations (1)–(3), we derive $M_{\mathrm{spec}} = 15 \pm 5 M_{\odot}$ and $M_0 = 12 \pm 2 M_{\odot}$, and $\log(L/L_{\odot}) = 4.26 \pm 0.04$, which are all in line with the rough classification described by the EW measurements and location in the CMD. The BEAST-based photometric parameters are $\log(T_{\mathrm{eff}}) = 4.49 \pm 0.05$ dex, $\log(g) = 4.12 \pm 0.68$ dex, and $M_{\mathrm{ini}} = 10.87 \pm 6.13 M_{\odot}$. The parameters are consistent with the spectroscopic measurement.

B10: B10 (F475W$_{\mathrm{VEGA}}$ = 21.7 and F336W$_{\mathrm{VEGA}}$ = 20.35) is the faintest star in our sample in the UV-optical CMD, but is brighter than several other stars in the optical-only CMD. Its spectrum is shown in Figure 8. We list its EW measurements in Table 3 and comment only on a subset of those features.

The spectrum shows the characteristics of an early B-type star with He I, Si III, and Si IV in absorption. Hα is in strong emission, while Hβ and Hδ show modest emission within their absorption feature. Hγ and Hε are both in absorption. B10 is not near any known H II region, suggesting the emission is not likely due to nebular contamination. Based on the observed EWs, we classify it as a B0.5 V (candidate Be) star.

We fit the full spectrum of B10 using The Payne and find $\log(T_{\mathrm{eff}}) = 4.43 \pm 0.02$ dex, $\log(g) = 3.92 \pm 0.11$ dex, and $v \sin i = 141 \pm 25$ km s$^{-1}$. Using Equations (1)–(3), we derive $M_{\mathrm{spec}} = 11 \pm 6 M_{\odot}$ and $M_0 = 11 \pm 1 M_{\odot}$, and $\log(L/L_{\odot}) = 4.24 \pm 0.04$ dex. These all are consistent with the EW-based spectroscopic type and location on the CMDs.

The photometric parameters yield $\log(T_{\mathrm{eff}}) = 4.47 \pm 0.03$ dex, $\log(g) = 4.02 \pm 0.07$ dex, and $M_0 = 12.2 \pm 1.21 M_{\odot}$.

B11: B11 (F475W$_{\mathrm{VEGA}}$ = 21.90 and F336W$_{\mathrm{VEGA}}$ = 20.25) is the second faintest star in the sample. Both in the optical-only and the UV-optical sample, it lies on the MS. Its spectrum is shown in Figure 8. We list its EW measurements in Table 3 and comment only on a subset of those features.

The spectrum shows the characteristics of an early B-type star. We find He I, Si II, Si III, and Si IV in absorption. Most Balmer lines are in absorption with variable amounts of emission. For Hα, the strong emission nearly cancels out the absorption feature. Hβ and Hδ show nearly no emission in their absorption features. Although Hγ and Hε are predominantly in absorption, there is some emission in the blue wings. Taking the EWs, including upper limits at face value, we determine that the spectral type is a B2 type star (Castro et al. 2008; Evans et al. 2015).

Since B11 spectrum shows an S/N of <10, we cannot reliably recover the spectroscopic parameters. The photometric parameters are $\log(T_{\mathrm{eff}}) = 4.27 \pm 0.04$ dex, $\log(g) = 3.81 \pm 0.06$ dex, and $M_{\mathrm{ini}} = 6.26 \pm 0.69 M_{\odot}$. The photometry is consistent with the star's position in the CMDs as well as the spectral type derived from the EW.

Given that the S/N is ∼10, we refrain from making any further speculations as to the origin of the emission.

B12: B12 is among the faintest stars in our sample (F475W$_{\mathrm{VEGA}}$ = 21.75 and F336W$_{\mathrm{VEGA}}$ = 20.19). Its locations in the optical and UV CMDs are typical of an MS star. Its spectrum is shown in Figure 8. We list its EW measurements in Table 3 and comment only on a subset of those features.

The spectrum shows the characteristics of an early B-type star. He I, Si III, and Si IV are detected in absorption. The Balmer lines show absorption features with emission for Hα, Hβ, and Hγ. Overall, the emission within the star appears to be predominantly stellar, since the star is not near any know gas region. The emission feature overall appear quite weak in comparison to the other stars with emission in this sample. Lastly, based the He EW ratios, the presence of Si IV, and absence of Mg II, we use the classification scheme of Castro et al. (2008) to conclude that the star is likely a B1.5V star.

We fit the full spectrum of B12 using The Payne. The parameters for the spectral fitting are listed in Table 4. For B12, we find $\log(T_{\mathrm{eff}}) = 4.44 \pm 0.05$ dex and $\log(g) = 4.00 \pm 0.10$ dex. Using Equations (1)–(3), we derive $M_{\mathrm{spec}} = 8 \pm 4 M_{\odot}$ and $M_0 = 11 \pm 2 M_{\odot}$, and $\log(L/L_{\odot}) = 4.07 \pm 0.05$ dex, which is all in line with the rough classification described by the EW measurements and location in the CMD. The photometric parameters yield $\log(T_{\mathrm{eff}}) = 4.34 \pm 0.05$ dex, $\log(g) = 3.86 \pm 0.08$ dex, and $M_{\mathrm{ini}} = 6.61 \pm 1.21 M_{\odot}$.

B13: B13 (F336W$_{\mathrm{VEGA}}$ = 20.25 and F475W$_{\mathrm{VEGA}}$ = 21.9) is among the faintest stars in the optical-CMD, but is much brighter than several other stars in the UV-optical CMD. Its location on the UV and optical CMDs indicates it is likely a late-type O-type. Its spectrum is shown in Figure 8. We list its EW measurements in Table 3 and comment only on a subset of those features.

The spectrum shows the characteristics of a typical late-type O-star: He I and He II are in absorption. Most Balmer lines appear in absorption with light contamination, except $H\alpha$, which is fully in emission. The low S/N (S/N∼ 12) makes it challenging to definitely classify this star. It is likely that B13 is a late O-type MS star. However, it may also be consistent with an early B-type star; we cannot rule this out. Based on the scheme applied thus far for EW, we determined the star is an O9.7-B0 dwarf.

From the full spectral fitting of B13, we find $\log(T_{\mathrm{eff}}) = 4.50 \pm 0.02$ dex, and $\log(g) = 4.64 \pm 0.06$ dex. Using Equations (1)–(3), we derive $M_{\mathrm{spec}} = 19 \pm 4 M_{\odot}$ and $M_0 = 12 \pm 1 M_{\odot}$, and $\log(L/L_{\odot}) = 4.07 \pm 0.08$. $M_0$ and $M_{\mathrm{ini}}$ are consistent, but $M_{\mathrm{spec}}$ appears to be too large; once again, this is likely due to the high $\log(g)$ value. The emission in the Balmer lines might affect the $\log(g)$, since the parameter is the most sensitive to it. Nevertheless, these values are in line with the rough classification described by the EW measurements and location in the CMD.

SED fitting yields $\log(T_{\mathrm{eff}}) = 4.50 \pm 0.01$ dex and $\log(g) = 4.37 \pm 0.02$ dex. Similarly to other O-stars, we find consistency in $\log(T_{\mathrm{eff}})$ and a slight discrepancy in $\log(g)$. $M_0$ and $M_{\mathrm{ini}}$ are consistent with one another.

Given that the star shows narrow emission features, poor spectroscopic fitting, and extended emission in the 2D spectra, we postulate that the star must have some nebular contamination.

B14: B14 (F475W$_{\mathrm{VEGA}}$ = 21.9 and F336W$_{\mathrm{VEGA}}$ = 20.40) is the faintest star in our sample in the optical-only CMD, but is brighter than several stars in the UV-optical CMD. In both cases, it lies on the MS. Its spectrum is shown in Figure 8. We list its EW measurements in Table 3 and comment only on a subset of those features.

The spectrum shows the characteristics of an early-to-mid B-type star, with emission in that He I features showing both absorption and emission. We also find Mg II, Si II, and Si III in absorption and [O II] in emission. B14 shows that Hα is primarily present in absorption. Furthermore, we find that Hβ, Hγ, and Hδ all show some potential emission in their wings. However, due to the low S/N (S/N∼ 15) of the spectrum, the weak emission in the Balmer lines could be artifacts of the data. Based on the classification scheme previously employed for EWs, we determine the star is B3 star.

We fit the full Keck spectrum of B14 using The Payne and find $\log(T_{\mathrm{eff}}) = 4.39 \pm 0.02$ dex, $\log(g) = 4.41 \pm 0.08$ dex, and $v\sin i = 80 \pm 15$ km s$^{-1}$. Using Equations (1)–(3), we derive $M_{\mathrm{spec}} = 24 \pm 5 M_\odot$ and $M_0 = 8 \pm 1 M_\odot$, and $\log(L/L_\odot) = 3.87 \pm 0.08$ dex. The spectroscopic $\log(g)$ value is likely too high, since it yields a high $M_{\mathrm{spec}}$ that is unreasonable for the star's position in the CMD (Figure 2). In comparison, the photometric parameters are $\log(T_{\mathrm{eff}}) = 4.42 \pm 0.06$ dex, $\log(g) = 4.12 \pm 0.06$ dex, and $M_{\mathrm{ini}} = 9.22 \pm 0.99 M_\odot$. $\log(T_{\mathrm{eff}})$ is consistent between the two methods, while $\log(g)$ shows a discrepancy.

## ORCID iDs

Maude Gull  https://orcid.org/0000-0003-3747-1394
Daniel R. Weisz  https://orcid.org/0000-0002-6442-6030
Peter Senchyna  https://orcid.org/0000-0002-9132-6561
Nathan R. Sandford  https://orcid.org/0000-0002-7393-3595
Yumi Choi  https://orcid.org/0000-0003-1680-1884
Ylva Götberg  https://orcid.org/0000-0002-6960-6911
Karoline M. Gilbert  https://orcid.org/0000-0003-0394-8377
Martha Boyer  https://orcid.org/0000-0003-4850-9589
Julianne J. Dalcanton  https://orcid.org/0000-0002-1264-2006
Puragra GuhaThakurta  https://orcid.org/0000-0001-8867-4234
Steven Goldman  https://orcid.org/0000-0002-8937-3844
Paola Marigo  https://orcid.org/0000-0002-9137-0773
Kristen B. W. McQuinn  https://orcid.org/0000-0001-5538-2614
Evan Skillman  https://orcid.org/0000-0003-0605-8732
Yuan-sen Ting  https://orcid.org/0000-0001-5082-9536
Benjamin F. Williams  https://orcid.org/0000-0002-7502-0597